\documentclass[
superscriptaddress,
reprint,
 amsmath,amssymb,
 aps,
 prl,
]{revtex4-1}

\usepackage[normalem]{ulem}
\usepackage{hyperref,xcolor}

\usepackage{physics}
\usepackage{graphicx}
\usepackage{dcolumn}
\usepackage{bm}
\usepackage{amsthm}

\newtheorem{theorem}{Theorem}
\newtheorem{prop}{Proposition}
\newtheorem{lemma}{Lemma}
\newtheorem{corollary}{Corollary}

\newcommand{\R}{\mathbb{R}}

\newcommand{\set}[1]{\mathsf{#1}}

\newcommand{\spc}[1]{\mathcal{#1}}

\def\d{{\rm d}}

\newcommand{\Supp}{\mathsf{Supp}}
\def\>{\rangle}
\def\<{\langle}

\newcommand{\st}[1]{\mathbf{#1}}
\newcommand{\bs}[1]{\boldsymbol{#1}}     

\newcommand{\map}[1]{\mathcal{#1}}

\newcommand{\St}{{\mathsf{St}}}

\begin{document}

\preprint{Draft Ver 8.2}

\title{The nonequilibrium cost of accurate information processing}
\author{Giulio Chiribella}
\email{giulio@cs.hku.hk}
\affiliation{QICI Quantum Information and Computation Initiative, Department of Computer Science, The University of Hong Kong, Pokfulam Road, Hong Kong SAR, China}
\affiliation{Department of Computer Science, University of Oxford, Parks Road, Oxford, UK}
\affiliation{Perimeter Institute for Theoretical Physics, 31 Caroline St North, Waterloo, ON N2L 2Y5, Canada}
\author{Fei Meng}
\email{fmeng@cs.hku.hk}
\affiliation{QICI Quantum Information and Computation Initiative, Department of Computer Science, The University of Hong Kong, Pokfulam Road, Hong Kong SAR, China}
\affiliation{Department of Physics, Southern University of Science and Technology, Shenzhen 518055, China}
\author{Renato Renner}
\email{renner@eth.ch}
\affiliation{Institute for Theoretical Physics, ETH Z\"urich}
\author{Man-Hong Yung}
\email{yung@sustech.edu.cn}
\affiliation{Department of Physics, Southern University of Science and Technology, Shenzhen 518055, China}
\affiliation{Shenzhen Key Laboratory of Quantum Science and Engineering, Shenzhen 518055, China}

\begin{abstract}

Accurate   information processing  is crucial  both in technology and in nature.   
To achieve it, any information processing  system  needs  an initial  supply of resources away from  thermal equilibrium. 
Here we establish a fundamental   limit  on the  accuracy achievable with  a given amount of nonequilibrium resources. 
The limit   applies to arbitrary information processing tasks and  arbitrary information processing systems subject to the laws of quantum mechanics. It  is easily computable and is expressed in terms of an entropic quantity, which we name  reverse entropy, associated to a time reversal of the information processing task under consideration.
The limit  is achievable for all deterministic classical computations and for all their quantum extensions. As an application, we establish the optimal tradeoff between nonequilibrium and accuracy for the fundamental tasks of storing, transmitting, cloning, and erasing information. 
Our results set a target for the design of new  devices approaching the ultimate efficiency limit,  and provide a framework for demonstrating  thermodynamical advantages of quantum devices over their classical counterparts.
  \end{abstract}

\maketitle

\section{Introduction}  

Many processes in nature depend on accurate processing of information.  For example, the development  of complex  organisms relies on the accurate replication of the information contained in their DNA, which  takes place with an error rate   estimated to be  less than one basis  per billion  \cite{mcculloch2008fidelity}.     

At the fundamental level,  information is stored into patterns that  stand out  from the thermal fluctuations of the  surrounding environment~\cite{wang2011self,england2013statistical}.
In order to achieve   deviations from thermal  equilibrium,  any information processing machine  needs  an initial  supply of  systems  in a non-thermal state~\cite{andrieux2008nonequilibrium,jarzynski2008thermodynamics}. 
For example,  
an ideal copy machine  for classical data requires  at least  a clean bit for every bit it copies~\cite{bennett1982thermodynamics,landauer1961irreversibility,leff2014maxwell}.
 For a general information processing task,  a fundamental  question is:  what is the minimum amount of nonequilibrium needed to achieve a target level of accuracy?   This question is especially prominent at the quantum scale, where many tasks cannot be  achieved  perfectly even in principle, as illustrated by the  no-cloning theorem~\cite{
wootters1982single,dieks1982communication}.

In recent years,  there has been a growing interest in the interplay between quantum information  and thermodynamics~\cite{goold2016role,vinjanampathy2016quantum,binder2018thermodynamics}, motivated both by fundamental questions~\cite{lloyd2000ultimate,sagawa2009minimal,linden2010small,parrondo2015thermodynamics,goold2015nonequilibrium} and by the experimental realisation  of new quantum devices~\cite{baugh2005experimental,toyabe2010experimental,vidrighin2016photonic}.   Research in this area led to   the development of resource-theoretic frameworks that can be used to study thermodynamics  beyond the macroscopic limit~\cite{janzing2000thermodynamic,horodecki2003reversible,brandao2013resource,horodecki2013fundamental,brandao2015second,brandao2015reversible,faist2015gibbs,gour2015resource,gour2018quantum}.  These frameworks have been   applied to characterise  thermodynamically allowed state transitions, to evaluate the work cost of logical operations~\cite{faist2015minimal,faist2018fundamental} and to study information erasure and work extraction in the quantum regime~\cite{del2011thermodynamic,aaberg2013truly,skrzypczyk2014work}.     {  From a different perspective, relations between accuracy and entropy production have been investigated in the field of stochastic thermodynamics \cite{seifert2018stochastic,barato2015thermodynamic,gingrich2016dissipation,barato2016cost,horowitz2020thermodynamic},   referring to specific  physical models   such as classical Markovian systems in nonequilibrium steady states.}



Here, we establish   a fundamental tradeoff  between accuracy and nonequilibrium, valid at the quantum scale and applicable  to arbitrary information processing tasks. 
 The main result is a limit on the accuracy, expressed in terms of an entropic quantity, which we call  {   the reverse entropy, associated to a time reversal of the  information processing task under consideration.  The limit  is attainable  in a broad class of tasks, including all deterministic classical  computations and all quantum extensions thereof.   For the task of erasing  quantum information, our limit  provides, as a byproduct, the ultimate accuracy achievable with a given amount of work.  
  For the tasks of  {   storage, transmission, and cloning of quantum information}, our results  reveal a thermodynamic advantage of   quantum setups over all classical setups  that measure the input and generate their output based only on the measurement outcomes.
 In the cases of storage and transmission, we show that  quantum machines  can break  the ultimate classical limit  on the amount of work required to achieve a desired level of accuracy.   This result
enables the demonstration of  work-efficient quantum memories and quantum communication systems outperforming all possible  classical setups. 

Our results establish a direct link between thermodynamic resources and the accuracy of information processing. They set an ideal  target for the design of new  devices,  and provide a framework for demonstrating a thermodynamic advantage of quantum devices in  fundamental tasks such as storing, copying, and transmitting information.}


\nopagebreak 

\section{Results}
{\bf   The nonequilibrium cost of accuracy.}
At the most basic level, the goal of information processing is to set up a desired relation between  an input and an output. 
For example, a deterministic classical  computation amounts to transforming a bit string  $x$ into another bit string  $  f(x)$, where $f$ is a given function.  
In the quantum domain, information processing tasks are often    associated to ideal state transformations  $\rho_x\mapsto  \rho_x'$, in which an input state described by a density operator $\rho_x$  has to be converted into a target output state described by another density operator $\rho_x'$, where $x$ is  a parameter in some given set $\set X$.    

Since every realistic  machine is  subject to imperfections, the physical realisations of an ideal  information processing task can have varying levels of accuracy.   Operationally, the accuracy can be quantified by performing a test on the output of the machine and by assigning a score to the outcomes of the measurement.   The resulting measure of accuracy is given by the expectation value of a suitable observable $O_x$, used to assess the closeness of the output to the target state $\rho_x'$. 
In the worst case over all possible inputs, the accuracy  achieved in a given task $\map T$ has the expression $\map F_{\map T} (\map M)  =  \min_{x}      \Tr[O_x  \map M (\rho_x)]$, where $\map M$ is the quantum channel   {   (completely positive trace-preserving map)} describing  the action of the machine.   {   Here,  the dependence of the input states $\rho_x$ and output observables $O_x$ on the parameter $x$ is fully general, and  includes in particular   cases where multiple observables are tested for the same input state.  The range of    values for the function $\map F_{\map T}$  depends on the choice of observables $O_x$: for example, if all the observables $O_x$ are projectors, the range of $\map F_{\map T}$   will be included in the interval $[0,1]$.   }

    \begin{figure}[t!]
	\includegraphics[width=0.5\textwidth]{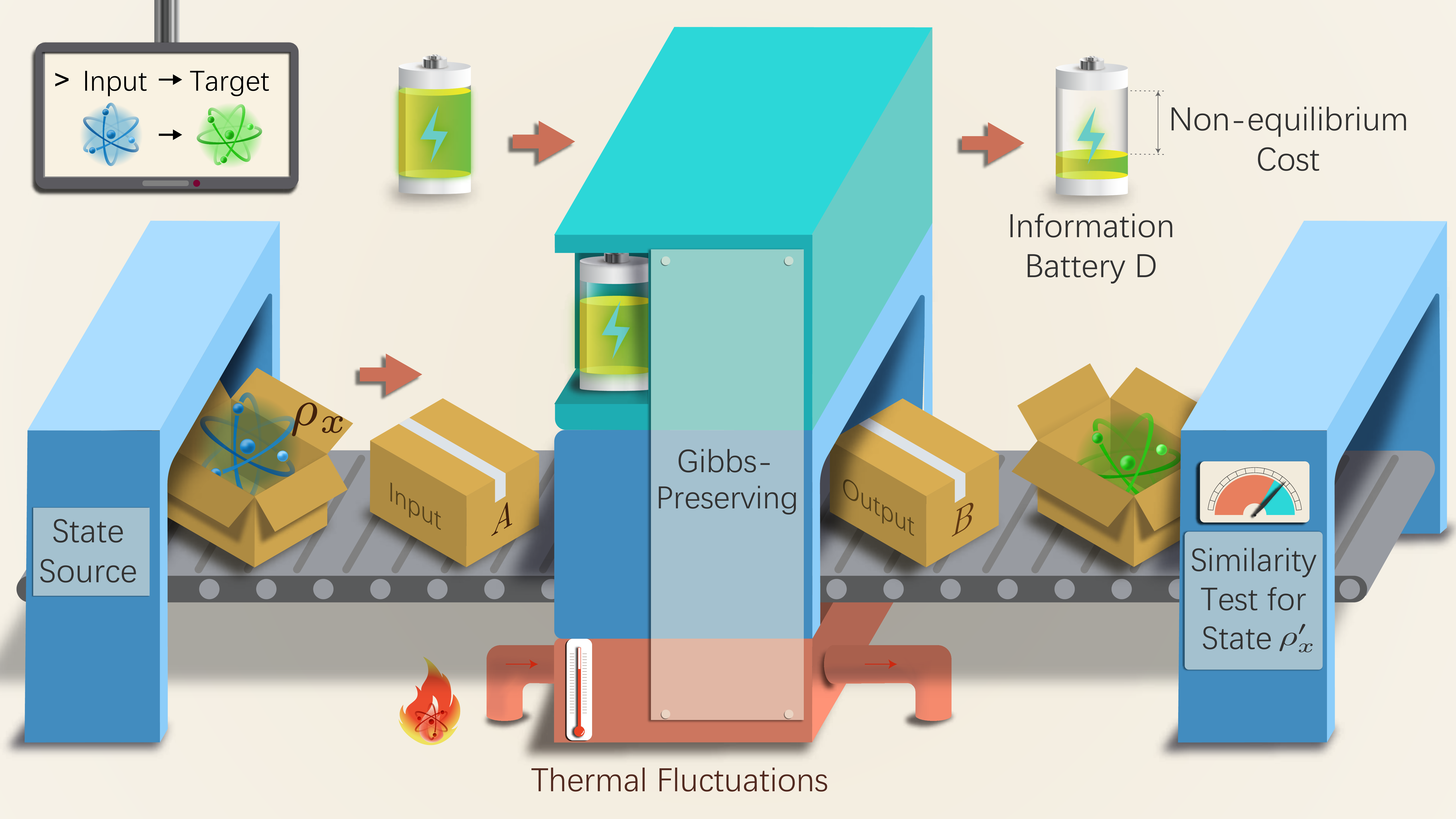}
	\caption{\textbf{The nonequilibrium cost of accuracy.}    A source generates a set of input states  for an information processing machine.  The machine  uses  an information battery (a supply of qubits initialised in a fixed pure state) and thermal fluctuations (a reservoir in the Gibbs state)  to transform  the input state $\rho_x$ into an approximation of the ideal target state $\rho_x'$.    Finally, the similarity between the output  and the target states is  assessed by a measurement.    The number of pure qubits consumed by the machine is the nonequilibrium cost that needs to be paid in order to achieve the desired level of accuracy.   }
	\label{fig:mod}
\end{figure}

Accurate information processing generally requires  an initial supply of systems away from equilibrium.    The amount of nonequilibrium required to implement a given task  can be rigorously quantified  in  a resource theoretic framework 
 where Gibbs states are regarded as freely available, and the only operations that can be performed free of cost are those that transform Gibbs states into Gibbs states~\cite{faist2015gibbs,faist2018fundamental}. These operations, known as {Gibbs preserving}, are  the largest class of processes  that maintain the condition of thermal equilibrium. 
   The initial nonequilibrium resources can be represented  in a canonical form  by introducing an  {information battery}~\cite{faist2015minimal,faist2018fundamental}, consisting of an array of qubits with degenerate energy levels.  
    The battery starts off with  some  qubits in a pure state (hereafter called the ``clean qubits''),  while  all the remaining  qubits are in the maximally mixed state. 
    To implement the desired information processing task, the machine will operate jointly on the input system and on the information battery, as illustrated in Figure~\ref{fig:mod}. 
    
The number of clean qubits required by a machine  is  an important measure of efficiency,  hereafter called the nonequilibrium cost. {   For a given quantum channel $\map M$,  the minimum nonequilibrium cost of any machine implementing channel $\map M$ (or some approximation thereof) has been evaluated in Refs. \cite{faist2015minimal,faist2018fundamental}. 
  Many information processing tasks, however, are not uniquely associated to  a specific quantum channel: for example, most  state transitions $\rho \mapsto \rho'$ can  be implemented by  infinitely many different quantum channels, which generally have different costs. When a  task can be implemented perfectly by more than one quantum channel, the existing results do not identify, in general, the minimum nonequilibrium cost that has to be paid for a desired level of accuracy.  
  Furthermore, there also exist  information processing tasks, such as quantum cloning \cite{wootters1982single,dieks1982communication}, that  cannot be perfectly achieved by any quantum channel.  In these scenarios, it is important to establish a  direct relation  between the accuracy achieved in the given task   and the minimum cost that has to be paid for that level of accuracy.   
Such a relation  would  provide a direct bridge between thermodynamics and abstract information processing, establishing a fundamental efficiency limit  valid for all machines allowed by quantum mechanics.   

In this paper we build concepts and methods for determining the nonequilibrium cost of accuracy  in a way that depends only on the information processing task under consideration, and not on a specific quantum channel. 
 Let us denote by $c  (  \map M,  \Pi_A)$   the  nonequilibrium cost   required for implementing a given channel $\map M$  on input states in the subspace specified by a projector $\Pi_A$.     We then define the nonequilibrium cost for achieving accuracy $F$ in a task $\map T$ as   $c_{\map T}  (F)   :=   \min \{  c(\map M,\Pi_A)~|~  \map F_{\map T} (\map M) \ge F\}$. Note that the  the specification   of the input subspace is included in  the task $\map T$.    In the following we  focus on tasks where the input subspace is invariant under time evolution,  namely $[\Pi_A,  H_A]=0$, where $H_A$ is the  Hamiltonian of the input system. Our main goal will be to evaluate $c_{\map T}  (F)$, the  nonequilibrium cost of accuracy. }

\medskip

  In Methods, we  provide an exact expression for  $c_{\map T}  (F)$. 
  The expression involves a semidefinite program, which  can be   solved numerically for low dimensional systems, thus providing the exact tradeoff between nonequilibrium and accuracy.    Still,   brute-force optimisation is intractable for high dimensional systems.   For this reason, it is crucial to have a computable bound that can be applied in a broader range of situations.    The central result of the paper is a  universal bound, valid for  all quantum systems and to all information processing tasks: the bound reads
\begin{align}\label{main}
c_{\map T}   (F) &  \ge   \kappa_{\map T}  +  \log  F \, ,
\end{align}
{  where  $\kappa_{\map T}:=  -\log  F_{\max}^{\map T_{\rm rev}}$ is an entropic quantity, hereafter called the reverse entropy,  and  $F^{\map T_{\rm rev}}_{\max}$ is the maximum accuracy allowed by quantum mechanics to a time-reversed information processing task  $\map T_{\rm rev}$, precisely defined in the following section   (see  Supplementary Note 1 for the derivation of Eq. (\ref{main})).   Note that the reverse entropy is a monotonically decreasing function of $F^{\map T_{\rm rev}}_{\max}$,  and becomes zero when the  time-reversed task  can be  implemented with unit accuracy. }

Eq.~{(\ref{main})}  can  be equivalently formulated as a limit on the accuracy attainable with a given budget of nonequilibrium resources:  for a given number of clean qubits {  $c$, the maximum achievable accuracy in the task $\map T$, denoted by $F_{\map T} (c)  :=  \max\{  \map F_{\map T}  (\map M)~|~   c(\map M,  \Pi_A) \le c \}$, satisfies the bound}
\begin{equation}\label{main_f}
	F_{\map T} (c)\leq     2^{       {c}   -   \kappa_{\map T}  }\, .
\end{equation}
This bound represents  an in-principle limit  on the performance of every information processing machine.
The bounds  (\ref{main}) and (\ref{main_f}) are achievable in a number of tasks, and have profound implications that will be discussed in the following sections.

\medskip 

{  {\bf   Time-reversed   tasks and reverse entropy.}      Here we discuss the notion of time reversal of an information processing task.  
Let us  start from  the simplest scenario, involving transformations of a fully degenerate  system into itself.   For  a state transformation task $\rho_x\mapsto \rho_x'$, we consider without loss of generality an   accuracy measure    where the    observables $O_x$ are positive operators, proportional to quantum states.     We then define   a time-reversed task  $\map T_{\rm rev}$, where the role of the input states $\rho_x$ and of the output observables $O_x$ are exchanged. The accuracy of a generic channel $\map M$ in the execution of the time-reversed task is specified by the reverse accuracy  $\map F_{\map T_{\rm rev}}  (\map M)   :  =  \min_x  \Tr  [ \rho_x   \map M   (O_x)]$.  Maximising over all possible channels, we  obtain  $F^{\map T_{\rm rev}}_{\max}$ and define  $\kappa_{\map T}  =  -\log F^{\map T_{\rm rev}}_{\max}$.   

For systems with non-trivial energy spectrum, we define the time-reversed task in terms of a  time reversal of   quantum operations introduced by Crooks  \cite{crooks2008quantum} and recently generalised in \cite{chiribella2021symmetries}.  In the Gibbs preserving context, this time reversal exchanges states with observables, mapping Gibbs states into  trivial observables (described by the identity matrix) and {\em vice-versa}. More generally,  the time reversal maps  the states $\rho_x$ into the observables  $\widetilde O_x :  =  \Gamma_A^{-\frac 12}    \rho_x   \Gamma_A^{-\frac 12}$  and the observables $O_x$ into the (unnormalised) states $\widetilde \rho_x: =   \Gamma_B^{\frac 12}    O_x   \Gamma_B^{\frac 12}$, where  $\Gamma_A$ and $\Gamma_B$ are the Gibbs states of the input and output systems, respectively. 
      The reverse accuracy of a channel $\map M$ is then defined as   $\map F_{\map T_{\rm rev}}(\map M)  :=  \min_x  \Tr  [  \widetilde O_x  \map M  (\widetilde \rho_x)]$. 
In   the Methods section,  we show that the reverse entropy can be equivalently written as }
  \begin{align}\label{thermocomp}
\kappa_{\map T}    = \max_{\st p}     H_{\min}  (A| B)_{\omega_{\map T, \st p}}  \, .
\end{align} 
where   $\st p  =  (p_x)_{x\in\set X}$ is a probability distribution, $\omega_{\map T, \st p}=  \sum_x  \,  p_x\,     \Gamma_A^{-\frac 12}   \rho_x^T    \Gamma_A^{-1/2} \otimes    \Gamma_B^{\frac 12}  O_x   \Gamma_B^{\frac 12} $ is an operator acting on the tensor product of the input and output systems,    $\rho_x^T$ is the transpose of the density matrix  $\rho_x$ with respect to the energy eigenbasis,    
  and $H_{\min}(A|B)_{\omega_{\map T, \st p}} :   =       -  \log   \min\{    \Tr[\Lambda_B]   ~|~      (I_A \otimes  \Lambda_B)  \ge   \omega_{\map T, \st p}   \}$ 
 is the conditional min-entropy~\cite{renner2004smooth,datta2009smooth,konig2009operational}.

 Crucially, the reverse entropy  depends only on the task under consideration, and not on a  specific quantum channel used to implement the task.   In fact, the reverse entropy  is well defined even for tasks that cannot be perfectly achieved  by any quantum channel,    as in the case of ideal quantum cloning, and even for tasks that are not formulated in terms of state transitions (see Methods). 

To gain a better understanding of the reverse entropy, it is useful    to evaluate  it in some special cases. Consider the case of a classical deterministic computation, corresponding to the evaluation of a  function $y = f(x)$.   In this case  the reverse entropy is   
\begin{align}\label{gibbsc}
\kappa_f  =   D_{\max}  (p_f  \|  g_B)
\end{align}
where   $D_{\max}  (p  \|  q)  {  =  \max_y   \,  p(y)/q(y)}$ is the max R\'enyi divergence between two probability distributions $p(y)$ and $q(y)$~\cite{renyi1961measures},  $g_B (y)$ is Gibbs distribution for the output system, and   $p_f(y)$ is the probability distribution of the random variable $y=  f(x)$, when $x$  is sampled from the Gibbs distribution (see Supplementary Note 2 for the derivation).  Eq.~(\ref{gibbsc}) shows that the reverse entropy of a classical computation is a measure of how much the computation transforms  thermal fluctuations into states  that deviate from thermal equilibrium.

{  In the quantum case, however, physical limits to the execution of the time-reversed task  can arise  even without any deviation from thermal equilibrium.} 
 Consider for example the transposition task $\rho_x  \mapsto \rho_x^T$~\cite{bu2000universal,horodecki2003limits,buscemi2003optimal,ricci2004teleportation,de2004contextual,lim2011experimental}, where $x$ parametrises all the possible pure states of a quantum system. 
  This transformation  does not generate any deviation from equilibrium as it  maps Gibbs states into Gibbs states. 
{   On the other hand, in the fully degenerate case the time-reversed task  is still transposition, and perfect transposition is forbidden by the laws of quantum mechanics \cite{bu2000universal,horodecki2003limits,buscemi2003optimal,ricci2004teleportation,de2004contextual,lim2011experimental}.   The maximum fidelity of an approximate transposition is $F_{\rm trans}   =  2/(d+1)$ for $d$-dimensional quantum systems, and therefore  $\kappa_{\rm trans}   =   \log [(d+1)/2]$. }

\medskip 

{\bf Condition for achieving  the limit.}   The appeal of the bounds (\ref{main}) and (\ref{main_f}) is that they are general and easy to use.  But  are they attainable?   To discuss  their attainability,  it is  important to first  identify the parameter range in which these bounds  are meaningful.    First of all,  the bound~(\ref{main}) is only meaningful when the desired accuracy does not exceed  the maximum accuracy  $F_{\max}$ allowed  by the laws of physics for the task $\map T$.  Similarly,  the bound~(\ref{main_f})  is  only meaningful if the initial amount of nonequilibrium resources does not go below  the  smallest nonequilibrium cost of  an arbitrary process acting on the given input subspace,  hereafter denoted by $c_{\min}$.  By maximising the accuracy over all quantum channels with minimum cost $c_{\min}$,  we then obtain a minimum value $F_{\min}$ below which reducing the accuracy does not result in any reduction of the nonequilibrium cost.

We now provide a criterion that guarantees the attainability of the bounds~(\ref{main}) and~(\ref{main_f})  in the full interval $[F_{\rm min}, F_{\max}]$. Since the two bounds are equivalent to one another, we will focus on bound~(\ref{main}). The condition for attainability  in the full interval $[F_{\rm min}, F_{\max}]$ is attainability  at the maximum value $F_{\max}$.  As we will see in the rest of the paper, this condition is satisfied by a  number of information processing tasks, notably including all classical computations and all quantum extensions thereof. 

\smallskip
\begin{theorem}\label{theo:achievability}
For every information processing task $\map T$ with $[  \Pi_A, H_A]=  0$, if the bound~(\ref{main}) is attainable for a value of the accuracy $F_0$, then   it is attainable for every value of the accuracy in the interval $[F_{\rm min}, F_0]$, with $F_{\rm min}  = 2^{ c_{\min}- \kappa_{\map T}}$. 
In particular, if the bound is attainable for the maximum accuracy $F_{\max}$,  then it is attainable for every  value of the accuracy in the interval $[F_{\rm min}, F_{\max}]$.  

\end{theorem}
In Supplementary Note 3 we prove the theorem by explicitly constructing a family of channels 
 that achieve the bound~(\ref{main}).

By evaluating the nonequilibrium cost of specific quantum channels,  one can  prove the attainability of the bound~(\ref{main}) for a variety of different tasks. For example, the bound~(\ref{main})  is attainable for every deterministic classical computation.    Moreover, it is achievable for every quantum extension of a classical computation:
on Supplementary Note 4 we show that 
for every value of the accuracy, the nonequilibrium cost is the same for the original classical computation and for its quantum extension, and therefore the achievability condition holds in both cases.

{  The  nonequilibrium cost  $c_{\map T}  (F)$ provides a fundamental lower bound to the amount of work that has to be invested in order to achieve accuracy $F$.    Indeed, the minimum work cost of a  specific channel $\map M$, denoted by $W (\map M,\Pi_A) $  can be  quantified by  the minimum number of clean qubits needed to implement the process  in a scheme like the one in Figure \ref{fig:mod}, with the only difference that Gibbs preserving operations are replaced by thermal operations, that is,  operations resulting from  a joint energy-preserving evolution  of the system together with auxiliary systems in the Gibbs state \cite{horodecki2013fundamental,brandao2013resource}.     Since  thermal operations are a proper subset of the Gibbs preserving operations~\cite{faist2015gibbs},  the restriction to thermal operations generally results into a larger number of clean qubits, and  the work cost  is lower bounded as  $W (\map M,\Pi_A)  \ge  k T  (\ln 2)  \,  c  (\map M,\Pi_A)$, where $k$ is the Boltzmann constant and $T$ is the temperature.  By minimising both sides over all channels that achieve accuracy $F$, we then get the bound $W_{\map T}  (F)  \ge kT (\ln 2)   \,  c_{\map T}  (F)$, where  $W_{\map T}  (F)  :  =  \min \{  W(\map M,  \Pi_A)  ~|~ F_{\map T}  (\map M) \ge F \}  $ is the  minimum work cost that has to be paid in order to reach accuracy $F$.

 The achievability of this bound is generally nontrivial, except for operations on  fully degenerate classical systems, wherein  the sets of thermal operations and Gibbs preserving maps coincide due to Birkhoff's theorem \cite{birkhoff1946tres}.  Another example is the task of erasing quantum states, corresponding to the state transformation $\rho_x  \mapsto |0\>\<0|$, where $\rho_x$ is an arbitrary state and $|0\>$ is the ground state.     In Supplementary Note 4, we show that  the bound $W_{\rm erase} (F) \ge kT  \ln 2 \,  c_{\rm erase} (F)$ holds  with the equality sign, and the minimum work cost of approximate erasure is given by  
\begin{align}
W_{\rm erase}   (F)  =   \Delta A  +    k T \,  \ln F    \, ,
\end{align}
where $\Delta A$ is the difference between the free energy of the ground  state  and the free energy of the Gibbs state, and the equality holds for every value of $F$ in the interval $[F_{\min}, F_{\max}]$, with $F_{\min}   =  e^{-\Delta A/(kT)}$ and $F_{\max} =1$.  
}

  \medskip

 \begin{figure}[ht!]
	\includegraphics[width=0.5\textwidth]{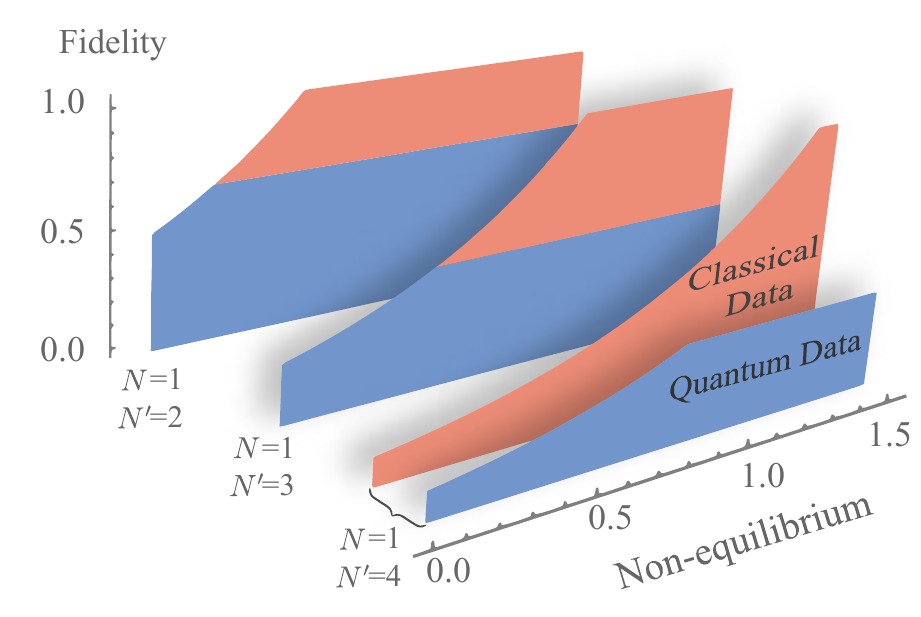}
	\caption{ {\bf Maximum cloning fidelity for a  given amount of  nonequilibrium resources.}  The optimal accuracy-nonequilibrium tradeoff is depicted  for $1\rightarrow2$, $1\rightarrow3$, and $1\rightarrow4$ cloning machines. The fidelities for copying classical (red region) and quantum data (blue region) are limited by the same boundary curve, except that the fidelity for the task of copying quantum data cannot reach to 1 due to the no-cloning theorem.}
	\label{fig:compare_fidelity_NC_QC}
\end{figure}

\medskip 

\textbf{Nonequilibrium cost of  classical  cloning.} 
Copying  is the quintessential example of an information processing task taking place in nature, its accurate implementation being crucial for processes such as DNA replication.  {  In the following, we will refer to  the copying of classical information as classical cloning.} 
   In abstract terms,  the classical cloning task is to  transform $N$ identical copies of a  pure  state picked from an orthonormal basis  into $N'\ge N$ copies of the same state. 
    Classically, this corresponds to the transformation $|x\>\<x|^{\otimes N}  \mapsto  |x\>\<x|^{\otimes N'}$, where $x$ labels  the  vectors of an orthonormal basis.  
The reverse entropy can be computed from Eq.~(\ref{gibbsc}), which gives 
\begin{equation}\label{Kclon} 
\kappa^{\rm C}_{\rm clon}  =   \frac{ \Delta N\,    \Delta A_{\max} }  {kT \,  \ln  2}   \, ,
\end{equation}
where   $\Delta N :  = N'  - N$ the number of extra copies, and $\Delta A_{\max}$   is the maximum  difference between the free energy of a single-copy pure state and the free energy of the single-copy Gibbs state.  Physically,  $\kappa^{\rm C}_{\rm clon}  $  coincides with the maximum amount of work  needed to generate    $\Delta N$  copies of a  pure state  from the thermal state~\cite{horodecki2013fundamental}.    

  

Since cloning  is a special case of a deterministic classical computation, the bound~(\ref{main})   is attainable, and the minimum  nonequilibrium cost of classical cloning  is 
\begin{align}\label{cloning}
c^{\rm C}_{\rm clon} (F)   =        \frac{{  \Delta N   \, \Delta A }}{kT\, \ln 2}  +  \log  F  \, .\end{align}   

This result generalises seminal results by Landauer and Bennett on the thermodynamics of classical cloning~\cite{landauer1961irreversibility,landauer1991information,bennett2003notes}, extending them from the ideal scenario  to realistic settings where the copying process is approximate.  {  For  systems with fully degenerate energy levels, one also has the equality $W_{\rm clon}^{\rm C}  (F)  =  kT\,  (\ln 2)  \, c^{\rm C}_{\rm clon}  (F)$, which  provides the minimum amount of work needed to replicate classical information with a target level of accuracy.  } 

   \medskip 
\textbf{Nonequilibrium cost of  quantum  cloning.} We now consider the task of  approximately cloning quantum information \cite{scarani2005quantum}.  {  The accuracy of quantum cloning is important both for foundational and practical reasons, as it is linked to the no signalling principle \cite{gisin1998quantum}, to quantum cryptography~\cite{scarani2005quantum},  quantum metrology~\cite{chiribella2013quantum},  and a variety of other quantum information tasks~\cite{fan2014quantum}.}  

Here   we consider arbitrary cloning tasks where the set of  single-copy states includes all energy eigenstates. This includes in particular the task of  universal quantum cloning \cite{hillery1997quantum,gisin1997optimal,werner1998optimal}, where the input states are arbitrary pure states.  The reverse entropy of universal quantum cloning is at least as large as the reverse entropy of  classical cloning: the bound $\kappa_{\rm clon}^{\rm Q}  \ge \kappa_{\rm clon}^{\rm C}$  follows immediately from Eq.~(\ref{thermocomp}), by  restricting the optimization to probability distributions that are concentrated on the eigenstates of the energy.


In Supplementary Note 5, we  show that {\em (i)}  the bound~(\ref{main}) is attainable for universal quantum cloning, and {\em (ii)}  $\kappa_{\rm clon}^{\rm Q} = \kappa_{\rm clon}^{\rm C}$.  These results imply that  classical and quantum cloning  exhibit exactly the same tradeoff  between accuracy and nonequilibrium: for every value of the accuracy,   the minimum nonequilibrium cost of information replication is given by Eq. (\ref{cloning}) both in the classical and in the quantum case.  In terms of accuracy/nonequilibrium tradeoff, the only difference between classical and quantum cloning is that the classical tradeoff curve goes all the way up to unit fidelity, while   the quantum  tradeoff curve stops at a maximum fidelity,
 which is strictly smaller than 1 due to the no-cloning theorem~\cite{wootters1982single,dieks1982communication}.

Considering the  differences between quantum and classical cloning, the fact that these two tasks  share the same tradeoff curve is quite  striking.  {   An insight into this  phenomenon comes from connection between the nonequilibrium cost and the time-reversed task  of cloning. For fully degenerate systems, the time-reversed task is to transform $N'$  copies of a state into $N\le N'$ copies of the same state, and in both cases it can be realised by discarding $N'-N$ systems.  The reverse accuracy of this task is the same for both classical and quantum systems, and  so is the reverse entropy.  
In the non-degenerate case, the analysis is more complex, but the conclusion remains the same. }

Although classical and quantum cloning share the same tradeoff curve,  in the following we will show that they  exhibit a fundamental difference in the way the tradeoff  is achieved: to achieve the fundamental limit,  cloning machines 
must use genuinely quantum strategies.

\begin{figure}[t!]
	\includegraphics[width=0.5\textwidth]{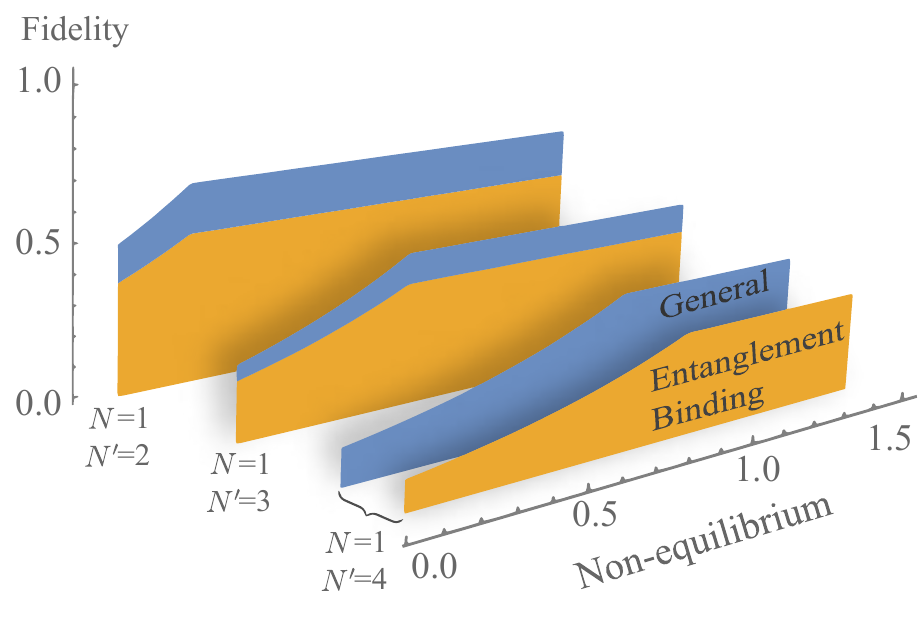}
	\caption{ {\bf Entanglement binding machines vs general quantum machines.}    The figure illustrates the accessible regions for the cloning fidelity for various values of $N$ and $N'$ in the case of qubits with degenerate Hamiltonian.   The values of the fidelity in the blue region are attainable by general quantum  machines, while  the values in the orange region are attainable by entanglement binding machines. The difference between the two regions indicates a thermodynamic advantage of general quantum machines over all classical machines.
	  }
	\label{fig:compare_fidelity_NC}
\end{figure}



\medskip

\textbf{Limit on the accuracy of classical  machines.}    Classical copy machines  scan the input copies and produce   replicas based on this information. Similarly, a classical machine for a general task can be modelled as a machine that measures the input and produces an output based on the measurement result.    When this approach is used at the quantum scale, it leads to a special class of quantum machines, known as entanglement breaking~\cite{horodecki2003entanglement}.  


Here we show that entanglement breaking machines satisfy a stricter bound. 
In fact, this stricter bound   applies not only to entanglement breaking machines, but also to  a broader class of machines,  called entanglement binding~\cite{horodecki2000binding}.  An entanglement binding channel is a quantum channel that degrades  every entangled state to a bound (a.k.a. PPT)  entangled state~\cite{peres1996separability,horodecki1997separability}. 
  In Methods, we show that the  minimum nonequilibrium cost over all  entanglement binding machines, denoted by $c_{\map T}^{{\rm eb}} (F)$,   must satisfy the inequality
 \begin{align}\label{binding}
c_{\map T}^{{\rm eb}}   (F) &  \ge   \max\{\kappa_{\map T},  \kappa_{\map T^*}\}   +  \log  F  \, ,
\end{align} 
where $\kappa_{\map T}$   is the reverse entropy of  the state transformation task $\rho_x\to \rho_x'$, and $\kappa_{\map T^*}$ is the reverse entropy of  the transposed task $\map T^*$, corresponding to the  state transformation $\rho_x  \mapsto (\rho_x')^T$.    This bound  can be used to demonstrate that a thermodynamic advantage of general quantum machines over all entanglement binding machines, including in particular all classical machines.   

\medskip 

{\bf Quantum advantage in  cloning.}   For quantum cloning, it turns out that no entanglement binding machine can achieve the optimal accuracy/nonequilibrium tradeoff. The reason for this is that  the reverse entropy of the transpose task  is  strictly larger than the reverse entropy  of the direct task, namely $\kappa_{\rm clon^*}  > \kappa_{\rm clon}$.  In Supplementary Note 6 we prove the  inequality  
\begin{align}\label{bindingbound}
\kappa_{\rm clon^*}   \ge   \kappa_{\rm clon}  +   \log  \frac{ d_{N+ N'}\,  e^{-\frac{N' \Delta  E}{kT} } }{d_{N'}}  \, ,
\end{align}
where $\Delta E$ is the difference between the maximum and minimum energy, and,   $d_K  =  (K+d-1)!/[ K!  (d-1)!]$ for $K  = N$ or $K=  N+ N'$.   Inserting this inequality into Eq.~(\ref{binding}), we conclude that    every entanglement binding machine necessarily  requires a larger number of clean qubits compared to the optimal  quantum machine.

When the energy levels are fully degenerate,  we show that the bounds~(\ref{binding}) and (\ref{bindingbound}) are exact equalities.  With this result at hand,  we can compare the exact performance of entanglement binding machines and general quantum machines, showing that the latter achieve a higher accuracy for every given amount of nonequilibrium resources. The comparison is presented in Figure \ref{fig:compare_fidelity_NC}.

Our result shows that entanglement binding machines are  thermodynamically inefficient for the task of information replication.  Achieving the ultimate efficiency limit requires machines that are able to preserve free ({\em i.~e.~}non-bound) entanglement.   This observation fits with the known fact that classical machines cannot achieve  the maximum copying accuracy allowed by quantum mechanics~\cite{gisin1997optimal,bruss1998optimal,werner1998optimal}.  Here we have shown that not only classical machines are limited in their accuracy, but also that, to achieve such limited accuracy, they require a higher amount of nonequilibrium  resources.  Interestingly, the thermodynamic advantage of general quantum machines vanishes in the asymptotic limit $N'\to \infty$, in which the optimal quantum cloning can be reproduced by state estimation~\cite{bae2006asymptotic,chiribella2006quantum,chiribella2010quantum}.
\medskip  

{  
{\bf Thermodynamic benchmark for  quantum memories and quantum communication.}  Quantum machines that preserve free entanglement also offer an advantage  in the storage and  transmission of quantum states, corresponding  to  the ideal  state transformation $\rho_x\mapsto\rho_x$ where $x$ parametrises the states of interest. In theory,   a noiseless   quantum machine   can  achieve  perfect accuracy  at  zero work cost. In practice, however,  the transmission is always subject to errors and inefficiencies,  resulting into nonunit fidelity and/or nonzero work.  For this reason, realistic experiments that aim to demonstrate genuine quantum transmission or storage  need criteria to demonstrate  superior performance with respect to all classical setups. A popular approach is to demonstrate an experimental  fidelity larger than  the maximum fidelity achievable by  classical schemes \cite{boschi1998experimental,braunstein1998teleportation,hammerer2005quantum}.  In the qubit case, the maximum classical fidelity is $F_{\max}^{\rm eb}= 2/3$ \cite{massar2005optimal}, and is often used as a benchmark for quantum communication experiments  \cite{li2022quantum,zhong2021deterministic,kurpiers2018deterministic}.  Here we provide a different benchmark,  in terms of the nonequilibrium cost needed to achieve a target fidelity $F$.   In  Supplementary Note 7, we show that  the minimum nonequilibrium cost  over all entanglement binding machines   for the storage/transmission of qubit states is   
\begin{align}\label{benchmark}
c^{{\rm eb}}_{\rm store/transmit}  (F) =  \log   \left[  F+  e^{ \frac{\Delta E}{kT} }  \,   \frac{       (2F-1)^2  }{1-F}   \right]   \, ,  \end{align} 
Eq.~(\ref{benchmark})   is valid for every  qubit Hamiltonian and for  every value of $F$ in the interval $[F^{{\rm eb} }_{\min} , F^{ {\rm eb} }_{\max}]$, with   $F^{{\rm eb} }_{\max}  = 2/3$ and  $F_{\min}^{{\rm eb} }=  (e^{ \frac{\Delta E}{kT}}   + 1)/(2   e^{ \frac{\Delta E}{kT}}  + 1)$.  The minimum cost   $c_{\rm store/store}^{{\rm eb}} (F)$ can be achieved by state estimation,  and therefore can be regarded as the classical limit on the nonequilibrium cost.  

For every $F>  F_{\min}$, the minimum nonequilibrium cost  (\ref{benchmark}) is strictly larger than zero for every  nondegenerate Hamiltonian. Since the nonequilibrium cost is a lower bound to the work cost, Eq.~(\ref{benchmark}) implies that every entanglement binding machine with fidelity $F$ requires at least $k T (\ln 2)  \,  c^{\rm eb}_{\rm store/transmit}  (F)$ work.   This value  can be used as a benchmark to certify genuine quantum information processing:   
 every realistic setup that achieves fidelity $F$  with less than $k T \ln  \left[  F+  e^{ \frac{\Delta E}{kT} }  \,    (2F-1)^2 /  (1-F)  \right]$ work will necessarily exhibit a performance that cannot be achieved by any classical setup.   Notably, the presence of a thermodynamic constraint (either on the nonequilibrium or on the work) provides a way to certify a quantum advantage even for noisy implementations of quantum memories and quantum communication systems with fidelity  below the classical fidelity threshold $F_{\max}  = 2/3$.     A generalisation of these results for higher dimensional systems is provided in Supplementary Note 7.
 
}

\section{Discussion} 
An important  feature  of our bound (\ref{main}) is that it applies also to state transformations that are forbidden by quantum mechanics, 
such as ideal quantum cloning or ideal quantum  transposition.  
   For state  transformations that can be exactly implemented, instead, it is interesting to compare our bound with related results in the literature.   

For exact implementations, the choice of accuracy measure  is less important,  and one can use  any measure  for which Eq. (\ref{main})   yields  a useful bound on the work cost.  For example, consider the problem of generating a  state $\rho$ from the equilibrium state. By choosing a suitable  measure of accuracy  (see Methods for the details), we find   that the nonequilibrium cost for the state transition $\Gamma  \mapsto  \rho$ is equal to $ D_{\max}  (\rho \|   \Gamma)$,  where {  $D_{\max}  (\rho \| \sigma)   :  =    \lim_{\alpha\to \infty}  D_\alpha  (\rho\|\sigma)$ is the max relative entropy,  $D(\rho\| \sigma)  :  =  \log \Tr[  \rho^\alpha \sigma^{1-\alpha}]/(\alpha-1) \, , \alpha\ge 0  $ being the the R\'enyi relative entropies.}   In this case,   the nonequilibrium cost coincides (up to a proportionality constant $kT \ln 2$) with the minimal amount of work needed to  generate the state $\rho$ without errors~\cite{horodecki2013fundamental}.  Similarly, one can consider the  task of extracting work from the state $\rho$, corresponding to the state transition $\rho \mapsto \Gamma$.  Ref.~\cite{horodecki2013fundamental} showed that the maximum extractable work is $   D_{\min}  (\<\rho \>  \|   \Gamma)\,  k T \ln 2$, {  where $D_{\min}  (\rho\|\sigma)  :=D_{0}   (\rho \|  \sigma)  $ is the  min relative entropy} as per Datta's definition~\cite{datta2009min} and $\< \rho\>$ is the time-average of $\rho$.   This value can also  be retrieved from our bound with a suitable choice of accuracy measure (see Supplementary Note  8 for the details).   Smooth versions of these entropic quantities naturally arise by ``smoothing the  task'', that is, by considering small deviation from the  input/output states that specify the desired state transformation   (see Methods).   
  
Our bound can also be applied to the task of information erasure with the assistance of a quantum memory~\cite{del2011thermodynamic}.  There, a machine has access to a system $S$ and to a quantum memory $Q$, 
and the goal is to reset system $S$ to a pure state $\eta_S$, without altering the local state of the memory.  When the initial states of system $SQ$ are drawn from a time-invariant subspace,  our bound (\ref{main})  implies that the work cost satisfies  the inequality $W /(kT  \, \ln  2)    \ge      D_{\max}  ( \eta_S\otimes  \gamma_Q  \|  \Gamma_{SQ})-  D_{\max}      (\widetilde \Gamma_{SQ}\|  \Gamma_{SQ})  $, where $\widetilde \Gamma_{SQ}$ is the quantum state obained by projecting the Gibbs state onto the input subspace, and $\gamma_Q  =  \Tr_S  [\widetilde \Gamma_{SQ}]$ is the marginal state of the memory.    The bound is tight, and, for degenerate Hamiltonians, it matches the upper bound from Ref.~\cite{del2011thermodynamic} up to logarithmic corrections in the error parameters (see Supplementary Note 9). 


{    Another interesting issue is to determine  when a given state transition $\rho \mapsto \rho'$ can be implemented without investing  work.  
For states that are diagonal in the energy basis, a necessary and sufficient condition was derived in  Ref.~\cite{brandao2015second},  adopting a framework where  catalysts are allowed. 
In this setting, Ref.~\cite{brandao2015second} showed that the state transition $\rho\mapsto  \rho'$ can be implemented catalytically without work cost if and only if  
\begin{align}\label{2laws}
D_\alpha  (\rho'  \|   \Gamma_B)  \le D_\alpha  (\rho  \|   \Gamma_A)  \qquad \forall  \alpha \ge 0\, .
\end{align}    These  conditions can be compared with our bound~(\ref{main}). In Methods, we show that, with a suitable choice of figure of merit, Eq. (\ref{main}) implies  the lower bound $W/(kT \ln 2)   \ge      \,   D_{\max}  (\rho' \|   \Gamma_B)   -   D_{\max}  (\rho \|   \Gamma_A)$ for the perfect execution of the state transition $\rho \mapsto \rho'$.    Hence, the work cost for the state transition $\rho\mapsto \rho'$ satisfies the bound $W/(kT\ln2)  \ge      D_{\max}  (\rho \|   \Gamma_A)-  D_{\max}  (\rho' \|   \Gamma_B) $, and the r.h.s. is nonpositive only if $  D_{\max}  (\rho' \|   \Gamma_B)\le  D_{\max}  (\rho \|   \Gamma_A)$.   The last condition is a special case of  Eq. (\ref{2laws}), corresponding to $\alpha \to \infty$.  Notably, this condition and Eq. (\ref{2laws}) are equivalent when the input and output states have well-defined energy, including in particular the case where the Hamiltonians of systems $A$ and $B$ are fully degenerate. {  Further discussion on the relation between quantum relative entropies and the cost of accuracy  is provided in Supplementary Note 10.}      

      }
    

While the applications  discussed in the paper focussed on one-shot tasks,  our results also apply to  the asymptotic scenario where the task is to implement the transformation $\rho_x^{\otimes n} \mapsto \rho_x^{\prime \otimes n}$ in the  large $n$ limit. 
 In Methods we consider the amount of nonequilibrium per copy required by this transformation, allowing for small deviations in the input and output states.   This setting leads to the definition of  a {\em smooth reverse entropy of a task}, whose value per copy is denoted by $\kappa_{\map T, \rm iid}$ and is  shown to satisfy the bound
  \begin{align}\label{iid} \kappa_{\map T,  \rm iid}  \ge  \max_x  \,   S(  \rho_x'\|  \Gamma_B  )   -  S(  \rho_x\|  \Gamma_A  ) \,,
 \end{align} where $S (\rho  \|  \sigma) :  =  \Tr [  \rho  (  \log \rho  - \log \sigma)]$ is the quantum relative entropy.

In the special case where the   state transformation $\rho_x\to \rho_x'$ can be implemented perfectly, and where  $(\rho_x)_{x\in\set X}$ is the set of all possible quantum states of the input system,    the r.h.s. of Eq.~(\ref{iid})   (times $kT \ln 2$)  coincides with the thermodynamic capacity introduced  by Faist, Berta, and Brand\~ ao in Ref.~\cite{faist2018thermodynamic}.   In this setting, the results of  Ref.~\cite{faist2018thermodynamic} imply that  the thermodynamic capacity  coincides with the amount of work per copy needed to implement  the transformation $\rho_x\to \rho_x'$.  Since the amount of work cannot be smaller than the nonequilibrium cost,  this result implies that our fundamental accuracy/nonequilibrium tradeoff is asymptotically achievable for all information processing tasks allowed by quantum mechanics. 

{   In a different setting and with different techniques, questions related to the  thermodynamical cost of physical processes have been studied in the field of stochastic thermodynamics  \cite{seifert2018stochastic}. Most of the works in this area  focus on the properties of  nonequilibrium steady states of  classical  systems with Markovian dynamics. An important result    is a tradeoff relation between the relative standard deviation of the  outputs associated to the currents in the nonequilibrium steady state and the overall entropy production \cite{barato2015thermodynamic,gingrich2016dissipation,barato2016cost,horowitz2020thermodynamic}.  This relation, called a thermodynamic uncertainty relation, is often  interpreted as a tradeoff between the precision of a process and  its thermodynamical cost. A difference with our work is that the notion of precision used in stochastic thermodynamics is not directly related to general information processing tasks.
  Another difference is that  thermodynamic uncertainty relations do not always hold for  systems outside the nonequilibrium steady state \cite{barato2016cost}, whereas our accuracy/nonequilibrium tradeoff applies universally to all quantum systems.   An interesting avenue of future research is to integrate  the information-theoretic methods developed in this paper with those of stochastic thermodynamics, seeking for concrete physical models that  approach the ultimate efficiency limits. }

\section{Methods}


{\bf General performance tests.} The  performance of a machine in a given information processing task can be operationally quantified by the probability to pass a  test~\cite{hammerer2005quantum,yang2014certifying,bai2018test}. 
 In the one-shot scenario, a general  test  $\map T$ consists in preparing states of a composite system $AR$, consisting of the input of the machine and an additional reference system. The machine is requested to act locally on system $A$, while the reference system undergoes the identity process, or some other {  (generally noisy)} process  {  $\map R_x$ implemented by the party that performs the test.}   Finally, the reference system and the output of the machine  undergo a joint measurement, described by a suitable observable. The measurement outcomes are regarded as the  score assigned to the machine.  The test  $\map T$ is then described by the possible {  triples $(\rho_x,  \map R_x , O_x)_{x\in\set X}$, consisting of an input state, a process on the reference system, and an output observable.     In the worst case over all possible triples,} one gets the accuracy $\map F_{\map T}   (\map M)  :  =  \min_x   \Tr[  O_x  \,  (\map M\otimes \map R_x)  (\rho_x)]$, where $\map M$ is the map describing the machine's action.   {   Note that  the dependence of the state $\rho_x$, transformation $\map R_x$, and measurement $O_x$ can be arbitrary, and that the parameter $x$ can also be a vector  $x  =  (x_1,  \dots,  x_n)$. For example, the input state $\rho_x$ could depend only on the subset of the entries of the vector $x$, while the output observable $ O_x$ could depend on all the entries, thus describing the situation where multiple observables are tested for the same input state.  }

 Performance tests provide a more general way to define information processing tasks.  Rather than specifying a desired state transformation $\rho_x\mapsto \rho_x'$, one can directly specify a test  that assigns a score to the machine.    The test can be expressed in a compact way in the Choi representation~\cite{choi1975completely}.   In this representation, the test is described by a set of operators $ (\Omega_x)_{x\in\set X}$, called the {\em performance operators}~\cite{bai2018test}, acting on the product of the input and output Hilbert spaces. The  accuracy of the test has the simple expression $ \map F_{\map T}  (\map M)  = \min_x  \Tr[M\, \Omega_x]$, where   $M:  =    ( \map I_A\otimes \map M)   (  |I_A\>\<  I_A|)$,  $|I_A\>  :=  \sum_i \,  |i\>\otimes |i\> $ is the Choi operator of channel $\map M$, and $\map I_A$ is the identity on  system $A$. 
  In the following we will take  each operator $\Omega_x$ to be positive semidefinite without loss of generality.    
\medskip 

{\bf Exact expression for the nonequilibrium cost.}      In Supplementary Note 1,  we show that the nonequilibrium cost of a general task $\map T$  can be evaluated with the expression $c_{\map T}  (F)       =  \max_{\st p}   c_{\map T,  \st p} (F) $,  where the minimum is over all probability distributions $\st p  =  (p_x)_{x\in\set X}$ and
\begin{align}\label{sdp1} 
c_{\map T, \st p}  (F)      =     \log     \max_{\begin{array}{c}   
     X_A\otimes I_B    + z\, \Omega_{ \st p}     \le      \Gamma_{A}'  \otimes Y_B   \\
          \Tr[\Gamma_B  Y_B]   \le 1  
      \end{array}   } \, \Tr[X_A]      +  z\,     F   \, ,
  \end{align}  
with  $\Omega_{\st p}   :=  \sum_x  \,  p_x \, \Omega_x$,  $\Gamma_A' : =  \Pi_A \Gamma_A  \Pi_A$. Here, the maximisation runs over all Hermitian operators $X_A$ $(Y_B)$ acting on system $A$  ($B$) and over all real numbers $z$.  

For every fixed probability distribution $\st p$, the evaluation  of $c_{\map T,\st p}  (F)$  is a semidefinite program~\cite{watrous2018theory}, and can be solved numerically for low dimensional systems.    
A simpler optimisation problem arises by setting $X_A=  0$, which provides the lower bound 
\begin{align} 
\nonumber 
c_{\map T, \st p}  (F)     & \ge     \log     \max_{\begin{array}{c}   
      z\, \Omega_{ \st p}     \le      \Gamma_{A}'  \otimes Y_B   \\
          \Tr[\Gamma_B  Y_B]   \le 1  
      \end{array}   } \,  z\,     F \\
       &  =  H  (A|B)_{\omega_{ \map T,\st p}}   +  \log F  \, . 
       \label{sdp2}
  \end{align}  
  (see Supplementary Note 1 for the derivation). 

\medskip 
{ {\bf Time-reversed tasks and reverse entropy.}    For a given task  $\map T$, implemented by operations with input $A$ and output $B$, we define a time-reversed task $\map T^{\rm rev}$, implemented by operations with input $B$ and output $A$.  For example, consider the case where the direct task is to transform pure states into pure states, according to a given mapping $\rho_x \mapsto \rho_x'$,  on a quantum system with fully degenerate energy levels,    and  the accuracy of the implementation measured by the fidelity $\map F_{\map T} (\map M)  =  \min_x  \Tr  [  \rho_x'  \map M  (\rho_x)]$.    In this case, the time-reversed task is to implement the transformation $\rho_x'\mapsto \rho_x$, using some channel $\map M$ with input $B$ and output $A$.  The accuracy is then given by the reverse fidelity   $\map F_{\map T_{\rm rev}}(\map M)  =  \min_x  \Tr  [  \rho_x  \map M  (\rho_x')]$.  More generally, we define the time-reversed task  $\map T_{\rm rev}$ in terms of a time reversal for quantum operations \cite{crooks2008quantum,chiribella2021symmetries}. The specific version of the time reversal used here  maps   the states $\rho_x$ into the observables  $\widetilde O_x :  =  \Gamma_A^{-\frac 12}    \rho_x   \Gamma_A^{-\frac 12}$  and the observables $O_x$ into the (unnormalised) states $\widetilde \rho_x: =   \Gamma_B^{\frac 12}    O_x   \Gamma_B^{\frac 12}$ \cite{chiribella2021symmetries}.    The reverse accuracy then becomes   $\map F_{\map T_{\rm rev}}(\map M)  :=  \min_x  \Tr  [  \widetilde O_x  \map M  (\widetilde \rho_x)]$. 

For a general information-processing  task with performance operators $  (\Omega_x)_{x\in\set X}$, we define the time-reversed task ${\map T}_{\rm rev}$ with performance operators $  (  \Omega_{x}^{\rm rev})_{x\in\set X}$ defined by   
\begin{align}\Omega_x^{\rm rev} : =       ( \Gamma_B^{1/2}  \otimes \Gamma_A^{-1/2}  )  E_{AB} \Omega^T_x  E_{AB} (\Gamma_B^{1/2} \otimes \Gamma_A^{-1/2} ) \,,
\end{align} 
where $E_{AB}: \spc H_A \otimes \spc H_B  \to \spc H_B \otimes \spc H_A$ is the unitary operator that exchanges systems $A$ and $B$.  The reverse accuracy of a generic quantum channel $\map M$ is then given by   $\map F_{\map T_{\rm rev}}  (\map M)  :  =  \min_x  \,  \Tr  [   \Omega_x^{\rm rev}    \,   M  ]$, where $M$ is the Choi operator of $\map M$.  The maximum of the reverse accuracy over all quantum channels can be equivalently expressed in terms of  a conditional min-entropy: indeed, one has
  \begin{align}
 \nonumber  F_{\map T_{\rm rev}}^{\max}   &  =   \max_{\map M}    \map F_{\map T_{\rm rev}}  (\map M)    \\
  \nonumber  &  =   \max_{M :  M\ge 0  \, ,  \Tr_A  [M]   =  I_B  }   \min_{x}    \Tr [ \Omega_{x}^{\rm rev}  \,  M  ] \\
 &  =   \max_{M :  M\ge 0  \, ,  \Tr_A  [M]   =  I_B  }   \min_{\st p}    \Tr [ \omega_{\map T, \st p}  \,  M  ]    \end{align}   
where   the minimum is over all probability distributions $\st p  = (p_x)$, and $\omega_{\map T,  \st p}     :   = ( \sum_x  \,  p_x  \,  \Omega_{x}^{{\rm rev}})^T $.  Using von Neumann's minimax theorem, we then obtain
\begin{align}
   \nonumber F^{\map T_{\rm rev}}_{\max}&  =   \min_{\st p}  \max_{M :  M\ge 0  \, ,  \Tr_A  [M]   =  I_B  }     \Tr [ \omega_{\map T, \st p}  \,  M  ]\\
  \nonumber  &  =   \min_{\st p}    2^{-      H_{\min}  (A|B)_{\omega_{\map T, \st p}}}    \\
   &  =     2^{-      \max_{\st p}  H_{\min}  (A|B)_{\omega_{\map T, \st p}}}    \, ,
\end{align}
where the second equality follows from the operational interpretation of the min-entropy \cite{konig2009operational}.     Taking the logarithm on both sides of the equality, we then obtain the relation 
$\kappa_{\map T}  :=    - \log  F^{\map T_{\rm rev}}_{\max}    =   \max_{\st p}  H_{\min}  (A|B)_{\omega_{\st p}} $, 
 corresponding to Eq. (\ref{thermocomp}) in the main text.   The bound (\ref{main}) then follows from the relation $c_{\map T}  (F)   =  \max_{\st p}  c_{\map T, \st p}$ and from Eq.  (\ref{sdp2}).     }

\medskip

{\bf Bounds on the reverse entropy.}    When the test  $\map T$ consists in the preparation of a set of states $(\rho_x)_{x\in\set X}$ of system $A$ and in the measurement of a set of observables $(O_x)_{x\in\set X}$ on system $B$, the reverse entropy  can  be lower bounded as 
\begin{align}\label{singlex}
\kappa_{\map T}  \ge  \max_x  \,   -\log \Tr[  \Gamma_B  \, O_x]   -   D_{\max }   (\rho_x\|\Gamma_A)  \, , 
\end{align}
with the equality holding when $|\set X| =1$ (see Supplementary Note 10 for the proof and for a discussion on the relation between the nonequilibrium cost of a state transformation task $\rho_x\mapsto \rho_x' , \forall x\in\set X$ and the nonequilibrium cost of the individual state transitions $\rho_x\mapsto \rho_x'$ for a fixed value of $x$).    

A possible choice of observable is $O_x=  P_x$, where $P_x$ is the projector on the support of the target state $\rho_x'$.  In this case, the  bound~(\ref{singlex}) becomes   
\begin{align}\label{dmin}
\kappa_{\map T}  \ge  \max_x  \,  D_{\min }    (\rho_x'\|  \Gamma_B)   -   D_{\max }   (\rho_x\|\Gamma_A) \,. 
\end{align}     An alternative choice of observables is   $O_x  =  \Gamma^{-1/2}  |\psi_x\>\<\psi_x|  \Gamma^{-1/2} /\| \Gamma^{-1/2}  \rho_x'  \Gamma^{-1/2} \| $, where $|\psi_x\>$ is the normalised eigenvector corresponding to the maximum eigenvalue of $ \Gamma_B^{-1/2 } \rho_x'   \Gamma_B^{-1/2 } $.  With this choice, the bound~(\ref{singlex}) becomes   $\kappa  \ge  \max_x  \,  D_{\max }    (\rho_x'\|  \Gamma_B)   -   D_{\max }   (\rho_x\|\Gamma_A)$, with the equality when $|\set X|=1$.  Combining this bound with Eq.~(\ref{main}), we obtain the following 
\begin{prop}
If   there exists a quantum channel $\map M$ such that $\map M (\rho_x)  = \rho_x'$ for every $x\in\set X$, then  its   nonequilibrium cost  satisfies the bound  $c  (\map M) \ge  \max_x  \,  D_{\max }    (\rho_x'\|  \Gamma_B)   -   D_{\max }   (\rho_x\|\Gamma_A)$.  
\end{prop}
The proposition follows from Eq. (\ref{main}) and from the fact that the channel $\map M$ has accuracy  $\map F  (\map M)=  \min_x  \Tr  [\map M (\rho_x)  O_x]  = 1$.

\medskip 

{\bf Smooth reverse entropy.}  For an information processing task $\map T$ with operators   $ (\Omega_x)_{x\in\set X}$, one can consider an approximate version, described by another task $\map T'$ with operators   $(\Omega'_x)_{x\in\set X'}$ that are close to $ (\Omega_x)_{x\in\set X}$  with respect to a suitable notion of distance.  One can then define the worst (best) case smooth reverse entropy  of the task $\kappa_{\map T, \epsilon}$ as the maximum (minimum) of $\kappa_{\map T'}$ over all tasks $\map T'$ that are within distance $\epsilon$ from the given task.  The choice between the worst case and the best case irreversibility depends on the problem at hand.  A best case irreversibility corresponds to introducing an error tolerance in the task, thus  discarding ``low probability events'' that would result in a higher cost~\cite{del2011thermodynamic,horodecki2013fundamental}.   Instead, a worst case irreversibility can be used to model noisy scenarios, where the input states may not be the ones in the ideal  information processing task. An example of this situation is the experimental implementation of quantum cloning,  where the input states may not be exactly pure. 


Smoothing is particularly useful in the asymptotic scenario.   Consider the test  $\map T_n$ that consists in preparing a multi-copy input state $\rho_x^{\otimes n}$ and measuring the observable $P_{x,n}$, where $P_{x,n}$ is the projector on the support of the target state $\rho_x^{\prime  \, \otimes n}$.  A natural approximation is to allow, for every $x\in  \set X$,   all inputs $\rho_{y,n}$ that are   $\epsilon$-close to $\rho_x^{\otimes n}$, and all outputs $\rho_{y,n}'$ that are $\epsilon$-close to $\rho_x^{\prime  \otimes n}$.  
 Choosing  $\kappa_{{\map T}_n, \epsilon }$ to be the worst case smooth reverse entropy of the task  $\map T_n$ , Eq.~(\ref{dmin}) gives the bound $\kappa_{\map T_n, \epsilon}  \ge  \max_x    D_{\min }^{\epsilon}    (\rho_{x}^{\prime \otimes n}  \|   \Gamma_B^{\otimes n})   -      D_{\max }^{\epsilon}    (\rho_{x}^{\otimes n} \|    \Gamma_A^{\otimes n}) $, where $D_{\min }^{\epsilon} $ and $D_{\max }^{\epsilon} $ are the smooth versions of $D_{\min }$ and $D_{\max } $ \cite{datta2009min}.   One can then define the regularised reverse entropy  of the task  as $\kappa_{\map T, \rm iid} :  =  \lim_{\epsilon\to 0}   \,   \sup_n  \,  \kappa_{\map T_n, \epsilon}/n$.  Using the relations 
   $ \lim_{\epsilon\to 0}   \,   \sup_n   D_{\min }^{\epsilon}    (\rho_{x}^{\prime \otimes n}  \|   \Gamma_B^{\otimes n})/n =     S    (\rho_{x}'  \|   \Gamma_B)  $  and    $ \lim_{\epsilon\to 0}   \,   \inf_n   D_{\max }^{\epsilon}    (\rho_{x}^{ \otimes n}  \|   \Gamma_A^{\otimes n})/n =     S   (\rho_{x}  \|   \Gamma_A)  $ \cite{datta2009min}  we finally obtain the bound  $\kappa_{\st T, \rm iid}  \ge  \max_x S   (\rho_{x}'  \|   \Gamma_B)   -   S   (\rho_{x}  \|   \Gamma_A)   $.   The quantity on the  r.h.s.  coincides with the thermodynamic capacity introduced  by Faist, Berta, and Brand\~ ao in Ref.~\cite{faist2018thermodynamic}, where it was shown that    the thermodynamic capacity  coincides with the amount of work per copy needed to implement the transformation $\rho_x\to \rho_x'$. Combining this result with our bounds, we obtain that the fundamental accuracy/nonequilibrium in Eq.  (\ref{main}) is asymptotically achievable for all transformations allowed by quantum mechanics.

 \medskip

\medskip 
 
{\bf Limit for entanglement binding channels.}  Entanglement binding  channels  generally  satisfy a more stringent limit than (\ref{main}).   The derivation of this strengthened limit is as follows: first, the definition of an entanglement binding channel  $\map P$  implies that the map $ {\map P}^{\rm PT}$ defined by ${\map P}^{\rm PT} (\rho):  = [    \map P(\rho) ]^T$ is a valid quantum channel.  Now, the nonequilibrium cost of the channels $\map P$ and $\map P^{\rm PT}$  is given by $D_{\max}   (   \map P  (\Pi_A  \Gamma_A \Pi_A \|  \Gamma_B) )$  and $D_{\max}   (   \map P^{\rm PT}  (\Pi_A  \Gamma_A \Pi_A \|  \Gamma_B) )$ (cf. Supplementary Note 1).    Since the max relative entropy satisfies  the relation $D_{\max}  (\rho\|\sigma) =  D_{\max} (\rho^T\| \sigma^T)$ for every pair of states $\rho$ and $\sigma$, the costs of $\map P$ and $\map P^{\rm PT}$  are equal.     

The second step is to note that the accuracy of the channel $\map P$  for the task specified by the performance operators $(\Omega_x)$ is equal to the accuracy of the channel $\map P^{\rm PT}$ for the task specified by the performance operators $(\Omega_x^{T_B})$, where $T_B$ denotes the partial transpose over system $B$.     Applying the bound~(\ref{main}) to channel  $\map P^{\rm PT}$, we then obtain the relation 
 \begin{align}
 \nonumber c  (\map P)   &    =    c  (\map P^{\rm PT})  \\
   & \ge    \kappa_{\map T^*}   + \log F \label{pptbound} \, ,
   \end{align} 
 where $\kappa_{\map T^*}$ is the reverse entropy of the transpose task $\map T^*$,   with performance operators $(\Omega_x^{T_B})$. 
Since entanglement binding channel is subject to both bounds (\ref{main}) and (\ref{pptbound}), Eq. (\ref{binding}) holds.

\section*{Data Availability}    The authors declare that the data supporting the findings of this study are available within the paper and in the supplementary information files. \\

\section*{Author Contributions}    All authors contributed substantially to the  development of the  research paper and to  the preparation of the paper.    GC and FM contributed equally. 

 \section*{Competing Interests}    The authors declare no competing interests. 
 
 \section{Acknowledgments}  
 
GC acknowledges a helpful discussion with Nilanjiana Datta on the quantum extensions of  R\'enyi relative entropies. FM acknowledges Yuxiang Yang, Mile Gu, and Oscar Dahlsten for helpful comments that helped improving the presentation.   This work was supported by the Hong Kong Research Grant Council through grants 17326616 and 17300918, and through 
the Senior Research Fellowship Scheme via SRFS2021-7S02,  by the Swiss National Science Foundation via grant  200021\_188541, by the National Natural Science Foundation of China through grants 11675136, 11875160 and U1801661,  by the Key R\&D Program of Guangdong province through grant 2018B030326001,  by the
Guangdong Provincial Key Laboratory through grant c1933200003,  the Guangdong Innovative and Entrepreneurial Research Team Program via grant 2016ZT06D348,  the Science, Technology and Innovation Commission of Shenzhen Municipality through grant KYTDPT20181011104202253.  
Research at the Perimeter Institute is supported by the Government of Canada through the Department of Innovation, Science and Economic Development Canada and by the Province of Ontario through the Ministry of Research, Innovation and Science.

\bibliography{energy_cloning.bib}

\onecolumngrid
\appendix

\bigskip
\bigskip
\begin{center}
	\Large{\textbf{Supplementary Notes.}}
\end{center}

\section*{Supplementary Note 1: Derivation of the accuracy-nonequilibrium tradeoff}

In this note we establish  a fundamental  tradeoff between accuracy and nonequilibrium in the execution of any given information processing task.  
The note consists of three parts. In the first part, we derive a closed-form expression for the nonequilibrium cost of a given quantum channel, building on results by Faist and Renner~\cite{faist2018fundamental}.   In the second part, we derive the optimal tradeoff curve as a semidefinite program. Finally, in the third part we introduce a  further constraint in the semidefinite program, which leads to the notion of reverse entropy  and to the bound~(5) in the main text.

\subsection*{The nonequilibrium cost of a given information processing task}

Here we discuss the basic settings and definitions used in our paper to evaluate the nonequilibrium of cost of a given information processing task.

Let us start from  the related problem of evaluating the cost of a specific quantum channel $\map M$.  
The minimum number of clean qubits  needed to realise  a channel $\map M$ using Gibbs preserving operations was derived by Faist and Renner in Ref.~\cite{faist2018fundamental}.   They considered   realisations where the machine  reproduces  the action of the desired channel  on a part of a given entangled state $|\Psi\>  \in  \spc H_A \otimes\spc H_R$, involving the input system $A$ and a reference system $R$.  
For approximate realisations    with error $\epsilon$,   the minimum number of clean qubits,  called the {\em nonequilibrium cost} in our paper,  was shown to be
\begin{align}\label{faist}
	c_{\epsilon}  (\map M,  |\Psi\>)  &=- \max_{ 
		\begin{array}{rl} 
			\map M'  (\Gamma_A)    &\le  2^{-\lambda}  \Gamma_B\\
			( \map M'  \otimes \map I_R  ) (|\Psi\>\<\Psi| )  & \approx_\epsilon  ( \map M  \otimes \map I_R  ) (|\Psi\>\<\Psi| )   \end{array}}     \lambda  \, ,   
\end{align}
where  the maximisation is over all quantum operations  (completely positive trace non-increasing maps) $\map M'$ with input system $A$ and output system $B$, and $ \approx_\epsilon$ denotes approximate equality with respect to a suitable distance measure.

{    In the following, we will set $\epsilon =  0$ and we will minimise the cost  $c_0  (\map M,|\Psi\>)$ over all channels $\map M$ that achieve  a desired level of accuracy in a given information processing task $\map T$. Specifically, we will evaluate the minimum cost 
	\begin{align}\label{cococost}
		c_{\map T}  (F) :  =  \min  \{  c_0  ( \map M  ,  |\Psi\> ) ~|~      \map F_{\map T}   (   \map M)  \ge  F    \} \, ,
	\end{align}  
	where   $ \map F_{\map T}   (   \map M)$  is the accuracy measure (as defined in the main text) and  $F$ is the desired level of accuracy.

	Here, the restriction  to exact realisations ($\epsilon=0$)  is  done without loss of generality, because the approximate implementation of the task $\map T$ is already taken into account  by the accuracy measure $ \map F_{\map T}   (   \map M)$.  
	
	It is worth stressing that our approach is different from the approach of most works in the literature, where one  fixes a channel $\map M$ and  asks what is the cost of implementing some $\epsilon$-approximation of $\map M$.    Following this approach, one might be tempted  to pick a channel $\map M$ that implements the  task $\map T$ perfectly, and then to minimise the cost over all channels in  an $\epsilon$-neighbourhood of $\map M$.  
	This approach, however, would not work in general. The problem is that the correspondence between information processing tasks and quantum channels is generally not one-to-one.  First, there exist tasks that cannot be implemented perfectly by any channel $\map M$, such as, for example, ideal quantum cloning.  Second,   there exist tasks that can be implemented perfectly by more than one channel. In those cases, the minimum cost in an $\epsilon$-neighbourhood of a specific channel  $\map M$ picked among those that achieve the task $\map T$ perfectly may not be equal to the minimum cost over all channels that achieve the task with error at most $\epsilon$.   This point is made clear by the following example. Consider the task of transforming the maximally mixed qubit state  $\rho= I/2 $ into the Gibbs state  $\rho'  =  \Gamma$, for  $\Gamma  =  2/3 \,  |0\>\<0|  +  1/3\,  |1\>\<1|$.   This task is achieved perfectly by the channel  $\map M_{\rm Gibbs}$ that maps every state into the Gibbs state. Channel $\map M_{\rm Gibbs}$ is a Gibbs preserving operation and, as such, it has zero cost.   Hence, the cost of the state transition $I/2 \mapsto \Gamma$ is zero.  On the other hand,  the state transition  $I/d \mapsto \Gamma$  can also be perfectly achieved  by any channel $\map M$ satisfying the relations $\map M(|0\>\<0| )   =     2/3\, |1\>\<1|  +   1/3 \,|0\>\<0|$ and $\map M(  |1\>\<1| )  =  |0\>\<0|$.  Any such channel $\map M$ is not Gibbs preserving and therefore has a strictly positive cost. Hence, if we just pick channel $\map M$ and ask what is the cost of $\epsilon$-approximating it,  we will generally get a positive cost, despite the fact that the cost of the state transition $I/2 \to \Gamma$ is zero.

	Instead of minimising the cost over all channels in an $\epsilon$-neighbourhood of a specific channel, here we minimise the cost over all physical implementations  that achieve a  accuracy  $F$ in the task $\map T$.  In this approach, the relevant minimisation problem is the one in Eq. (\ref{cococost}).   Indeed,   every implementation of the task  $\map T$  realises---by definition without error---a specific channel $\map M$.  The minimum cost over all implementations that realise channel $\map M$ without error is   $c_0(\map M, |\Psi\>)$.   Hence, our goal is to  minimise $c_0  (\map M,|\Psi\>)$ over all channels $\map M$ that achieve accuracy $F$.  These settings correspond to the minimisation problem in Eq. (\ref{cococost}).

	Let us now discuss the choice of the state $|\Psi\>$ in Eq. (\ref{cococost}). For $\epsilon=0$, the condition  $( \map M'  \otimes \map I_R  ) (|\Psi\>\<\Psi| )   = ( \map M  \otimes \map I_R  ) (|\Psi\>\<\Psi| )$  in Eq. (\ref{faist})  is equivalent to
	\begin{align}\label{sameonpia}
		\map M'  (\Pi_A \rho  \Pi_A)   =  \map M(\Pi_A \rho  \Pi_A) \, ,  \qquad \forall \rho \in\St (\spc H_A) \, ,   
	\end{align}
	where $\Pi_A$ is the projector on the support of the marginal state $\rho_A  : =   \Tr_R  [\Psi\>\<\Psi|]$.   
	Hence, the cost does not depend specifically on the state $|\Psi\>$, but only on the projector $\Pi_A$. From now on, we will denote by  
	\begin{align}
		c  (\map M, \Pi_A  )    := c_{\epsilon=0}    (\map M,  |\Psi\>)
	\end{align} the nonequilibrium cost that has to be paid for an  exact realisation of the channel $\map M$ upon input states in the support of $\Pi_A$.  
}



With this notation, we have the following: 
\begin{prop}
	The nonequilibrium cost of a channel $\map M$  upon inputs in the support of $\Pi_A$ 
	is upper bounded as  
	\begin{align}\label{easybound}
		c(\map M, \Pi_A)  \le  D_{\max}   (   \map M  (\Pi_A \Gamma_A \Pi_A)  \|   \Gamma_B ) \,,
	\end{align} 
	where $D_{\max}   (  \rho\| \sigma  ) :  = \log  \left\|     \sigma^{-\frac 12}  \rho  \sigma^{-\frac 12} \right\|$ is the  max relative entropy  \cite{datta2009min}.  When the  support of  $\Pi_A$  is invariant under the group of  time translations $U_t:  =  e^{-i t H_A/\hbar}$, $t\in \R$ (equivalently, when $[ \Pi_A ,  H_A]  =  0$),  the equality sign holds  and one has 
	\begin{align}\label{easyfaist}
		c (\map M,\Pi_A)      =  D_{\max}   (   \map M  (\Pi_A \Gamma_A \Pi_A)  \|   \Gamma_B ) 
		\, .
	\end{align}
\end{prop}

{\bf Proof.}  For $\epsilon =  0$, Eq.~(\ref{faist}) reads 
\begin{align}\label{aa}
	c  (\map M, \Pi_A)  
	&= \min_{ 
		\begin{array}{rl} 
			\map M'   (\Pi_A  \rho\Pi_A)    & =   \map M  (\Pi_A\rho \Pi_A )  \, , \forall \rho \in \St (A)   \end{array}}  
	\log\left \|   
	\Gamma_B^{-\frac 12}\map M'  (\Gamma_A)      \Gamma_B^{-\frac 12}\right\|
\end{align}

Choosing the quantum operation $\map M'$ defined by $\map M' (\rho)   :  =  \map M  (   \Pi_A  \rho \Pi_A )$, Eq.~(\ref{aa}) yields the inequality~$c  (\map M,\Pi_A)   \le    \log\left \|   
\Gamma_B^{-\frac 12}   \map M  (   \Pi_A  \rho \Pi_A )        \Gamma_B^{-\frac 12}\right\|  \equiv   D_{\max}   (   \map M  (\Pi_A \Gamma_A \Pi_A)  \|   \Gamma_B )$, thus proving Eq.~(\ref{easybound}).

If the support of $\Pi_A$ is time-invariant,  then $\Pi_A$  commutes with the Hamiltonian of system $A$ and  the Gibbs state can be written as $\Gamma_A   =  \Pi_A  \Gamma_A  \Pi_A   + (I_A  - \Pi_A) \, \Gamma_A \, (I_A  - \Pi_A)   $.     
Hence, we have the bound 
\begin{align}\label{cc}
	\nonumber \left \|   
	\Gamma_B^{-\frac 12}\map M'  (  \Gamma_A )         \Gamma_B^{-\frac 12}  \right\|      &  =     \left \|   
	\Gamma_B^{-\frac 12}\map M'  ( \Pi_A \Gamma_A \Pi_A)         \Gamma_B^{-\frac 12}  +   \Gamma_B^{-\frac 12}\map M'  ( (I_A  -\Pi_A) \Gamma_A (I_A  - \Pi_A))         \Gamma_B^{-\frac 12}  \right\|   \\
	\nonumber    &\ge    \left \|   
	\Gamma_B^{-\frac 12}\,  \map M'  ( \Pi_A \Gamma_A \Pi_A)       \,   \Gamma_B^{-\frac 12}\right\|\\
	&  =\left \|   
	\Gamma_B^{-\frac 12}\map M  (  \Pi_A\Gamma_A\Pi_A)      \Gamma_B^{-\frac 12}\right\| \, ,
\end{align}
where the  inequality is due to the relation $\|  A + B\|  \ge \|  A \|$ valid for arbitrary positive operators $A$ and $B$, and the second equality is due to the condition~(\ref{sameonpia}).   
Taking the logarithm on both sides, and minimising over $\map M'$, we obtain the inequality $c(\map M,\Pi_A)  \ge   D_{\max}   (   \map M  (\Pi_A \Gamma_A \Pi_A)  \|   \Gamma_B )$. 
\qed

\subsection*{Exact expression for  the accuracy-nonequilibrium tradeoff}

Here we provide an exact expression for the amount of nonequilibrium needed to achieve a desired level of accuracy in a given task.    

The problem is to find the quantum channel (completely positive trace-preserving map) $\map M$ that has minimum nonequilibrium cost among all  the channels that  attain accuracy at least $F$.  
{   To get started, we recall the expression for accuracy of a channel, defined in terms of a general performance test,  as in the Methods section. For a task $\map T$ defined by a set of input states $(\rho_x)_{x\in\set X}$ and  a set of output observables $(O_x)_{x\in\set X}$, the worst-case accuracy of the channel $\map M$ is defined as $\map F_{\map T} (\map M):  =   \min_x  \Tr  [O_x \map M (\rho_x)]$.    In the Choi representation, the accuracy can be expressed as $ \map F_{\map T} (\map M)   =  \min_x \Tr [  M   \Omega_x]$, with $\Omega_x   =   \rho_x^T\otimes O_x$.  
	In general, a performance test $\map T$ is specified by a set of performance operators $(\Omega_x)_{x  \in  \set X}$ and the accuracy is given by  $\map F_{\map T} (\map M)   =  \min_x \Tr [  M   \Omega_x]$.

	Now, consider the average accuracy with respect to a probability distribution $\st p  =  (p_x)_{x\in \set X}$, defined as 
}
\begin{align}
	\map F_{\map T,  \st p}   (\map M) :=  \Tr[  M \,  \Omega_{\st p}] \,,
\end{align} 
where  $\Omega_{\st p}  : =  \sum_x   \,   p_x\,  \Omega_x$ is the average performance operator associated to the given task. 
Note that the performance depends only on the projection of the Choi operator on the support of $\Omega_{\st p}$.    Defining the marginal operator  $\omega_A  :=  \Tr_B [\Omega_{\st p}]$,  the projector $\Pi_A$ onto the support of $\omega_A$, and the projected Choi operator $\widetilde M  :=  (  \Pi_A \otimes I_B  )  \, M\,   (\Pi_A\otimes I_B)$,   we have the  relation $\map F_{\st p}  (\map M) = \Tr [\widetilde M \,  \Omega_{\st p}]$, meaning that the accuracy depends only on  $\widetilde M$, rather than on the full operator $M$.

The nonequilibrium cost for the implementation of the channel $\map M$ on the support of $\Pi_A$ is given by Eq.~(\ref{easyfaist}).   In turn, the norm  in Eq.~(\ref{easyfaist})  can be expressed as 
\begin{align}\label{normsdp}
	\left\|   \Gamma_B^{-\frac{1}{2}}  \,  \map M  (\Pi_A \Gamma_A \Pi_A)\,    \Gamma_B^{-\frac 12} \right\|   =   \min_{ \map M  (\Pi_A \Gamma_A \Pi_A)  \le \lambda \,     \Gamma_B } \,  \lambda \, .
\end{align}
Now, recall that the action of a quantum channel on a given operator $\rho$ (not necessarily a quantum state) is given by $\map M (\rho)   =  \Tr_A [    ( \rho^T\otimes I_A )  \, M]$, where $\rho^T$ denotes the transpose of $\rho$ with respect to a fixed bases, here chosen to consist of energy eigenstates.  Choosing $\rho  = \Pi_A \Gamma_A \Pi_A$, we then obtain the equality  
\begin{align}
	\map M  (\Pi_A \Gamma_A \Pi_A)  =  \Tr_A  [    (I_B \otimes \Pi_A  \Gamma_A  \Pi_A)  \, M]  \, ,
\end{align}   
where we used the fact that $\Gamma_A $ and $\Pi_A$ are diagonal in the energy eigenbasis, and therefore $\Gamma_A^T  =  \Gamma_A$ and $\Pi_A^T  = \Pi_A$.  

Hence, the minimum  nonequilibrium cost  for achieving average  accuracy at least $F$, denoted by $c_{\map T, \bf p}    (F)  $, can be written as 
\begin{align}\label{sdpmother}
	\nonumber  c_{\map T,  \bf p}    (F)       &=   \min\{    c(\map M,  \Pi_A) ~|~  \map F_{\map T,\st p}  (\map M)  \ge F\} \\
	&  = \log     \,  \min_{\begin{array}{c}   
			M\ge 0  \, ,      \Tr_B [M]  =  I_A \\
			\lambda \, \Gamma_B  \ge  \Tr_A  [    ( \Pi_A  \Gamma_A  \Pi_A  \otimes I_B)  \, M]  \\
			\Tr[  M    \, \Omega_{\st p} ]  \ge  F  
	\end{array}   } \, \lambda   \, .
\end{align}
This minimisation problem is a semidefinite program and can be solved efficiently with existing software packages.

We now connect minimisation of the average cost to the minimisation of the worst case cost: 
\begin{prop}\label{prop:minmax}
	For every test $\map T$, one has the equality $c_{\map T}  (F)  =  \max_{\st p}  c_{\map T, \st p}  (F)$ for every $F\in  [F_{\min},  F_{\max}]$.  
\end{prop} 
{\bf Proof.}    The inequality 
\begin{align}\label{bastissima}
	c_{\map T} (F) \ge \max_{\st p}  c_{\map T, \st p}  (F)
\end{align} is immediate from the fact that the  worst case accuracy cannot be larger than the average    accuracy.   We now show the converse inequality.  To this purpose, we recall the definition 
\begin{align}\label{fc}
	F_{\map T}   (c)  :   =  \max_{\map M:  ~  c(\map M  ,  \Pi_A) \le c}    F_{\map T}  (\map M)      =    \max_{\map M:  ~  c(\map M  ,  \Pi_A) \le c}    \min_{\st p}    F_{\map T,\st p}   (\map M)  \, .
\end{align} 
and use the equality 
\begin{align}
	\nonumber 
	F_{\map T}   (c)   &  =\max_{\map M:  ~  c(\map M  ,  \Pi_A) \le c}    \min_{\st p}    F_{\map T,\st p}   (\map M)    \\ 
	\nonumber   &  =      \max_{\begin{array}{c}   
			M\ge 0  \, ,      \Tr_B [M]  =  I_A \\
			\Tr_A  [    ( \Pi_A  \Gamma_A  \Pi_A  \otimes I_B)  \, M]       \le 2^c\, \Gamma_B   \end{array}   }  \min_{\st p}   \, \Tr  [ M  \,  \Omega_{\st p} ]     \\
	\nonumber  & =    \min_{\st p}  \max_{\begin{array}{c}   
			M\ge 0  \, ,      \Tr_B [M]  =  I_A \\
			\Tr_A  [    ( \Pi_A  \Gamma_A  \Pi_A  \otimes I_B)  \, M]    \le 2^c\, \Gamma_B   \end{array}   } \, \Tr  [ M  \,  \Omega_{\st p} ]  \label{exchangeminmax}\\
	&  =\min_{\st p}   F_{\map T, \st p}  (c)   \, , 
\end{align} 
where we defined 
\begin{align}
	F_{\map T, \st p}  (c)   :  =  \max_{ \map M: ~  c(  \map M,  \Pi_A) \le c}  F_{ \map T,\st p  }  (\map M) \, . 
\end{align} 
The exchange of the minimum and maximum in Eq. (\ref{exchangeminmax}) is possible thanks to von Neumann's minimax theorem.   

For every fixed value of $c$, the equality $F_{\map T}  (c)=   \min_{\st p}   F_{\map T,\st p}  (c)$ implies that  there exists a probability distribution $\st p_0$ such that $F_{\map T, \st  p_0 }  (c) =   F_{\map T}  (c)$.  
Now, let   $\map M$  be an arbitrary  channel such that  $F_{\map T, \st p_0}  (\map M) \ge  F_{\map T} (c)  =:  F$.    By definition, the cost of this channel must satisfy $c(\map M, \Pi_A)  \ge  c$.    
Hence, we obtain  
\begin{align}
	c_{\map T,  \st p_0}   (F)   =   \min_{\map M  :~  F_{\map T, \st p_0}  ( \map M )    \ge F }    c  (\map M,  \Pi_A)  \ge  c    \,.
\end{align} 
On the other hand, the definition $F:  =  F_{\map T}  (c)$ implies   $c  =  c_{\map T}  (F)$ for every $F \in  [F_{\min},  F_{\max}]$. Hence, we obtained  the inequality  $c_{\map T,  \st p_0}   (F)   \ge    c_{\map T}(  F)$, and therefore
\begin{align}\label{bastissimissima}
	\max_{\st p}   c_{\map T, \st p}  (  F)    \ge  c_{\map T}   (F) \,. 
\end{align}
The thesis then follows from Eqs. (\ref{bastissima}) and (\ref{bastissimissima}). \qed

\medskip 

Summarising, we have derived the  following expression for the nonequilibrium cost: 
\begin{align}\label{ancoraqui}
	c_{\map T}  (F)  =  \max_{\st p}  c_{\map T, \st p}  (F)   =   \log      \max_{\st p}  \,  \min_{\begin{array}{c}   
			M\ge 0  \, ,      \Tr_B [M]  =  I_A \\
			\lambda \, \Gamma_B  \ge  \Tr_A  [    ( \Pi_A  \Gamma_A  \Pi_A  \otimes I_B)  \, M]  \\
			\Tr[  M    \, \Omega_{\st p} ]  \ge  F  
	\end{array}   } \, \lambda   \, .     
\end{align}  
\subsection*{Lower bound on the nonequilibrium cost}  

For every fixed probability distribution $\st p$, the optimisation over $M$ in Eq. (\ref{ancoraqui}) is a semidefinite program.    This semidefinite program   (\ref{sdpmother}) admits a dual formulation~(see {\em e.g.} \cite{watrous2018theory} for the basics of the duality theory), which yields the bound 
\begin{align}\label{sdp} \min_{\begin{array}{c}   
			M\ge 0  \, ,      \Tr_B [M]  =  I_A \\
			\lambda \, \Gamma_B  \ge  \Tr_A  [    (  \Pi_A  \Gamma_A  \Pi_A\otimes I_B )  \, M]  \\
			\Tr[  M\,  \Omega_{\st p} ]  \ge  F  \, .
	\end{array}   } \, \lambda      \ge   \max_{\begin{array}{c}   
			X_A \otimes I_B  + z\, \Omega_{\st p}     \le     \Pi_A  \Gamma_A  \Pi_A\otimes Y_B   \\
			\Tr[\Gamma_B  Y_B]   \le 1  
	\end{array}   } \, \Tr[X_A]      +  z\,     F 
\end{align}  
The inequality is in fact an equality, because the above program satisfies the condition of strong duality~\cite{watrous2018theory}, although this fact will not be used in the following. 

We now introduce a simplification  in the dual program~(\ref{sdp}).   The simplification  consists in  restricting the maximisation to triples $(X_A,Y_B,z)$ where the operator $X_A$ is set to zero.  This constraint leads to the new maximisation problem, whose optimal value is below the optimal value on the right-hand-side of ~(\ref{sdp}).   Explicitly, one has
\begin{align}\label{uno}
	\max_{\begin{array}{c}   
			X_A \otimes I_B + z\, \Omega_{\st p}     \le    \Pi_A  \Gamma_A  \Pi_A\otimes I_B    \\
			\Tr[\Gamma_B  Y]   \le 1  
	\end{array}   } \, \Tr[X]      +  z\,     F  \ge   F \,  \left(  \max_{\begin{array}{c}   
			z\, \Omega_{\st p}     \le     \Pi_A  \Gamma_A  \Pi_A \otimes Y_B   \\
			\Tr[\Gamma_B  Y_B]   \le 1  
	\end{array}   } \,      z\right)   \, . 
\end{align}  

The new maximisation problem  admits a closed-form solution. To find it,  we use the change of variables $\sigma_B : = \Gamma_B^{1/2} \, Y_B \, \Gamma_B^{1/2}$.     With this change of variable, the inequality  $ z\, \Omega_{\st p}     \le      \Pi_A  \Gamma_A  \Pi_A \otimes Y_B  $ becomes $ z\, \Omega_{\st p}     \le    \Pi_A  \Gamma_A  \Pi_A\otimes  \Gamma_B^{-1/2} \sigma_B \Gamma_B^{-1/2} $.   In turn, this inequality is equivalent to  $    (  \Gamma_A \otimes  \Gamma_B^{-1})^{1/2} \, \Omega_{\st p}  \,  (    \Gamma_A \otimes  \Gamma_B^{-1})^{1/2}    \le   I_A \otimes Y_B/z$.  
Finally, defining the operator $\Lambda_B   =  \sigma_B/z$, we obtain the condition    $ I_A \otimes \Lambda_B \ge     (  \Gamma_A \otimes  \Gamma_B^{-1})^{1/2}   \, \Omega_{\st p}  \,    (  \Gamma_A \otimes  \Gamma_B^{-1})^{1/2}     =:   \omega_{\map T,\st p}$, where the subscript $\map T$ stresses the dependence of the operator $ \omega_{\map T,\st p}$ on the task $\map T$.   
Hence, we obtained the equality
\begin{align}\label{due}
	\nonumber    \max_{\begin{array}{c}   
			z\, \Omega_{\map T, \st p}     \le     \Pi_A  \Gamma_A  \Pi_A  \otimes Y_B \\
			\Tr[\Gamma_B  Y_B]   \le 1  
	\end{array}   } \,      z    &=    \left(  \min_{\begin{array}{c}   
			I_A \otimes  \Lambda_B     \ge    \omega_{\map T,  \st p} 
	\end{array}   } \,      \Tr[\Lambda_B]   \right)^{-1}\\
	&  =  2^{ H_{\min}(A|B)_{\omega_{\map T, \st p}}}  \, .
\end{align}  
Combining Eqs.~(\ref{sdpmother}),~(\ref{uno}), and~(\ref{due}), we  obtain the bound 
\begin{align}
	c_{\map T,  \bf p} (F)  \ge       H_{\min} (A|B)_{\omega_{\map T, \st p}}     +  \log F  \, , 
\end{align}
which follows from Eq.~(\ref{easyfaist}).    Maximising the l.h.s.  over the choice of probability distribution $\st p$ and using Proposition \ref{prop:minmax}  we then  obtain 
\begin{align}
	c_{\map T} (F)  \ge       H_{\min} (A|B)_{\omega_{\map T, \st p}}     +  \log F  \qquad \forall \st p \, .
\end{align}
Finally, we maximise the r.h.s. over all probability distributions, thus obtaining
\begin{align}\label{eq:main}
	c_{\map T}  (F ) \ge       \kappa_{\map T}      +  \log F  \, ,
\end{align}
where     $\kappa_{\map T}  =  \max_{\st p} H_{\min} (A|B)_{\omega_{\map T, \st p}}$ is the reverse entropy defined in the main text.

Eq.~(\ref{eq:main}) can be  reformulated as a bound on the  maximum accuracy achievable with a given amount of nonequilibrium resources, namely
\begin{align}\label{eq:main_f}
	F_{\map T}  (c)\le     2^{  c-\kappa_{\map T}} \, .
\end{align}

\section*{Supplementary Note 2:   thermodynamic irreversibility of classical deterministic computations}

For the computation of a function $f:  A\to B,  x\mapsto  f(x)$, where $A$ and $B$ are two finite sets,   one has 
\begin{align}
	\nonumber \omega_{\map T,  \st p}       &=  \sum_{x  \in  A}    \, p_x   \, \frac {g_{B}  (f(x))}{g_{A}  (x)} \,   |x\>\<x| \otimes  |f(x)\>\<f(x)  |    \\
	&   =      \sum_{y \in  f(A)}        \,  g_{B}  (y)  \,   \left(   \sum_{x :   f(x)  =  y}    \, \frac {p_x}{g_{A}(x)} \,  |x\>\<x|\right)    \otimes  |y\>\<y|     \, ,
\end{align} 
where $g_A(x)$ and $g_B(y)$ are the Gibbs probability distributions for the input and the output, respectively. Hence, the condition $I_A \otimes  \Lambda_B  \ge   \omega_{\map T, \st p}$ is equivalent to  
\begin{align}
	\<y|  \Lambda_B  |y\>   \ge     g_{B}  (y)  \,     \max_{x :   f(x)  =  y}    \, \frac {p_x}{g_{A}  (x)} \,    \qquad\forall y\in  f(A) \, .
\end{align}

Choosing the operator $\Lambda_B$  that achieves the equality, we obtain  
\begin{align}
	H_{\min}(A|B)_{\omega_{\map T, \st p}}   =  -   \log \left(     \sum_{y\in  f(A)}  \, g_{B}  (y)  \,     \max_{x :   f(x)  =  y}    \, \frac {p_x}{g_{A}  (x)}\right) \, .
\end{align}

Now, we need to maximise the above expression over all probability distributions $\st p =  (p_x) $.    First, note that the maximum is attained by probability distributions satisfying the condition 
\begin{align}
	\frac {p_x}{g_{A}  (x)}         =    \frac {p_{x'}}{g_{A}  (x')}  \qquad \forall x,x':    \,  f(x) =  f(x') \, .
\end{align}
Second, we define the probabilities $q_y  :  = \sum_{x:  f(x)   =  y}    p_x$ and $p_f(y)   :=   \sum_{x :  f(x)  =  y} \,  g_{A}  (y)$.  
With this notation, we have the equality 
\begin{align}
	\nonumber \max_{\st p} \,  H_{\min}(A|B)_{\omega_{\map T, \st p}}       &   =   \max_{\st q}       -   \log \left(     \sum_{y\in  f(A)}  \, g_{B}  (y)  \,      \, \frac {q_y}{p_{f}  (y)}\right)  \\
	\nonumber  &  =     -   \log \left(     \min_{y\in  f(A)}  \,     \, \frac { g_{B}  (y) }{p_{f}  (y)}\right)   \\
	&  =  D_{\max}    (    p_f  \|   g_{B})   \, .
\end{align}
Hence, the reverse entropy of the classical computation of $f$ is 
\begin{align}\label{thermof}
	\kappa_f  =  D_{\max}    (    p_f  \|   g_{B})     \, .
\end{align}

\section*{Supplementary Note 3: proof of the attainability criterion}

In this note,  we prove  the attainability criterion given in the main text: specifically, we show that, if the bound~(\ref{eq:main}) is saturated for the maximum accuracy $F = F_{\rm  max}$, then the bound is tight for all values of the accuracy in the interval $F \in [F_{\min}, F_{\rm max}]$.  

\subsection{Evaluation of $F_{\min}$}
By definition,   $F_{\min}$ is the maximum worst-case fidelity achieved by a physical process with minimum nonequilibrium cost on the input subspace, namely  $F_{\min}  : = \map F_{\map T}(c_{\min})$ where 
\begin{align}c_{\min}   : = \min_{\map M}  \,  c(\map M, \Pi_A)   
\end{align} 
is the minimum nonequilibrium cost over all possible quantum channels acting on the support of $\Pi_A$.   The minimum nonequilibrium cost can be readily evaluated, as shown  in the following lemma:  
\begin{lemma}\label{lem:cmin}
	The mininimum nonequilibrium cost for the realisation of an quantum channel upon inputs in the support of a projector $\Pi_A$ with $[\Pi_A,H_A]=0$     is 
	\begin{align}\label{cmin}
		c_{\min}  = \log \Tr[\Pi_A \,  \Gamma_A ]\,,
	\end{align}  and it  is achieved by every quantum channel $\map M$ satisfying the condition  
	\begin{align}\label{cccc}
		\map M   (\widetilde \Gamma_A)  =  \Gamma_B \,, \qquad  {\rm with}  \qquad  \widetilde{\Gamma}_A  :  =  \frac{  \Pi_A  \Gamma_A  \Pi_A}{\Tr [ \Pi_A  \Gamma_A  \Pi_A]} \, .
	\end{align}
\end{lemma} 

{\bf Proof. } For an arbitrary channel $\map M$,   Eq.~(\ref{easyfaist}) yields the expression
\begin{align}
	\nonumber c(\map M,\Pi_A)  &    =  D_{\max}   (   \map M  (\Pi_A \Gamma_A \Pi_A)  \|   \Gamma_B )    \\
	\nonumber &  = \log    \left\| \Gamma_B^{-\frac 12}   \,    \map M (  \Pi_A  \, \Gamma_A  \,  \Pi_A) \,  \Gamma_B^{-\frac 12}  \right\|   \\
	\nonumber   &  =  \log \Tr  [  \Pi_A \, \Gamma_A \, \Pi_A]      +   \log      \left\| \Gamma_B^{-\frac 12}   \,    \map M ( \widetilde \Gamma_A)   \,  \Gamma_B^{-\frac 12}  \right\|  \\
	&   = \log \Tr  [  \Pi_A \, \Gamma_A ]    +  D_{\max}  (\map M (\widetilde \Gamma_A) \,\|\,   \Gamma_B)  \,  \,.
\end{align} 

Since the max relative entropy is non-negative, the above expression implies the bound $c(\map M, \Pi_A)  \ge  \Tr[\Pi_A \,  \Gamma_A]$, valid for every quantum channel $\map M$, and the equality holds if and only if $D_{\max}  (\map M (\widetilde \Gamma_A) \, \|\,   \Gamma_B)  =  0$, that is, if and only if condition~(\ref{cccc}) is satisfied.     \qed 

\medskip

We now consider the set of all channels with minimum nonequilibrium cost $c_{\min}$, and, among them, we search for  the channel with maximum accuracy.  
\begin{lemma}\label{lem:fmin}
	If the bound~(\ref{eq:main}) is attainable for a given value $F=  F_{0}$, then $F_{\min}= 2^{c_{\min}  - \kappa_{\map T}}$.  
\end{lemma}

{\bf Proof. }   
Note that, in general one has the relation
\begin{align}
	F_{\min}   & :  =  \max_{ \map M:   c (\map M,  \Pi_A) \le c_{\min}}   \map F_{\map T} (\map M)  \equiv \max_{ \map M:   c (\map M,  \Pi_A)  =  c_{\min}}     \map F_{\map T} (\map M)  \le    \max_{ \map M:   c (\map M,  \Pi_A)  =  c_{\min}}      \le 2^{c  (\map M,\Pi_A)  - \kappa_{\map T}} = 2^{c_{\min}  - \kappa_{\map T}}  =:  F_*   \,,   \label{minbound}
\end{align}   
where the inequality follows from Eq.~(\ref{eq:main_f}).  Our goal is to show that the bound  $F_{\min}\le F_*$  holds with the equality sign whenever the bound~(\ref{eq:main}) is attainable for a given value $F=  F_{0}$.  

Let  $\mathcal{M}_{0}$ be a quantum channel that achieves the bound ~(\ref{eq:main}) at $F  =  F_{0}$, namely 
\begin{equation}\label{satmax}
	\map F_{\map T} (\map M_0)   =  F_0 \qquad {\rm and}  \qquad c(\mathcal{M}_{0},\Pi_A) = \kappa_{\map T}+ \log  F_{0} \, . 
\end{equation}  

Let us consider the case   $F_{0}  = F_*$.  In this case, we have  
\begin{align}\label{qwerty}
	2^{c (\map M_0,  \Pi_A)  -  \kappa_{\map T}}  =  F_0   = F_*  =    2^{  c_{\min}  -  \kappa_{\map T}} \, ,
\end{align} 
which implies $c  (\map M_0, \Pi_A)   =   c_{\min}$.   Since $F_{\min}$ is the maximum accuracy achieved by quantum channels with cost $c_{\min}$, we conclude that $F_*  =  \map F_{\map T} (\map M_0)   \le  F_{\min}$. Combined with the Eq. (\ref{minbound}), this bound implies $F_*=  F_{\min}$.

Now, suppose that  $F_{0}  >  f$.  Consider the parametric family of quantum channels $\map M_F$ of the form 
\begin{equation}\label{MF}
	\mathcal{M}_F := p_F \, \mathcal{M}_{0} + (1 - p_F)\, \chi_B \, \Tr_A \,   \qquad  F \in  [F_{\min},  F_{0}] \, ,
\end{equation}
where $\chi_B$ is a fixed quantum state  (to be determined later) and $p_F  :  =  F/ F_{0}$. 

Note that the accuracy of the channel $\map M_F$ is at least $F$,  as one has
\begin{align}\label{Fatleast}
	\map  F_{\map T}(\map M_F)    \ge   p_F  \,  F_{0}       =  F     
\end{align}
(the bound following from the fact that the performance operators  are nonnegative). 

We now set $F$ to $F_*$, and  choose the state $\chi_B$ so that  the channel $\map M_{F_*}$ has minimum cost.  The minimum cost condition~(\ref{cccc})  implies 
\begin{align}\label{bbbb}
	\chi_B  =  \frac{\Gamma_B  -   p_{F_*} \,   \map M_{0} (\widetilde \Gamma_A)}{   1  -  p_{F_*}  }  \,. 
\end{align}
This expression is well-defined because $p_{F_*}  =  F_*/F_0 $ is strictly smaller than 1.

We now show that Eq.~(\ref{bbbb}) defines a valid quantum state.    It is immediate to see that the operator $\chi_B$ has unit trace.  It remains to show that  $\chi_B$ is positive semidefinite. To this purpose, note that the condition $\chi_B\ge 0$ is equivalent to  $\Gamma_B  \ge p_{F_*}    \,  \map M_{0}  (\widetilde \Gamma_A)$,      which  is equivalent to  $ I_B  \ge   p_{F_*}  \,  \Gamma_B^{-\frac 12}  \, \map  M_{0}  (\widetilde \Gamma_A)  \,  \Gamma_B^{-\frac 12}$. In turn, this condition is equivalent to 
\begin{align}\label{aaaa}
	1 \ge   p_{F_*}  \,  \left\|\Gamma_B^{-\frac 12}  \, \map  M_{0}  (\widetilde \Gamma_A)  \,  \Gamma_B^{-\frac 12}\right\|  \, .
\end{align} 
We now show that Eq.~(\ref{aaaa}) is satisfied.    Inserting the definition of the state $\widetilde \Gamma_A$  [Eq.~(\ref{cccc})] into Eq.~(\ref{aaaa}), we obtain   
\begin{align}
	\nonumber  p_{F_*}  \,  \left\|\Gamma_B^{-\frac 12}  \, \map  M_{0}  (\widetilde \Gamma_A)  \,  \Gamma_B^{-\frac 12}\right\|    & =   \frac{p_{F_*}}{\Tr[\Pi_A \, \Gamma_A\,  \Pi_A]}    \,  \left\|\Gamma_B^{-\frac 12}  \, \map  M_{0}  (\Pi_A \,\Gamma_A  \,\Pi_A)  \,  \Gamma_B^{-\frac 12}\right\|\\
	&  =  \frac{p_{F_*}\,  2^{c (\map M_{0}, \Pi_A) }}{2^{c_{\min}}}     \,,
\end{align} 
having used Eqs.~(\ref{easyfaist}) and~(\ref{cmin}).   Then, inserting the definition $p_{F_*}: =  F_*/F_0 = 2^{  c_{\min}  -\kappa_{\map T}}/  F_0$ and the relation $2^{  c(\map M_0,\Pi_A)}  =  F_0  2^{\kappa_{\map T}}$  (following from  Eq.~(\ref{satmax})) in the right hand side, we obtain  
\begin{align}
	\nonumber  p_{F_*}  \,  \left\|\Gamma_B^{-\frac 12}  \, \map  M_{0}  (\widetilde \Gamma_A)  \,  \Gamma_B^{-\frac 12}\right\|     &  =  \frac{2^{c_{\min}- \kappa_{\map T}}}{  F_{0}}\,      \,         \frac{ F_{0} \, 2^{\kappa_{\map T}}   }{  2^{c_{\min}}}   \\
	&  = 1   \,,
\end{align} 
Hence, condition~(\ref{aaaa}) is satisfied.  

In summary,  the operator $\chi_B$ is a valid quantum state, and therefore the map $\map M_{F}$ defined in Eq.~(\ref{MF}) is a valid quantum channel.  In particular,  $\map M_{F_*}$ is a valid quantum channel.   The nonequilibrium cost of $\map M_{F_*}$ is $c_{\min}$ and its accuracy is at least $F_*$.  Since $F_{\min}$ is the maximum accuracy achievable with $c_{\min}$ clean qubits, we have the inequality $F_{\min}  \ge  F_*  = 2^{c_{\min}  - \kappa_{\map T}}  $.   On the other hand, Eq.~(\ref{minbound}) gives the bound  $F_{\min }  \le  2^{c_{\min}  - \kappa_{\map T}}   $.  Hence, we conclude that the equality $F_{\min}  =   2^{c_{\min} - \kappa_{\map T}}$ holds. \qed  

\medskip 
\begin{corollary}
	If the input subspace is the whole Hilbert space (i.e. $\Pi_A  =  I_A$),  then   the equality $F_{\min}  = F_{\max}^{\rm rev}$ holds, where $F_{\max}^{\rm rev}$ is the accuracy of the time-reversed task defined in the Methods section of the main text.
\end{corollary} 
{\bf Proof.}  By  Lemma \ref{lem:fmin}, the minimum fidelity is $F_{\min}  =  2^{  c_{\min}  -  \kappa_{\map T}}$.  If the input subspace is the whole Hilbert space, then the projector $\Pi_A $ is the identity operator, and Lemma  \ref{lem:cmin} yields $c_{\min} = 0$.    Hence, $F_{\min}  =  2^{-\kappa_{\map T}}$.  Recalling the definition of the reverse entropy $\kappa:  = \log  (1/F_{\max}^{\rm rev})$, one finally obtains the equality $F_{\min} = F_{\max}^{\rm rev} $. \qed
\subsection*{Proof of the attainability criterion}

\begin{theorem}
	If the bound~(\ref{eq:main}) is attainable at $F=  F_{0}$, then it is attainable for every $F  \in  [F_{\min},  F_{0}]$, with $F_{\min}  =  2^{c_{\min}  -\kappa_{\map T}}$. 
\end{theorem}
{\bf Proof.}  Let  $\map M_{0}$ be the channel that saturates the bound~(\ref{eq:main})  at $F=  F_{0}$, and let $\map M_F$ be the channel defined in Eq.~(\ref{MF}).    The nonequilibrium cost of $\map M_F$, given by Eq.~(\ref{easyfaist}),  is 
\begin{align}
	\nonumber c (\map M_F,  \Pi_A) & =  \log\norm{\Gamma_B^{- \frac{1}{2}}
		\mathcal{M}_F (\Pi_A\,  {\Gamma}_A  \,\Pi_A) \Gamma_B^{-
			\frac{1}{2}}}\\
	&  = \log \Tr [  \Pi_A \,  \Gamma_A]     +    \log\norm{\Gamma_B^{- \frac{1}{2}}
		\mathcal{M}_F (\widetilde{\Gamma}_A  ) \Gamma_B^{-
			\frac{1}{2}}} \, ,  \label{eeee}
\end{align}
where $\widetilde \Gamma_A$ is defined as in Eq.~(\ref{cccc}).  

Using the definition of $\map M_F$, we obtain  
\begin{align}
	\nonumber \mathcal{M}_F (\widetilde{\Gamma}_A  )      &=   p_F  \, \map M_{0}   (\widetilde{\Gamma}_A  )     +  (1-p_F)  \,  \chi_B  \\
	&  =   p_F  \, \map M_{0}   (\widetilde{\Gamma}_A  )     +  (1-p_F)  ~ \frac{\Gamma_B -  p_{F_*}  \,  \map M_{0}    (\widetilde{\Gamma}_A  )  }{  1-  p_{F_*}} \, , 
\end{align}   
the second equality following from Eq.~(\ref{bbbb}).  Rearranging the terms, we  obtain  
\begin{align}
	\mathcal{M}_F (\widetilde{\Gamma}_A  )      &  =\frac{ 1- p_F }{1-p_{F_*}}  \, \Gamma_B +  \frac{p_F  -   p_{F_*}}{1-p_{F_*}}\,  \map M_{0}  (\widetilde{\Gamma}_A  )  \, , 
\end{align}
and therefore,  
\begin{align}
	\nonumber  \norm{\Gamma_B^{- \frac{1}{2}}
		\mathcal{M}_F (\widetilde{\Gamma}_A  ) \Gamma_B^{-
			\frac{1}{2}}}      & =\left\|   \frac{ 1- p_F }{1-p_{F_*}}  \,   I_B +  \frac{p_F  -   p_{F_*}}{1-p_{F_*}}\,  \Gamma_B^{-\frac 12}\, \map M_{0}  (\widetilde{\Gamma}_A  ) \,  \Gamma_B^{-\frac 12}   \right\|\\
	\nonumber 			&   =   \frac{ 1- p_F }{1-p_{F_*}}     +   \frac{p_F  -   p_{F_*}}{1-p_{F_*}}\,  \left\| \Gamma_B^{-\frac 12}\, \map M_{0}  (\widetilde{\Gamma}_A  ) \,  \Gamma_B^{-\frac 12}   \right\|  \\
	\nonumber &   =   \frac{ 1- p_F }{1-p_{F_*}}     +   \frac{p_F  -   p_{F_*}}{1-p_{F_*}}\,   \frac{ \left\| \Gamma_B^{-\frac 12}\, \map M_{0}  (\Pi_A \, {\Gamma}_A  \, \Pi_A) \,  \Gamma_B^{-\frac 12}   \right\|}{\Tr[  \Pi_A \, \Gamma_A]  }\\
	&  =  \frac{ 1- p_F }{1-p_{F_*}}     +   \frac{p_F  -   p_{F_*}}{1-p_{F_*}}\,   2^{c  (\map M_{0}, \Pi_A)  - c_{\min}   } \, ,
\end{align}
where the last equation follows from Eqs.~(\ref{cmin}) and~(\ref{easyfaist}).   Using Eq.~(\ref{satmax}) and~the definition $F_* = 2^{c_{\min}-\kappa_{\map T}}$, we then obtain  
\begin{align}
	\nonumber  \norm{\Gamma_B^{- \frac{1}{2}}
		\mathcal{M}_F (\widetilde{\Gamma}_A  ) \Gamma_B^{-
			\frac{1}{2}}}      
	\nonumber		&  =  \frac{ 1- p_F }{1-p_{F_*}}     +   \frac{p_F  -   p_{F_*}}{1-p_{F_*}}\,   \frac{  F_{0}}{F_*}\\
	\nonumber		&  =   \frac{ 1- p_F }{1-p_{F_*}}     +   \frac{p_F  -   p_{F_*}}{1-p_{F_*}}\,   \frac{ 1}{p_{F_*}}\\
	\nonumber &  =  \frac{p_F}{p_{F_*}} \\
	&  =  \frac F{F_*} \, .\label{ffff}
\end{align}
Inserting Eqs.~(\ref{cmin}) and~(\ref{ffff})   into Eq.~(\ref{eeee}), we finally obtain  
\begin{align}
	\nonumber c (\map M_F,  \Pi_A) & =    c_{\min}    +  \log    \frac{F}{F_*}   \\
	&  =   \kappa_{\map T}  +   \log F \, ,
\end{align}
where the last equality follows from the definition $F_*:  =  2^{c_{\min}-\kappa_{\map T}}$.  

Summarising, the channel $\map M_F$ has  nonequilibrium cost $\kappa_{\map T}+  \log F$, and accuracy at least $F$ (by Eq.~(\ref{Fatleast})).   
To conclude, we observe that  channel $\map M_F$ has accuracy $ \map{F}_{\map T}  (\map M_F)  =  F$. Indeed, we know from Eq.~(\ref{Fatleast}) that the accuracy satisfies the inequality $ \map{F}_{\map T}  (\map M_F)  \ge  F$. On the other hand, the bound~(\ref{eq:main}) implies the inequality  
\begin{align}
	\nonumber   \kappa_{\map T}  +   \log F  &   =  c (\map M_F,  \Pi_A)  \\
	&  \ge  \kappa_{\map T} + \log \map{F}_{\map T}  (\map M_F)   ,
\end{align}      
which implies  $\map {F}_{\map T}   (\map M_F)    \le  F$. 

Summarising, the channel $\map M_F$ has  nonequilibrium cost $\kappa_{\map T}+  \log F$, and accuracy $F$ (by Eq.~(\ref{Fatleast})).  Hence, it saturates the bound~(\ref{eq:main}). \qed

\section*{Supplementary Note 4: Attainability  results}

{\bf Classical deterministic computations.}    Here we show that the bound  (\ref{eq:main}) is attainable for every classical deterministic computation.  Let $ f:  \,    \{1,\dots,  d_A\} \to    \{1,\dots,  d_B\}  \, , x\mapsto y=  f(x)$ be a function between two finite sets.  Mathematically, a classical process that evaluates the function $f$ without errors can be represented as a quantum channel $\map M_f$, whose action on a generic input state $\rho$ is given by 
\begin{align}\label{Cf}
	\map M_f (\rho)   =  \sum_x     \<  x |\rho |x\>  \,    |f(x)\>\<f(x)|  \, .
\end{align}  
For an approximate realisation of erasure, corresponding to a different channel $\map M$, we consider the fidelity $F_f(\map M)  =\min_x  \<f(x)|  \map M(|x\>\<x|)  |f(x)\>$ as the figure of merit. Operationally, the fidelity corresponds to the probability that the channel $\map M$ computes the correct value of $f(x)$, in the worst case over all possible inputs $x$.   

\begin{prop}
	The nonequilibrium cost for approximately computing the function $f$ with fidelity $F$ is $c_f (F)=  \kappa_f  + \log F$, where $\kappa_f  =D_{\max}  (p_f \|  g_B)$ is the reverse entropy derived in Supplementary Note 2. The equality holds for every $F$ in the interval $[F_{\min},F_{\max}]$, with $F_{\min}  = 2^{c_{\min}-\kappa_f}$ and $F_{\max}=1$. 
\end{prop}
{\bf Proof.} Applying the channel to the Gibbs state $\Gamma_A$, we obtain  
\begin{align}
	\map M_f(\Gamma_A)   =  \sum_{y \in  f(A)}  \,   p_f (y) \,  |y\>\<y|  \qquad p_f(y)  :  =    \sum_{x:  f(x)  =  y}  \,    g_A  (x) \, .  
\end{align}
Hence, the nonequilibrium cost of channel $\map M_f$, given by Eq.~(\ref{easyfaist}) is  
\begin{align}\label{costofmf}
	c  (  \map M_f, \Pi_A) =  D_{\max}  (  \map M  (\Gamma_A)\|  \Gamma_B)  =  D_{\max}  (p_f \|  g_B) \,.
\end{align}  
Comparing this equation with Eq.~(\ref{thermof}), we obtain the equality 
\begin{align}\label{minni}
	c  (  \map M_f,\Pi_A)  =  \kappa_{f} \, ,
\end{align}   
which implies
\begin{align}
	c_f (1)   \le     c  (  \map M_f,\Pi_A)    =  \kappa_{f}   \,, 
\end{align}
because the channel $\map M_f$ has unit accuracy.  

On the other hand, evaluating the bound  (\ref{eq:main})  at $F=1$ yields the converse inequality 
\begin{align}
	c_f ({1})  \ge  \kappa_{f}    \, .  
\end{align} 
Hence, the equality $c_f ({1})  = \kappa_{f}$ holds.  

Summarising,   the bound  (\ref{eq:main}) holds  at $F=1$ with the equality sign. 
Hence, Theorem \ref{theo:achievability} implies that the bound  (\ref{eq:main})  holds for every $F$ in the interval $[F_{\min},  1]$.  \qed

\medskip

{\bf Quantum extensions of classical computations.}   Here we show that the bound  (\ref{eq:main}) is attainable for every quantum extension  (in the sense defined precisely in the following) of a classical computation.

Let $ f:  \,    \{1,\dots,  d_A\} \to    \{1,\dots,  d_B\}  \, , x\mapsto y=  f(x)$ be a function between two finite sets, and let $\widehat{\map M}_f$ be a quantum channel from a $d_A$-dimensional input system $A$ to a $d_B$-dimensional output system $B$, satisfying the condition
\begin{align} 
	\widehat{\map M}_f (|x\>\<x|)  =  |f(x)\>\<f(x)| \qquad \forall x\in  \{1,\dots, d_A\} \, .
\end{align} 
We call the channel $\widehat{\map M}_f$ a {\em quantum extension} of the function $f$.  

Now, consider the  task of transforming the input state $\rho_x$ into the target output state $\rho_x'  = \map M_f (\rho_x)$, where the input state  $\rho_x$ is picked from a set that includes  the classical states $\{|x\>\<x|\}_{x=1}^{d_A}$.      When the output states  $\rho_x'$ are mixed, we assume that the figure of merit  $\widehat{\map F}_f$ for the state transformation task $\rho_x\mapsto \widehat{\map M}_f  (\rho_x)$ has been chosen  in such a way that the channel $\widehat{\map M}_f$  has accuracy $\widehat{\map F}_f  (\widehat{\map M}_f)   =1$.    When these conditions are satisfied, we call the state transformation task  $\rho_x\mapsto \widehat{\map M}_f  (\rho_x)$ a quantum extension of the classical task of computing the function $f$. 

\begin{prop}\label{prop:quantumextensioncost}
	Let $\widehat {\map M}_f$ be a quantum extension of the function $f$, and let the state transformation task $\rho_x\mapsto  \widehat {\map M}_F (\rho_x)$ be a quantum extension of the task of computing the function $f$.      The quantum extension specified by $\widehat {\map M}_f$ has
	\begin{enumerate}
		\item  reverse entropy  $\widehat \kappa_{f} =  \kappa_{f}$, where $\kappa_f$ is the reverse entropy of the computation of $f$, derived in Supplementary Note 2, and
		\item nonequilibrium cost $\widehat c_f (F)  =  \kappa_f  +  \log F$.  
	\end{enumerate}
	The second equality holds for every $F$ in the interval $[F_{\min},F_{\max}]$, with $F_{\min}  = 2^{c_{\min}-\kappa_f}$ and $F_{\max}=1$. 
\end{prop}

{\bf Proof.} Since the set of input states for the  task $\rho_x \mapsto   \map M_f (\rho_x)$ includes the  classical states $(|x\>\<x|)_{x=1}^{d_A}$, the reverse entropy of the quantum-extended task, denoted by $\kappa_{\widehat{\map M}_f}$,    is generally larger than the reverse entropy of the original classical computation  task, namely
\begin{align}
	\widehat \kappa_{f }  \ge  \kappa_f     \, .
\end{align}
Evaluating the bound  (\ref{eq:main})  at $F=1$, we then obtain the  inequality 
\begin{align}\label{pluto}
	\widehat c_{f}  (1)  \ge  \widehat \kappa_{f}  \ge  \kappa_f    \, .  
\end{align}

On the other hand, the nonequilibrium cost of the channel $\widehat{\map M}_f$ is exactly the same as the nonequilibrium cost of the channel $\map M_f$ in Eq.~(\ref{Cf}), because the two channels act in the same way on the Gibbs state.   Hence, we have the equality 
\begin{align}
	c (\widehat{\map M}_f, \Pi_A) =    c (\map M_f,  \Pi_A)  =   \kappa_f  \, ,
\end{align}
where the second equality follows from  Eq. (\ref{minni}).  

Since the channel $\widehat{\map M}_f$ achieves the state transformation task  $\rho_x\mapsto  \widehat{\map M}_f  (\rho_x)$ with unit accuracy, the above equality implies the bound  
\begin{align}\label{pippo}
	\widehat c_{f}  (1)    \le   c (\widehat{\map M}_f, \Pi_A) =   \kappa_f    \,.
\end{align}

Comparing Eqs. (\ref{pluto}) and (\ref{pippo}), we  obtain the following chain of inequalities 
\begin{align}
	\kappa_f    \le \widehat \kappa_{f}  \le    \widehat c_f (1)  \le    c (\widehat{\map M}_f,\Pi_A) =   \kappa_f \, , 
\end{align} 
which imply the equalities  
\begin{align}\label{kappaextkappaf}
	\widehat \kappa_{f}  =   \kappa_f     
\end{align}
and
\begin{align}
	\widehat c_{f}  (1)  = \widehat \kappa_{f} \, . 
\end{align}
The second equality implies that  the bound  (\ref{eq:main}) holds  at $F=1$ with the equality sign. 
Hence, Theorem \ref{theo:achievability} guarantees that the bound  (\ref{eq:main})  holds for every $F$ in the interval $[F_{\min},  1]$, and $\widehat c_f (F)   = \widehat \kappa_{f}   +  \log F   \equiv  \kappa_f + \log F$. \qed  

\medskip

{ 
	{\bf Erasure of quantum states.}    Here we show that the work cost for  the approximate erasure of a $d$-dimensional quantum system to the ground state is  
	\begin{align}\label{wferasure}
		W_{\rm erase} (F )  =    \frac{\Delta A}{kT}    +  \ln F  \, .
	\end{align}
	where $\Delta A$ is the difference between the free energy of the ground state and the free energy of the Gibbs state.  
	
	The erasure  task corresponds to the  state transformation $\rho  \mapsto  |0\>\<0|$, where $\rho$ is an arbitrary pure state.  As the figure of merit for approximate erasure,  we consider the erasure fidelity $F_{\rm erase}   (\map M)   =   \min_\rho   \<  0 | \,  \map M   (\rho)  \, |0\>$, where $\map M$ is the quantum channel used to  implement the erasure task.  
	
	\begin{lemma}\label{lem:kappaerasure}
		The reverse entropy of the erasure task is $\kappa_{\rm erase}  =  \frac{\Delta A}{kT\,  \ln 2}  $.
	\end{lemma}
	
	{\bf Proof.}  The state transformation $\rho  \mapsto  |0\>\<0|$ is  the quantum extension of the classical deterministic computation $f_0 :    \,  x  \mapsto  f(x)  =  0$, where the classical input $x\in  \{0,\dots,  d-1\}$ is encoded in the computational basis state $|x\>$,  as discussed in the previous section of this Supplementary Note.    Hence, the reverse entropy of the erasure task can be computed with Eqs. (\ref{kappaextkappaf}) and (\ref{thermof}), which give
	\begin{align}
		\nonumber \kappa_{\rm erase}     &  =  D_{\max}  (p_{f_0} \|  g_B)   \\
		\nonumber &  =  \log \frac1{  g_B(0)}\\
		\nonumber &  =   \log [ e^{E_0/(kT)}  \,  Z]  \\
		\nonumber &  =      \frac{E_0}{kT} \,  \log e    + \log Z \\ 
		\nonumber &  =  \frac{   A_0  -  A_{\rm Gibbs}}{kT\,  \ln  2}   \\
		&  =   \frac{\Delta A}{kT\,  \ln 2}  \, , 
	\end{align}
	where  $A_0$  (respectively, $A_{\rm Gibbs}$) is the free energy of the state $|0\>\<0|$   (respectively, $\Gamma$), the free energy of a generic state $\rho$ being defined as $A (\rho)   :=   E(\rho)    -  kT  \,  S(\rho)$  with $E(\rho) = \Tr  [   H  \,  \rho]$ and $S(\rho)   =   -    \Tr[  \rho  \ln \rho] $.     \qed
	
	\medskip  
	
	\begin{lemma}\label{lem:workerasure}
		The nonequilibrium cost and the work cost of erasure are  
		\begin{align} c_{\rm erase} (F)  = \kappa_{\rm erase}   +  \log F   \qquad {\rm and}  \qquad W_{\rm erase} (F)  =  kT  \ln2\,  c_{\rm erase}  (F) \,,
		\end{align} 
		respectively
		These expressions hold for every $F$ in the interval $ [F_{\min},  F_{\max}]$, with $F_{\min}  = 2^{-\kappa_{\rm erase}}=   e^{-\Delta A/(kT)} $ and $F_{\max}=1$. 
	\end{lemma}
	
	{\bf Proof.}  By Proposition \ref{prop:quantumextensioncost},  every quantum extension of a classical deterministic computation has cost $c_f (F)  =  \kappa_f + \log F$.     In the special case of erasure,  the minimum nonequilibrium cost is
	\begin{align}\label{noneqerase}
		c_{\rm erase} (F) =  \kappa_{\rm erase}  +  \log F  \, .
	\end{align}
	The quality holds for every $F$ in the interval $[F_{\min},  F_{\max}]$, with $F_{\max}=1$  and $F_{\min}  = 2^{c_{\min}-\kappa_{\rm erase}} =  2^{-\kappa_{\rm erase}}  =   e^{-\Delta A/(kT)}$  (having used Lemma \ref{lem:fmin}) for the expression of $F_{\min}$, plus the fact that the input subspace for the erasure task is the whole Hilbert space, and therefore the minimum nonequilibrium cost is   $c_{\min}=0$ by Lemma \ref{lem:cmin}).

	We now prove the equality $W_{\rm erase}  (F)  =  kT   \ln 2\,  c_{\rm erase} (F) $.  
	To this purpose, we consider the one-parameter family of quantum channels $\map M_F$ defined in Eq. (\ref{MF}). In the specific case of erasure, the expression of the channel $\map M_F$ is  
	\begin{align}\label{specialMF}     
		\map M_F  (\rho)   =  \rho_F \,\quad \forall \rho
	\end{align}   
	with 
	\begin{align}
		\rho_F  :=    F  \,  |0\>\<0|  +  (1-F)  \,  \chi    \qquad {\rm and} \qquad     \chi  :=  \frac{\Gamma  -   F_{\min} \,   |0\>\<0|  }{   1  -  F_{\min  } } \,.
	\end{align}
	(see Lemma \ref{lem:fmin} for the proof that $\chi$ is a valid quantum state).  
	
	The proof of Theorem \ref{theo:achievability} shows that the channel $\map M_F$ has fidelity at least $F$ and satisfies the equality
	\begin{align}
		c (\map M_F,  \Pi_A)   =   c_{\rm erase}  (F)  \, . 
	\end{align} 
	On the other hand, Eq. (\ref{specialMF}) implies that the nonequilibrium cost of the channel $\map M_F$  is  
	\begin{align}
		c(\map M_F ,  \Pi_A )=   D_{\max}   (   \map M_F  (\Gamma) \,\| \, \Gamma )   =   D_{\max}  (\rho_F \|  \Gamma) \, .
	\end{align}
	
	One way to realise the channel $\map M_F$ in  Eq. (\ref{specialMF}) is to prepare the state $\rho_F$ and to swap it with the state of the input.  The work cost of this realisation is equal to the work cost of generating the state $\rho_F$, which is given by    $ kT\, \ln 2 \, D_{\max}  (\rho_F \|  \Gamma) $~\cite{horodecki2013fundamental}. Since $W_{\rm erase} (F)$  is the minimum work cost of erasure, we have the bound 
	\begin{align}
		W_{\rm erase} (F) &\le      kT\, \ln 2 \, D_{\max}  (\rho_F \|  \Gamma)     =  kT \, \ln 2\,  c (\map M_F,  \Pi_A)    =   kT\, \ln 2  \,  c_{\rm erase}  (F)   \, .
	\end{align}
	On the other hand, the nonequilibrium cost times $kT \ln 2$ is a lower bound to the work cost. Hence, we obtained the equality $W_{\rm erase}  (F)  =   kT\, \ln 2  \,  c_{\rm erase}  (F)$.  \qed
	
	\medskip 
	
	Combining Lemmas \ref{lem:kappaerasure} and \ref{lem:workerasure} we finally obtain Eq. (\ref{wferasure}).}

\section*{Supplementary Note 5: the nonequilibrium cost of quantum cloning} 
Here we establish that  {\em (i)} the reverse entropy of quantum cloning is equal to the reverse entropy of classical cloning, and {\em (ii)} the bound~(4) in the main text is attainable for quantum cloning.  

Our strategy is to evaluate the nonequilibrium cost of the optimal quantum cloner~\cite{werner1998optimal}, and to infer from it the value of the reverse entropy, and the attainability of the bound~(4) in the main text.  

\subsection*{The nonequilibrium cost of the optimal cloner}  Here we calculate the nonequilibrium cost of the optimal universal quantum cloning machine $\mathcal{M}_{\rm opt}$ by Werner~\cite{werner1998optimal}, which transforms $N$ copies of an arbitrary pure state $\psi_x$ to $N'\ge N$ approximate copies of the same state. The cloner is described by the following quantum channel, 
\begin{equation}\label{cloner}
	\mathcal{M}_{\rm opt}(\psi_x^{\otimes N})=\frac{d_N}{d_{N'}}P_{N'}\, \left(\psi_x^{\otimes N} \otimes {I}^{\otimes N'-N}\right)\,  P_{N'} \, ,
\end{equation}
where $P_{k}$ is the projector onto the totally symmetric subspace of the $k$ tensor product of $k$ systems, and  $d_k :  = \Tr[  P_k]$ is the dimension of the totally symmetric subspace.  

Here we consider  non-interacting $d$-level systems, each with the same individual Hamiltonian $H=\sum_{i=0}^{d-1} E_i \ket{i}\bra{i}$. We label the eigenvalues in increasing order, with $E_0\le E_1 \le  \cdots  \le E_{d-1}  =:  E_{\max}$. 

In the following, we will denote by $\Gamma  := e^{-\beta   H}/\Tr[e^{-\beta  H}]$ ($Z  :  =  \Tr[  e^{-\beta H}]$) the single-particle Gibbs state (partition function), by  $E_{\rm Gibbs}  :  =  \Tr [  H \Gamma]$, $S_{\rm Gibbs}  =  -  \Tr[  \Gamma \,  \ln \Gamma]$, and $A_{\rm Gibbs}  : =  E_{\rm Gibbs}  -  kT   \,  S_{\rm Gibbs}=   -  kT \ln  Z$, the Gibbs state energy, entropy, and free energy, respectively.   

\begin{theorem}[nonequilibrium cost of the optimal  universal cloner]	\label{thm:NCofthe coherent quantum cloner}
	The nonequilibrium cost of the optimal universal cloner $\mathcal{M}_{\rm opt}$ in Eq.~(\ref{cloner}) is
	\begin{equation}
		c(\mathcal{M}_{\rm opt}, \Pi_A)= \frac{\Delta N     \,  \Delta A_{\max} }{k T \ln 2} + \log  F_{\max} \ ,
		\label{eq385}
	\end{equation}
	where $F_{\max} :=\log  \frac{d_N}{d_{N'}}$ is the optimal cloning  fidelity,   $\Delta N  := N' -N$ is the number of extra copies, and  $\Delta A_{\max}  := A_{\max}   -   A_{\rm Gibbs}$  is the difference between  the maximum free energy  of a single-copy state, given by $A_{\max} :=  E_{\max}$, and the free energy    of the single-copy Gibbs state, given by $A_{\rm Gibbs}  :=  -k T  \ln  Z$.  
\end{theorem}

{\bf Proof.}  For  the universal cloner, the input subspace is the totally symmetric subspace, and therefore $\Pi_A  =  P_N$. Hence,  the nonequilibrium cost~(\ref{easyfaist}) reads 
\begin{align}
	c(\map M_{\rm opt}, P_N)     =   \log \left\|  \,\left(\Gamma^{\otimes N'} \right)^{-1}    \,    \map M_{\rm opt}  \left( P_N  \, \Gamma^{\otimes N} \,  P_N \right) \,    \left(\Gamma^{\otimes N'} \right)^{-1}   \right\| \, . 
\end{align}
The projected Gibbs state $P_N  \, \Gamma^{\otimes N} \,  P_N$ can be written as  
\begin{equation}
	P_N  \, \Gamma^{\otimes N} \,  P_N = \frac{1}{Z^N} \sum_{\vec{n}   \in  \set S_{N, d}} e^{- \beta
		\vec{n} \cdot \vec{E}} \, \ket{N,  \vec{n}} \bra{N, \vec{n}} \ ,
\end{equation} 
where $\ket{N, \vec{n}}$ denotes the normalised symmetric state with occupation number $n_j$ in the $j$-th mode,   $\vec{n} := (n_0, \dots,n_{d-1})$ a partition of $N$  into $d$    nonnegative integers,  $\set S_{N,  d}$ is the  set of all such partitions, and $\vec E  =  (E_0,  \dots  , E_{d-1})$ is the vector of single-system energy eigenstates.  

The quantum cloner transforms the input state $|N,  \vec n\>$ into  
\begin{align}
	\nonumber		   \mathcal{M}_{\rm opt}(\,  |N, \vec n  \>\<N,\vec n|  ) &=  \frac{d_N}{d_{N'}} P_{N'} \,  \left (   |N, \vec n  \>\<N,\vec n| \otimes
	{I}^{\otimes \Delta  N}\right) \, P_{N'}\\
	&=     \frac{d_N}{d_{N'}} \sum_{\vec k  , \vec l  \in  \set S_{ N', d }}  \,   |N', \vec k\> \<  N',  \vec l |   \quad     \<N',  \vec k|  \, \left(  |N, \vec n  \>\<N,\vec n|  \otimes
	{I}^{\otimes \Delta  N}\right)   \,    |N', \vec l\>   
	\,  ,
\end{align}
with
\begin{align}
	\nonumber 	\<N',  \vec k|  \, (  |N, \vec n  \>\<N,\vec n| \otimes
	{I}^{\otimes \Delta  N})   \,    |N', \vec l\>     &  =  \sum_{\vec r \in \set S_{\Delta  N, d}}  \,   \<N',  \vec k|  \, \left(\,   |N, \vec n  \>\<N,\vec n| \otimes  | \Delta  N,  \vec r \, \>\< \Delta  N , \vec r\,  | \, \right)   \,    |N', \vec l\>    \\
	\nonumber 		&  =   \delta_{\vec k,  \vec l} ~   \left| \,\left( \<N,\vec n| \otimes  \< \Delta  N , \vec k- \vec n | \,\right)  |N',\vec k\>  \right|^2  \\
	& =   \delta_{\vec k,  \vec l} ~   \frac{   \begin{pmatrix}  \vec k \, \,   \\  \vec n \end{pmatrix}}{ \begin{pmatrix}  N'  \\ N \end{pmatrix}} \qquad     \begin{pmatrix} \vec  k  \, \, \\  \vec n \end{pmatrix}  :=  \prod_{j=0}^{d-1} \,    \begin{pmatrix}  k_j  \\  n_j \end{pmatrix} \, .
\end{align}
Hence, we have 
\begin{align}
	\map M_{\rm opt}  \left(  P_N  \, \Gamma^{\otimes N} \,  P_N\right)  = \frac{1}{Z^N} \frac{d_N}{d_{N'}}   \,   \sum_{\vec{n}   \in  \set S_{N, d}}  \sum_{\vec n'   \in  S_{N',  d }} e^{- \beta
		\vec{n} \cdot \vec{E}}  \,   \frac{     \begin{pmatrix}  \vec n'  \, \, \\  \vec n \end{pmatrix}}{ \begin{pmatrix}  N'  \\ N \end{pmatrix}}  \, \ket{N',  \vec{n}'} \bra{N', \vec{n}'} \, ,  
\end{align}
and 
\begin{align}
	\left(\Gamma^{\otimes N'} \right)^{-\frac 12} \, \map M_{\rm opt}  \left(  P_N  \, \Gamma^{\otimes N} \,  P_N\right) \left(\Gamma^{\otimes N'} \right)^{-\frac 12}    = Z^{ \Delta  N   }   \,      \frac{d_N}{d_{N'}}   \,   \sum_{\vec n'   \in  S_{N',  d }}  \, \left(   \sum_{\vec{n}   \in  \set S_{N, d}}  e^{ \beta\,  (\vec n'  -  
		\vec{n} ) \cdot \vec{E}}  \,   \frac{   \begin{pmatrix}  \vec n'  \\  \vec n \end{pmatrix}}{ \begin{pmatrix}  N'  \\ N \end{pmatrix}} \right) \, \ket{N',  \vec{n}'} \bra{N', \vec{n}'} \, .  
\end{align}
The norm of the above operator is given by 
\begin{align}
	\nonumber \left\|  \left(\Gamma^{\otimes N'} \right)^{-\frac 12} \, \map M_{\rm opt}  \left(  P_N  \, \Gamma^{\otimes N} \,  P_N\right) \left(\Gamma^{\otimes N'} \right)^{-\frac 12} \right\|   = Z^{ \Delta  N   }   \,      \frac{d_N}{d_{N'}}   \,   \max_{\vec n'   \in  S_{N',  d }}  \, \left(   \sum_{\vec{n}   \in  \set S_{N, d}}  e^{ \beta\,  (\vec n'  -  
		\vec{n} ) \cdot \vec{E}}  \,   \frac{   \begin{pmatrix}  \vec n'  \\  \vec n \end{pmatrix}}{ \begin{pmatrix}  N'  \\ N \end{pmatrix}} \right)    \, .
\end{align}
Note that the binomial coefficient in the numerator of the r.h.s. is nonzero only if all the entries of the vector  $\vec n'  -\vec  n$ are nonnegative.   Under this condition, we have the inequality $e^{\beta  (   \vec n'  -\vec n)  \,  \vec E}   \le   e^{\beta   \Delta N   \,E_{\max}}$, and therefore the norm is upper bounded as  
\begin{align}
	\nonumber \left\|  \left(\Gamma^{\otimes N'} \right)^{-\frac 12} \, \map M_{\rm opt}  \left(  P_N  \, \Gamma^{\otimes N} \,  P_N\right) \left(\Gamma^{\otimes N'} \right)^{-\frac 12} \right\| 
	&  \le Z^{ \Delta  N   }   \,      \frac{d_N}{d_{N'}}   \,  e^{ \beta\,  \Delta N \, E_{\max} }  \,        \max_{\vec n'   \in  S_{N',  d }}  \, \left(   \sum_{\vec{n}   \in  \set S_{N, d}}    \frac{   \begin{pmatrix}  \vec n'  \\  \vec n \end{pmatrix}}{ \begin{pmatrix}  N'  \\ N \end{pmatrix}} \right)  \\
	&=     Z^{ \Delta  N   }   \,      \frac{d_N}{d_{N'}}   \,  e^{ \beta\,  \Delta N \, E_{\max} }   \, ,
\end{align}
where the last equality follows from the Chu-Vandermonde identity.  The upper bound is attained by choosing the vector $\vec n'$ with $n_{d-1}'   =  N'$ and $n_j= 0  \,   , \forall j  \not  =  d-1$.   

Summarising, we obtained the equality  
\begin{align}
	\nonumber  c(\map M_{\rm opt},  P_N)  &=       \Delta  N    \,  \log Z   +    \log \frac{d_N}{d_{N'}}    +    \beta\,  \Delta N \, E_{\max}  \,  \log e  \\
	\nonumber  &  =  \beta  \, \log e \,  \Delta  N    \,   ( E_{\max}   -  A_{\rm Gibbs}  )     +  \log F_{\max} \\
	&  \equiv  \frac{ \Delta  N  \,   \Delta  A_{\max}}{  kT  \,  \ln 2}   +  \log F_{\max}  \, .
\end{align}
\qed  

\subsection*{The reverse entropy of quantum cloning}  

Using the result of the previous subsection, we now show that the reverse entropy of quantum cloning coincides with the reverse entropy of classical cloning.  From the main text, we have the bound 
\begin{align}\label{uffa1}
	\kappa_{\rm clon}^{\rm Q}   \ge  \kappa_{\rm clon}^{\rm C}    =   \frac{ \Delta  N  \,   \Delta  A_{\max}}{  kT  \,  \ln 2}  \, .
\end{align}
On the other hand, applying the bound~(\ref{eq:main}) to the optimal cloning channel $\map M_{\rm opt}$ yields the inequality 
\begin{align}\label{uffa}
	c(\map M_{\rm opt}, \Pi_A)  \ge   c_{\rm clon}  (F_{\max})  \ge    \kappa_{\rm clon}^{\rm Q}   +    \log F_{\max} \, .
\end{align}
Substituting Eq.~(\ref{eq385}) into Eq.~(\ref{uffa}), we then obtain the bound 
\begin{align}\label{uffa2}\kappa_{\rm clon}^{\rm Q} \le   \frac{ \Delta  N  \,   \Delta  A_{\max}}{  kT  \,  \ln 2}  \equiv  \kappa_{\rm clon}^{\rm C} \,.
\end{align}   
Hence, we obtained the equality      $ \kappa_{\rm clon}^{\rm Q}   =   \kappa_{\rm clon}^{\rm C}$.  

\subsection*{Achievability of the lower bound~(\ref{eq:main})}  
We conclude the section by showing that the bound~(\ref{eq:main}) is attainable for every value of the fidelity in the interval $[F_{\min},  F_{\max}]$.  

For the optimal cloning channel,  the bound~(\ref{eq:main})  reads $c(\map M_{\rm opt},  \Pi_A)  \ge    \frac{ \Delta  N  \,   \Delta  A_{\max}}{  kT  \,  \ln 2} 
+    \log F_{\max}  $.      On the other hand, Eq.~(\ref{eq385}) shows that the bound is achieved with the equality sign.    In other words, the bound  ~(\ref{eq:main}) is attainable at $F  =  F_{\max}$.  Using the attainability criterion provided in the main text, we can then conclude that the bound  ~(\ref{eq:main}) is attainable for every value of $F$ in the interval $[F_{\min},  F_{\max}]$.  

\section*{Supplementary Note 6: cloning with entanglement binding  machines}

\subsection*{Bound on the reverse entropy of the transpose cloning task}  

Here we consider the task of transpose cloning, which consists in transforming $N$ copies of a pure quantum state $\rho_x$ into $N'$ copies of its transpose state $\rho_x^T$.  To estimate the reverse entropy of transpose cloning, we  use the expression $\kappa_{\rm clon^*}   =   \max_{\st p}  H(A|B)_{\omega_{{\rm clon^*},  \st p}}$ and  we fix the prior probability distribution  $\st p$   to be the normalised Haar measure  $p(\d x)$.  With this choice, the operator $\omega_{{\rm clon^*},\st p}$ is 
\begin{align}
	\nonumber \omega_{{\rm clon^*},\st p}  &  = \int  p(\d x) ~   ( \Gamma_A^{-1}  \otimes \Gamma_B)^{1/2} \,     \underbrace{\rho_x^T  \otimes  \cdots \otimes \rho_x^T}_{N+N'~{\rm times}}\,  ( \Gamma_A^{-1}  \otimes \Gamma_B)^{1/2}   \\
	&  =  ( \Gamma_A^{-1}  \otimes \Gamma_B)^{1/2} \,   \frac{ P_{N+N'}}{d_{N+N'}} \,  ( \Gamma_A^{-1}  \otimes \Gamma_B)^{1/2} \, , 
\end{align}
where $P_{N+ N'}$ is the projector on the symmetric subspace of $N+N'$ systems, and $d_{N + N'}  =  \Tr[P_{N+N'}]$ is the dimension of the symmetric subspace. 

Now, the condition $I_A\otimes   \Lambda_B\ge \omega_{{\rm clon^*},\st p}$ is equivalent to 
\begin{align}\label{tobesat}\Gamma_A  \otimes \Gamma_B^{-1/2}  \Lambda_B  \Gamma_B^{-1/2}  \ge \frac{P_{N+N'}}{d_{N+N'}} \, .
\end{align}  
Recall that $\Gamma_A  =  \Gamma^{\otimes N}$, and notice that one has 
\begin{align}
	\Gamma_A  \ge  g_{\min}^N   \,   I_A \, ,   
\end{align} 
where $g_{\min}   = e^{-\beta E_{\max}}/Z$ is the smallest probability in the Gibbs distribution.   Then, the condition~(\ref{tobesat}) is satisfied by the operator 
\begin{align}\Lambda^{\min}_B  := \frac{\Gamma_B^{1/2}  P_{N'}  \Gamma_B^{1/2}  }{ g_{\min}^N   \,  d_{N+N'}} \, ,
\end{align}   
where $P_{N'}$ is the projector on the symmetric subspace of $N'$ systems. 

Hence, the reverse entropy satisfies the condition 
\begin{align}
	\nonumber \kappa_{\rm clon^*}   & = -  \min_{I_A\otimes   \Lambda_B\ge \omega_{{\rm clon}*, \st p} } \, \log \Tr[\Lambda_B]\\
	\nonumber &   \ge  -\log  \Tr [\Lambda_B^{\min}]  \\
	&  =   \log   \frac{g_{\min}^N  \,  d_{N+N'}}{\Tr[  \Gamma_B   P_{N'}]}\\
	&  \ge   \log  \frac{  g_{\min}^N  \,  d_{N+N'}}{  g_{\max}^{N'}  \, d_{N'}}  \, ,
\end{align}
where $g_{\max}=  e^{-\beta  E_{\min}}/Z$ is the maximum probability in the Gibbs state, and $d_{N'} =  \Tr[P_{N'}]$ is the dimension of the symmetric subspace of $N'$ systems.  

Rearranging the terms, we finally obtain the inequality 
\begin{align}
	\nonumber \kappa_{\rm clon^*}   &\ge \log  g_{\min}^{N-N'}       +   \log  \frac{  d_{N+N'}   \,  (g_{\min}/g_{\max})^{N'}}{d_{N'}} \\
	\label{transposeclon}&  =  \kappa_{\rm clon}   +   \log  \frac{  d_{N+N'}   \,  e^{-\beta  N' \Delta E}}{d_{N'}} \,, 
\end{align} 
with $\Delta  E:=  E_{\max} -  E_{\min}$. 

The above inequality implies that the nonequilibrium cost of the transpose cloning task is lower bounded as 
\begin{align}\label{ebclon}
	c_{\rm clon^*}  (F)  \ge      \kappa_{\rm clon}   +  \log F  +   \log  \frac{  d_{N+N'}   \,  e^{-\beta  N'\,  \Delta E}}{d_{N'}}   \, . 
\end{align}
This bound  applies to  all entanglement binding machines for the task of quantum cloning, due to the general argument  shown in the main text.  

\subsection*{  Achievability of the bounds~(\ref{transposeclon}) and~(\ref{ebclon}) for fully degenerate systems}

Here we show that the the bounds~(\ref{transposeclon}) and~(\ref{ebclon}) are exactly achievable for fully degenerate systems, corresponding to $\Delta E=  0$.  

Let $\map E_{\rm opt}$ be the optimal state estimation channel \cite{chiribella2010quantum}, whose action on the symmetric subspace is defined by 
\begin{align}
	\map E_{\rm opt}   (  P_N \rho  P_N )    :  = d_N  \,    \int  \d \psi   \,      \Tr[  \psi^{\otimes N}  \,  P_N \rho  P_N ]  ~  \psi^{\otimes N'} \, ,
\end{align}
where $\psi:  = |\psi\>\<\psi|$ denote the projector on a generic pure state $|\psi\>$, and $\d \psi$ is the normalised unitarily invariant measure on the set of pure states.  

Then, define the channel  $\map E_{\rm opt^*}$ via the relation 
\begin{align}
	\map E_{\rm opt^*}  (\rho)  :  =    \left[\map E_{\rm opt}  (\rho)\right]^T \qquad \forall \rho \, .
\end{align} 
This channel achieves the optimal fidelity $F_{\max}  =  d_N/d_{N+N'}$ for the transpose cloning task, which coincides with the optimal fidelity for implementing the original cloning task via state estimation \cite{chiribella2010quantum}.

For the channel $\map E_{\rm opt^*}$, we have the bound
\begin{align}
	\nonumber 	c (\mathcal{E}_{{\rm opt}},  \Pi_A)   & \ge   c_{\rm clon^*}  (F_{\max})\ge    \kappa_{\rm clon^*}  + \log F_{\rm max}   \ge  (N' - N) \log  d + \log \frac{d_{N+N'}}{d_{N'}} +
	\log  \frac{d_N}{d_{N+N'}}  \\
	\label{chip}	& =(N' - N) \log  d + \log \frac{d_{N}}{d_{N'}}    \, , 
\end{align}
where the second inequality follows from  the bound~(\ref{ebclon}) with $\Delta E =  0$. 

On the other hand,  the nonequilibrium cost of the channel $\map E_{\rm opt^*}$ is 
\begin{align}\nonumber 
	c (\map E_{\rm opt^*},  \Pi_A)   &=  \log \left\|    d^{N'- N}    \,  \map E_{\rm opt^*}  (P_N)     \right\|\\
	\nonumber 
	\nonumber &   =\log \left\|    d^{N'- N}    \,    \frac{  d_N}{d_{N'}} P_{N'}     \right\|\\ 
	&   =  (N'-N) \,\log d    +  \log  \frac{  d_N}{d_{N'}}    \, .  \label{chop}
\end{align}  
Combining Eqs. (\ref{chip}) and (\ref{chop}), we then obtain the equalities 
\begin{align}\label{donald}
	c_{\rm clon^*}  (F_{\max})  =   \kappa_{\rm clon^*}  + \log F_{\rm max} 
\end{align}
and
\begin{equation}
	\kappa_{\rm clon^*}= (N' - N) \log  d +  \log  \frac{  d_{N + N'}}{d_{N'}}  \, .
\end{equation}
In particular, Eq. (\ref{donald}) implies that the bound (\ref{eq:main}) for transpose cloning is attained at $F=  F_{\max}$.  Hence, Theorem \ref{theo:achievability} implies that the bound  (\ref{eq:main}) for transpose cloning is attained for every value of the fidelity in the interval $[F_{\min}, F_{\max}]$, namely  
\begin{align}
	c_{\rm clon^*}  (F_{\max})  =   \kappa_{\rm clon^*}  + \log F_{\rm max} 
\end{align}
Incidentally, we observe that the one-parameter family of quantum channels $\map M_F$ defined in the proof of Theorem \ref{theo:achievability} consists of entanglement breaking channels.  Hence, the optimal accuracy/nonequilibrium tradeoff for transpose cloning is achieved by an entanglement breaking (and, in particular, an entanglement binding) channel for every value of $F$. 

{  \section*{Supplementary Note 7:  state transmission and state transposition with entanglement binding machines}
	
	Here we establish a bound on the nonequilibrium cost of entanglement binding (EB) machines in  the state transmission task $\rho_x \mapsto \rho_x$, where $\rho_x$ is an arbitrary pure state.   The bound is derived from a bound on the nonequilibrium cost  of arbitrary quantum machines  for  the state transposition task $\rho_x \mapsto \rho_x^T$.  This task is a special case of the transpose cloning task considered in the Supplementary Note 6.   
	
	The logic of the derivation is as follows. As discussed in Methods,    the minimum nonequilibrium cost of EB machines achieving fidelity $F$ in a given task $\rho_x \mapsto \rho_x'$  coincides with the minimum nonequilibrium cost  of  EB machines achieving fidelity $F$ in the transpose task $\rho_x\mapsto \rho_x^{\prime \, T}$.  In turn,  the minimum cost of EB machines is  lower bounded by the minimum cost   of arbitrary quantum machines, mathematically described by trace-preserving completely positive linear maps. Hence, we have the bound  
	\begin{align}
		c_{\rm transmit  , \,  eb}   (F)  \equiv  c_{\rm transpose  , \,  eb}    (F)  \ge  c_{\rm transpose} (F)
	\end{align}
	where $c_{\rm transmit  , \,  eb} (F) $     ($c_{\rm transpose  , \,  eb}  (F)$)  is the minimum nonequilibrium cost needed to achieve fidelity $F$ for state transmission  (state transposition) with EB  machines, and $c_{\rm transpose} (F)$  is the minimum nonequilibrium cost needed to achieve fidelity $F$ for state transposition with arbitrary quantum machines. 
	
	The main result of this section is the following bound 
	\begin{align}\label{cftransp}
		c_{\rm transpose} (F)  \ge   \log  \left[  \frac  {   (d^2  F - d)^2}{4 \gamma  (1-F)}  +  F\right]\, , 
	\end{align}
	with  $\gamma:  =    \left(\sum_{m=1}^{d-1}   \,   e^{  -  \frac{ \Delta E_m}{2kT}}\right)^2$, $\Delta E_m:  =  E_m  - E_{m-1}$, and with  the  eigenvalues of the Hamiltonian  ordered so that $E_0  \le E_1 \le  \dots \le E_{d-1}$.   
	This bound implies  that the nonequilibrium cost is strictly larger than zero whenever the fidelity satisfies the condition
	\begin{align}\label{minimumftransp}
		F>   \frac{ d+  2\sqrt{\gamma}}{d^2  +  2\sqrt \gamma}    =:  F_* \, .  
	\end{align}
	For qubits, we will show that the bound  (\ref{cftransp}) is achievable for every Hamiltonian and for every value of the fidelity between $F_*  =  (1+  \sqrt\gamma)/(2 +  \sqrt \gamma)  \equiv F_{\min}$   (the fidelity corresponding  to the minimum  nonequilibrium cost $c_{\min}  = 0$) and $F_{\max}  = 2/3$ (the maximum fidelity allowed by quantum mechanics,  corresponding to the maximum nonequilibrium cost  $c_{\max}  =   \log\left(\gamma^{-1}  + 2)/3 \right)$).

	For $d=2$, this bound coincides with Eq. (11)  in the main text.

	\subsection*{State transposition with bounded nonequilibrium resources}  
	Our strategy to derive the bound (\ref{cftransp}) is to consider the maximisation of the transposition fidelity under a constraint on the nonequilibrium resources.  We define the maximum fidelity as
	\begin{align}  
		F_{\rm transpose}   (c)   : =   \max_{\map M  ~:~   c (\map M,  \Pi_A)  \le c}  \,  \min_{\rho_x}     \,  \Tr [  \rho_x^T \, \map M(\rho_x)]    \,,  \label{Fstarc}  
	\end{align}  
	where the minimisation runs over all pure states $\rho_x   =  |\psi_x\>\<\psi_x|$, $x$ being some parametrisation of the unit sphere.        
	To derive a bound on $F_{\rm transpose}   (c) $, we observe that the maximisation in Eq. (\ref{Fstarc}) can be restricted without loss of generality to quantum channels $\map M$ satisfying the covariance property $ \map M  \circ  \map U_{\bs \theta}   =   \map U^T_{\bs \theta} \circ \map M$, where $\bs \theta  =  (\theta_0,\dots,  \theta_{d-1})  \in  [0,2\pi]^{\times d}$ is a vector of $d$ phases,  $\map U_{\bs \theta}$  ($\overline {\map U}_{\bs \theta}$)   is the unitary channel defined by $\map U_{\bs \theta}  (\rho)  :=  U_{\bs \theta}  \rho  U_{\bs \theta}^\dag \, , \forall \rho$  ($\overline{\map U}_{\bs \theta}  (\rho)  :=  \overline U_{\bs \theta}  \rho  \overline U_{\bs \theta}^T \, , \forall \rho$), and $U_{\bs \theta}   :  =  \sum_m \,   e^{-i\theta_m}  \,  |m\>\<m|$  ($\overline U_{\bs \theta}   :  =  \sum_m \,   e^{i\theta_m}  \,  |m\>\<m|$)  (see e.g. \cite{chiribella2014optimal}).     The reduction to covariant channels can be made without loss of generality, because for every given channel $\map M$,  the covariant channel 
	\begin{align}
		\map M'   :  =  \int_{0}^{2\pi}  \frac{  \d  \theta_0}{2\pi} \cdots  \int_{0}^{2\pi}  \frac{  \d  \theta_{d-1}}{2\pi}     \map U_{\bs \theta}^T \circ \map M \circ \map U_{\bs \theta} \, ,
	\end{align} 
	satisfies the conditions  $F_{\rm transpose}  (  \map M')  \ge F_{\rm transpose}(\map M)$ and $c(\map M', \Pi_A) \le c (\map M,  \Pi_A)$.
	
	In the Choi representation, the covariance condition  $ \map M  \circ  \map U_{\bs \theta}   =   \map U^T_{\bs \theta} \circ \map M\,, \forall   \bs \theta$ is equivalent to the commutation condition  $[  M  , U_{\bs \theta} \otimes U_{\bs \theta}]=  0  \, , \forall  \bs \theta$, where $M$ is the Choi operator of $\map M$  (see e.g. \cite{chiribella2005extremal}).   Using this commutation, the Choi operators of covariant quantum channels (completely positive trace preserving maps) can be characterised  
	as those with the following block diagonal form
	\begin{align}
		\nonumber M  =&  \sum_{m=0}^{d-1}\,   p_{mm}  \,  |m\>\<m|  \otimes |m\>\<m|   \\
		&  +      \sum_{m=1}^{d-1}  \sum_{n:  \,  n<m}  \,  \Big( p_{mn} \,   |m\>\<m|  \otimes |n\>\<n|    +    p_{nm}  \,       |n\>\<n|  \otimes |m\>\<m|   + c_{mn}  \,  |m\>\<n|  \otimes |n\>\<m|  +   \overline c_{mn}  \,  |n\>\<m|  \otimes |m\>\<n|   \Big) \,,  \label{phasecovariantchoi}
	\end{align}   
	where $p_{mn}$ are probabilities satisfying the normalisation $\sum_m  p_{mn}  = 1 \, \forall n$, and  $c_{mn}$ are complex coefficients satisfying the condition $|c_{mn}|^2  \le  p_{mn}  p_{nm}$.  
	
	For a generic state $|\psi\> =  \sum_m  \,  \psi_m  \,  |m\>$, the fidelity of the covariant channel $\map M$ satisfies the bound
	\begin{align}
		\nonumber F_\psi   (\map M)   &:  =  \Tr[ \,  ( |\psi\>\<\psi|)^T    \map M  (|\psi\>\<\psi|)  ] \\
		\nonumber   &=\< \overline \psi|  \<\overline \psi|    M   | \overline \psi\>  |\overline \psi\>  \\
		&=    \sum_{m=0}^{d-1}    p_{mm}  \,  |\psi_m|^4     +  \sum_{m=1}^{d-1}  \sum_{n<m}  \,   |\psi_m|^2  |\psi_n|^2  \,   \big(   p_{mn}  +  p_{nm}    +  c_{mn}  +  \overline c_{mn}   \big)    \label{phasecovariantfid1} \\
		&\le    \sum_{m=0}^{d-1}    p_{mm}  \,  |\psi_m|^4     +  \sum_{m=1}^{d-1}  \sum_{n: \,   n<m}  \,   |\psi_m|^2  |\psi_n|^2  \,   \big(  p_{mn}  +  p_{nm}    +  2\sqrt{  p_{mn}  p_{nm}}   \big) \, ,     \label{phasecovariantfid}
	\end{align}
	which can be achieved with the appropriate choice of coefficients  $c_{mn}   =  \sqrt{p_{mn}  p_{nm}}$.  
	In particular, a computational basis state $|m\>$ gives fidelity $F_m  (\map M)   =  p_{mm}$, while an equatorial state, with $|  \psi_m|  =  1/d \, \forall m$, gives fidelity  
	\begin{align}
		\nonumber  F_{\rm equatorial}  (\map M) &=     \frac{ \sum_{m=0}^{d-1}    p_{mm} +   \sum_{m}  \sum_{n:  \,n<m}     p_{mn}  +  p_{nm}    +  2\sqrt{  p_{mn}  p_{nm}}  }{d^2}  \\
		& =  \frac{ d  + 2   \sum_{m=1}^{d-1}  \sum_{n: \,    n<m}  \sqrt{  p_{mn}  p_{nm}}       }{d^2}
		\, .\end{align} 
	
	The above fidelities, maximised over all channels with nonequilibrium cost bounded by $c$,  provide upper bounds to the worst case fidelity $F_{\rm transpose}   (c)$. To introduce the nonequilibrium constraint, we observe that 
	\begin{align}
		\Gamma^{-\frac 12}\map M  (\Gamma) \Gamma^{-\frac 12}  =  \Gamma^{-\frac 12} \Tr_{A}  [   (  I  \otimes \Gamma^T)  M]  \Gamma^{-\frac 12}=  \sum_{m,n}  \,   p_{mn}  \frac{ g_n}{g_m} \,  |m\>\<m|   \, ,     
	\end{align} 
	where $g_m  =  e^{-  E_m/(kT)}/Z$ are the eigenvalues of the Gibbs state, ordered so that $g_0  \le  g_1 \le  \cdots \le g_{d-1}$.  Hence, the nonequilibrium cost (\ref{easyfaist}) is given by
	\begin{align}\label{phasecovcost}
		c  (\map M,  \Pi_A)  =  \max_m   \,   \log \left( \sum_n  p_{mn}  \frac{ g_n}{g_m}\right)   \, .
	\end{align}
	Note that, for every $m\ge1$,  one has the bounds 
	\begin{align}
		\sum_{n :  n<  m}    p_{mn}        \le   \frac{ g_m }{g_{m-1}}  \,  \left( \sum_{n :  n<  m}    p_{mn} \, \frac{ g_n}{g_m} \right) \le  \frac{ g_m }{g_{m-1}}  \,  \left( \sum_{n :  n \not = m}    p_{mn} \, \frac{ g_n}{g_m} \right) \le    \frac{ g_m }{g_{m-1}}  \,     \Big(  2^{c (\map M,  \Pi_A)}   -   p_{mm}\Big) 
	\end{align}
	and 
	\begin{align}
		\sum_{n :  n<  m}   \sqrt{ p_{mn} p_{nm}} \le \sqrt{   \left(  \sum_{n :  n<  m}    p_{mn}    \right) \,   \left(  \sum_{n :  n<  m}    p_{nm}    \right)   }      \le   \sqrt{  \frac{ g_m }{g_{m-1}}  \,     \Big(  2^{c (\map M,  \Pi_A)}   -   p_{mm}\Big)   \, (1- p_{mm})} \, .   
	\end{align}
	Later we will see that this bound is achievable for qubits (but generally not for higher dimensional systems).  
	Using this bound, the equatorial fidelity is upper bounded as 
	\begin{align}
		F_{\rm equatorial}  (\map M)  \le  \frac{ d  +  2  \sum_{m  = 1}^{d-1}  \sqrt{  \frac{ g_m }{g_{m-1}}  \,     \Big(  2^{c (\map M,  \Pi_A)}   -   p_{mm}\Big)   \, (1- p_{mm})}  }{d^2}  \label{equatorial2}
	\end{align}
	
	Now, suppose that a channel $\map M_0$ has nonequilibrium cost less than $c$, namely $c (\map M_0,  \Pi_A) \le c$.    Recall that $p_{mm}$ is equal to the fidelity on the computational basis state $|m\>$, and therefore it is lower bounded by the worst case fidelity $F_{\rm transpose}   (c) $.     
	Hence, we have the upper bound  
	\begin{align}
		\nonumber F_{\rm transpose}   (c)    &\le  F_{\rm equatorial}  (\map M_0)  \le  \frac{ d  +  2  \sum_{m =1}^{d-1}  \sqrt{  \frac{ g_m }{g_{m-1}}  \,     \Big(  2^{c}   -  F_{\rm transpose}   (c)   \Big)   \, (1- F_{\rm transpose}   (c) )}  }{d^2}  \\
		& \le   \frac{ d  +  2  \sqrt{    \gamma  \,     \Big(  2^{c}   -  F_{\rm transpose}   (c) \Big)   \, (1- F_{\rm transpose}   (c) )}  }{d^2}  \,,
	\end{align}
	having defined $\gamma:  =    \left(\sum_{m=1}^{d-1}   \,   \sqrt{  g_m/g_{m-1}}\right)^2$.    Solving the above inequality in  $F_{\rm transpose}   (c) $, we obtain the bound 
	\begin{align}
		F_{\rm transpose}   (c)  \le  \frac{d^3  -  2\gamma   (1+  2^c)+  \sqrt{  \left[   d^3  -  2\gamma   (1+  2^c)  \right]^2    -  (d^4-  4\gamma)  (d^2  -  4\gamma 2^c)   }}{d^4  - 4\gamma} \, ,  
	\end{align}
	valid for $F_c  \le 1/d$.  
	Solving the inequality in $c$, instead, we obtain the bound 
	\begin{align}
		c  \ge  \log\left[ F_{\rm transpose}   (c)   +  \frac{   \left(d^2\, F_c-d\right)^2}{  4\gamma  (1-F_c)}  \right] \, ,    
	\end{align}
	valid for $F_{\rm transpose}   (c)  \le 1/d$.  
	This bound  is equivalent to Eq.  (\ref{cftransp}).
	
	\subsection*{Achievability of the bound (\ref{cftransp})  for qubits}

	For $d=2$, the bound (\ref{cftransp}) reads  
	\begin{align}\label{cftranspqubit}
		c_{\rm transpose}   (F) \ge   \log\left[ F  + e^{  \frac{ \Delta E}{kT} } \,  \frac{   \left(2 F-1\right)^2}{    (1-F)}  \right]  \, , \qquad \forall F \ge \frac 12 \, .
	\end{align}
	We now show that the bound holds with the equality sign for all possible Hamiltonians, and for all values of  $F$ in the interval $[F_{\min},  F_{\max}]$, with 
	\begin{align}
		F_{\min}    =\frac{e^{  \frac{ \Delta E}{kT} }+1}{2\, e^{  \frac{ \Delta E}{kT} }  + 1}   \qquad {\rm and}  \qquad F_{\max}  = \frac 23 \, .
	\end{align}
	(cf. Eq. (\ref{minimumftransp}) and the  discussion following it). 
	
	To prove the achievability of the bound,  we consider the quantum channel $\map M$ with Choi operator
	\begin{align}\nonumber
		M & =   p_{00}  \,  |0\>\<0|  \otimes |0\>\<0|   +   p_{11}  \,  |1\>\<1|  \otimes |1\>\<1|   \\
		&  +       p_{01} \,   |0\>\<0|  \otimes |1\>\<1|    +    p_{10}  \,       |1\>\<1|  \otimes |0\>\<0|   + \sqrt{p_{01}  \,  p_{10}}  \,\Big(  |0\>\<1|  \otimes |1\>\<0|  +   \,  |1\>\<0|  \otimes |0\>\<1|   \Big)\,   \label{defoptimalm} \, ,
	\end{align}
	with 
	\begin{align}
		p_{11}  =  F \, , \qquad  p_{01}   =  1-F   \,, \qquad  p_{10}  =   \frac{  (2F-1)^2}{1-F}  \,, \qquad  {\rm and}  \qquad p_{00}   =  \frac{3F-4F^2}{1-F}    \, . 
	\end{align}
	These parameters define a completely positive trace-preserving map  whenever $p_{00}  \ge 0$, that is, whenever  $F \le 3/4$.    In particular, they define a valid quantum channel whenever $F  \le  2/3  \equiv  F_{\max}$.

	For the channel  $\map M$,  the nonequilibrium cost, given by Eq. (\ref{phasecovcost}),  is
	\begin{align}
		\nonumber c  (\map M, \Pi_A)    &=  \max    \left\{    \log \left( p_{00}   +    p_{01}  \frac{ g_1}{g_0} \right)  , \,    \log \left( p_{11}   +    p_{10}  \frac{ g_0}{g_1} \right)  \right\}   \\
		&  =  \max\left\{  \log \left[   \frac{3F-4F^2}{1-F}  + \frac{ g_1}{g_0} \,  (1-F)\right],  ~   \log  \left[ F   +   e^{\Delta E}  \frac{  (2F-1)^2}{1-F} \right] \right\} 
	\end{align}
	For every $  F  \ge   \left(e^{\Delta E/(kT)}  + 1  \right)/   \left(2 \, e^{\Delta E/(kT)}  + 1  \right)  \equiv F_{\min}$, one has the equality  
	\begin{align}\label{98}
		c  (\map M, \Pi_A)    &=  \log  \left[ F   +   e^{\Delta E/(kT)}  \frac{  (2F-1)^2}{1-F} \right]  \qquad \forall F  \in   [F_{\min}  , F_{\max}]    \, .
	\end{align}
	We now show that the worst case fidelity of the channel $\map M$ is $F$.  The fidelity of this channel on a generic state $|\psi\>  =  \psi_0  \, |0\>  +  \psi_1\,  |1\>$ is given by Eq. (\ref{phasecovariantfid1})  and yields the relation 
	\begin{align}
		F_\psi   (\map M)   =    \sum_{m=0}^{1}    p_{mm}  \,  |\psi_m|^4     +    \,   |\psi_0|^2  |\psi_1|^2  \,   \big(  \sqrt{  p_{01}}  +  \sqrt{  p_{10}}   \big)^2   \, ,    
	\end{align}
	Note that one has  $\sqrt{  p_{01}}  +  \sqrt{  p_{10}}     =  (2F-1)/\sqrt{1-F}  +  \sqrt{1-F}=   F/\sqrt{1-F} $, and therefore 
	\begin{align}
		F_\psi   (\map M)   =    F \,   |\psi_1|^4   +  \frac{  3F  -  4F^2}{1-F}       |\psi_0|^4          +    \,   |\psi_0|^2  |\psi_1|^2  \,\frac{  F^2}{1-F} \, .    
	\end{align}
	For every $F  \le 2/3   \equiv F_{\max}$, the minimum of $F_\psi   (\map M)$ is attained for $|\psi\>  =  |1\>$, whence one has 
	\begin{align}\label{100}
		F_{\rm transpose}  (\map M)   =   \min_{|\psi\>}   F_\psi   (\map M)   =  F   \qquad \forall F \le F_{\max}   \, .
	\end{align}
	Summarising,  Eqs. (\ref{98}) and (\ref{100}) imply that the bound (\ref{cftranspqubit}) is achievable for every  Hamiltonian and for every value of $F$ in the interval $[F_{\min},  F_{\max}]$.
	\subsection{Bound on the nonequilibrium cost of state transmission/state transposition with entanglement binding channels} 
	The bound (\ref{cftranspqubit}), valid for arbitrary quantum channels, implies the bound   
	\begin{align}\label{cfeb}
		c_{\rm transmit,  eb}  (F)   \equiv  c_{\rm transpose,  eb}   (F) \ge   \log\left[ F  + e^{  \frac{ \Delta E}{kT} } \,  \frac{   \left(2 F-1\right)^2}{    (1-F)}  \right]  \, ,
	\end{align}
	on the nonequilibrium cost of every EB channel that transmits or transposes quantum states with  fidelity $F$.  
	We now show that this bound holds with the equality sign for every Hamiltonian and  for every value of the fidelity between $F_{\min}$ and $F_{\max}$.  Indeed, the  channel $\map M$ defined through Eq.  (\ref{defoptimalm}) is entanglement binding for every $F \in  [F_{\min},  F_{\max}]$:  to check this, it is enough to evaluate the partial transpose of the Choi operator $M$ on the output system, which is given by 
	\begin{align}\nonumber
		M^{T_B}   & =   p_{00}  \,  |0\>\<0|  \otimes |0\>\<0|   +   p_{11}  \,  |1\>\<1|  \otimes |1\>\<1|   \\
		&  +       p_{01} \,   |0\>\<0|  \otimes |1\>\<1|    +    p_{10}  \,       |1\>\<1|  \otimes |0\>\<0|   + \sqrt{p_{01}  \,  p_{10}}  \,\Big(  |0\>\<1|  \otimes |0\>\<1|  +   \,  |1\>\<0|  \otimes |1\>\<0|   \Big)\, .
	\end{align}
	The Choi operator is positive if and only if $  p_{01} \,  p_{10} \le  p_{00}  \,  p_{11}$, that is, if and only if    $(2F-1)^2  \le   \frac{3F^2-4F^3}{1-F} $.    This inequality is satisfied in the interval $[  (5-  \sqrt 5)/10  ,  (5+ \sqrt 5)/10]$, which contains the interval $[F_{\min}, F_{\max}]$. 
	
	Summarising, the transposed Choi operator $M^{T_B}$ is positive for every   $F\in  [F_{\min}, F_{\max}]$.  Hence, the channel $\map M$ is entanglement binding.  Furthermore, we observe that, since $M$ is a two-qubit operator, the Peres-Horodecki criterion \cite{peres1996separability,horodecki1997separability} implies that $M$ is separable. In turn, separability of $M$ implies that the channel $\map M$ is entanglement-breaking \cite{horodecki2003entanglement}, or equivalently, that $\map M$ is a measure-and-prepare channel, of the form 
	\begin{align}
		\map M  (\rho)   =  \sum_{i=1}^k \,   \Tr[  P_i\, \rho]\,   \sigma_i \, ,
	\end{align} 
	where $(P_i)_{i=1}^k$ are positive operators representing a quantum measurement, and $(\sigma_i)_{i=1}^k$ are quantum states.  Operationally, this means that the channel $\map M$ can be realised by performing a measurement on the input, and re-preparing the output in the state $\sigma_i$.  
}

\section*{Supplementary Note 8: bound on work extraction}

Here we show how the result of Ref.~\cite{horodecki2013fundamental} on work extraction can be retrieved from our main bound~(\ref{eq:main}). To this purpose, it is useful to review the framework of  Ref.~\cite{horodecki2013fundamental}, where the allowed operations on  system $S$ are obtained from a joint energy-preserving unitary evolution that couples system $S$ with a heat bath $B$ in the thermal state, and with a work register $R$, initially in an energy eigenstate. 
In this framework, obtaining work $\Delta W$ means transforming the state of the work register from  an energy eigenstate with energy $W_{\rm out}$ to an energy
eigenstate with energy $W_{\rm in}$, with $   W_{\rm in}- W_{\rm out} = \Delta W$.    

Crucially, the effective evolution from system $S$ to the composite system $SR$, consisting of the system and the work register, is a quantum channel $\map N$ satisfying the covariance property   $    \map N (  U_t  \rho  U_t^\dag )   =      (U_t \otimes V_t)\, \map N(  \rho)\,    (U_t \otimes V_t)^\dag  $ where $U_t=  e^{-it   H_S/\hbar}$ and $V_t=  e^{-it   H_R/\hbar}$ are  the time evolution operators for the system and for the work register, respectively, and  $t \in  \R$ is an arbitrary evolution time.

Now,  consider the task of extracting work from  an initial state $\rho$.    Since the channel $\map N$ is covariant, the amount of work extracted from $\rho$ is equal to the amount of work extracted from $\rho_t  :=  U_t  \rho U_t^\dag$, and, in turn, is equal to the amount of work extracted from the average state  
\begin{align}
	\<\rho\>     =   \lim_{T\to \infty}    \frac 1{T}   \, \int_{-T/2}^{T/2}    \d t\,   U_t  \rho    U_t^{\dag} \, .
\end{align}

One of the results of Ref.~\cite{horodecki2013fundamental} is that the maximum work extractable from $\rho$ is given by the min relative entropy $D_{\min}  (\<\rho\>  \|  \Gamma)=  \Tr [\Pi  \,  \Gamma]$, where $\Pi$ is the projector on the support of $\<\rho\>$.    In the following, we will retrieve this result  from our main bound~(\ref{eq:main}), by constructing a suitable test.

Let $\map M  (\cdot)  := \Tr_R  [\map N(\cdot)] $ be the quantum channel representing the effective evolution of the system in the transformation that extracts work   $\Delta W$.   
Without loss of generality, we  can assume that  every state $\sigma$ with support contained in the support of $\<\rho\>$ is mapped to the thermal state.   Now, consider the test $\map T$ consisting in applying  channel $\map M$  to the  state $\sigma  :=  \Pi  \Gamma \Pi/\Tr[\Gamma\,  \Pi]$, and then measuring  an observable $O$. The exact choice of observable will turn out to be irrelevant. 

The accuracy measure defined by this test is   \begin{align}
	\map F_{\map T}   (\map M) :=    \Tr [ O  \,    \map M  (  \sigma)]  \equiv \Tr[  O\, \Gamma] \, .
\end{align}  
On the other hand, the reverse entropy of the task specified by the input state $\sigma$ and by the observable $O$ is   \begin{align}
	\nonumber   \kappa_{\map T}  &=   -\log \Tr  [  O\,\Gamma ]        -  D_{\max}  ( \sigma \|  \Gamma ) \\
	\nonumber &  =   -\log \Tr  [  O\,\Gamma ]        - \log  \|   \Gamma^{-1/2} \sigma   \Gamma^{-1/2} \| \\
	\nonumber &  =   -\log \Tr  [  O\,\Gamma ]        -     \log  \left \| \frac{\Pi}{  \Tr[  \Pi \, \Gamma]} \right\|   \\
	\nonumber  &  =   -\log \Tr  [  O\,\Gamma ]    +        \log   \Tr[  \Pi \, \Gamma]  \\
	& =    -\log \Tr  [  O\,\Gamma ]    -  D_{\min}  (\<\rho\>  \|  \Gamma)          \, ,
\end{align}    
where the first equality follows from Eq.~(13) in the main text, the third equality follows from the fact that $\Pi$ and $\Gamma$ commute, and  the last equality follows from the definition of $ D_{\min}$.    
Hence, our main bound~(\ref{eq:main}) becomes $c_{\map T} (F)  \ge  \log  F   + \kappa_{\map T} =  -  D_{\min}  (\<\rho\>  \|  \Gamma)$.   
Since $kT  \ln2 \,  c_{\map T}  (F)$ is a lower bound to the work cost, the work cost of channel $\map M$ is upper  bounded by $ - kT \, \ln 2  \,   D_{\min}  (\<\rho\>  \|  \Gamma)$, meaning that at most work $kT \, \ln 2  \,   D_{\min}  (\<\rho\>  \|  \Gamma)$ can be extracted.

\section*{Supplementary Note 9: erasure with the assistance of a quantum memory}

Consider the task of erasing a quantum system $S$ with the assistance of a quantum memory $Q$~\cite{del2011thermodynamic}.  The task is to reset the state of system $S$ to a fixed pure state $\eta_S$, while preserving the state of the memory $Q$, possibly including  its correlations to an external reference system $R$.  Mathematically, the task can be concisely described as:  transform a given pure state $|\Psi\>_{SQR}$ into the state $\eta_S \otimes \rho_{QR}$, with $\rho_{QR}:  = \Tr_{S} [|\Psi\>\<\Psi|_{SQR}]$, by operating only on the system $S$ and on the memory $Q$. In the following, we will first discuss  the exact case, and then consider its approximate version, making connection with the results of~\cite{del2011thermodynamic}.

In the exact case, the erasure task is equivalent to implementing the state transformation $\rho_{x, SQ}  \mapsto  \eta_S \otimes \rho_{x,Q}$,  $\rho_{x, Q}:  = \Tr_S [\rho_{x,SQ}]$, for every input state $\rho_{x,SQ}$ with support contained in the support of $\rho_{SQ}: = \Tr_R [|\Psi\>\<\Psi|_{SQR}]$.      Let us denote  by $\Pi_{SQ} $ the projector on the support of $\rho_{SQ}$.   Since we are  interested in bounding the work cost, we will assume without loss of generality that the support of $\rho_{SQ}$ is invariant under time translations, that is, $[\Pi_{SQ},  H_{SQ}]=  0$, where $H_{SQ}$ is the joint Hamiltonian of the system and the memory. If this condition is not satisfied, the argument in Supplementary Note 7 shows that replacing the state $\rho_{SQ}$ with its time average $\<\rho_{SQ}\>$ does not affect the work cost.  

Let $\map M$ be any quantum channel that achieves perfect erasure for the states in the support of $\rho_{SQ}$.   Proposition 1 in the main text then implies that the nonequilibrium cost of $\map M$ satisfies the bound  $c  (\map M, \Pi_{SQ})  \ge  D_{\max}   (  \eta_S \otimes \rho_{x,Q} \|  \Gamma_{SQ})     -   D_{\max}    (\rho_{x,SQ}  \|  \Gamma_{SQ})$ for every state $\rho_{x,SQ}$ with support in the given subspace.    In particular, let us choose the input state 
\begin{align}   \widetilde \Gamma_{SQ}  : =  \frac{  \Pi_{SQ}  \Gamma_{SQ}  \Pi_{SQ}}{\Tr[\Pi_{SQ}  \Gamma_{SQ}]} 
	\, , 
\end{align}
where $\Pi_{SQ}$ is the projector on the subspace containing the possible input states.   
With this choice, the bound becomes  $c  (\map M, \Pi_{SQ})  \ge  D_{\max}   (  \eta_S \otimes \gamma_Q \|  \Gamma_{SQ})     -   D_{\max}    (\widetilde \Gamma_{SQ}  \|  \Gamma_{SQ})$, with $\gamma_Q : =  \Tr_S [ \widetilde \Gamma_{SQ}]$.  
In fact, explicit calculation from Eq.~(\ref{easyfaist}) shows that the inequality holds with the equality sign.  Hence, the work cost for implementing the channel $\map M$ upon the support of $\Pi_{SQ}$, denoted by $W(\map M,  \Pi_{SQ})$, satisfies the bound
\begin{align}\label{dwork}\frac {W  (\map M,  \Pi_{SQ})}{kT\ln 2}  \ge  D_{\max}   (  \eta_S \otimes \gamma_Q \|  \Gamma_{SQ})     -   D_{\max}    (\widetilde \Gamma_{SQ}  \|  \Gamma_{SQ})\, .
\end{align}

When the Hamiltonian of system $SQ$ is completely degenerate, the bound becomes 
\begin{align}
	\nonumber  \frac {W  (\map M,  \Pi_{SQ})}{kT\ln 2}  &\ge   \log \|  \gamma_Q\|     + \log  \Tr [\Pi_{SQ}]  \\
	\nonumber &  =  \log \left\|  \Tr_S  [\Pi_{SQ}] \right\|  \\
	&  =:     H_0  (S|Q)_{\rho_{SQ}} \, ,  \label{ddeg}
\end{align} 
where $  H_0  (S|Q)_{\rho_{SQ}}$ is the  conditional   R\'enyi entropy of order  $\alpha=0$.

We now consider  a relaxation of the erasure task where the input state may slightly differ from $\rho_{SQ}$,  and the operation performed by the machine may slightly differ from the ideal erasure operation.     To make connection with the results of  \cite{del2011thermodynamic}, here we take the system to have fully degenerate Hamiltonian.

To define the relaxation, we consider an input state $\widehat \rho_{SQ}$ which deviates from $\rho_{SQ}$ by at most $\epsilon$ with respect to the purified distance~\cite{tomamichel2010duality}, defined as  $P(  \rho,\widehat \rho\,)   :=    \sqrt{ 1  -  F(\rho,\widehat \rho\, )} $, where $F  (\rho ,  \widehat \rho\, )  : = \left(  \|\sqrt{\rho}  \sqrt{\widehat \rho}\,  \|_1  +  \sqrt{(1-\Tr[\rho])(1-\Tr[\widehat\rho \, ])}\right)^2$  is the (generalised) fidelity and $\|  O \|_1  :  =  \Tr [\sqrt{O^\dag O}]$ is the trace norm.  Moreover, we consider a quantum channel  $\widehat {\map M}$ that  implements an approximate  erasure on a purification $\widehat \Psi_{SQR}$ of the state $\widehat  \rho_{SQ}$, that is  
\begin{align}\label{distance}  T   (    ( \widehat{ \map M}  \otimes \map I_R)  (\widehat \Psi_{SQR})  ,   \eta_S \otimes  \Tr_S [ \widehat \Psi_{SQR}]  )   \le \delta \,,
\end{align} where $T  (\rho,\sigma)  :  =  \|  \rho  -\sigma\|_1/2$ is the trace distance.  

For the task of implementing the approximate erasure $\widehat{\map M}$ on the approximate input state $\widehat \rho_{SQ}$, Proposition 1 in the main text implies that the nonequilibrium cost for fully degenerate Hamiltonians  is lower bounded as   $c(\widehat{\map M},  \widetilde \Pi_{SQ})  \ge D_{\max}  (  \widehat{\map M}  ( \widehat \pi_{SQ})  \| I_{SQ}/d_{SQ} )  -    D_{\max}  ( \widehat \pi_{SQ}  \|    I_{SQ}/d_{SQ} )$, where  $d_{SQ}$ is the dimension of system $SQ$, and $\widehat \pi_{SQ}:=    \widehat \Pi_{SQ}/\Tr[\widehat \Pi_{SQ}] $ is the normalised quantum state proportional to the projector $\widehat \Pi_{SQ}$ on the support of $\widehat \rho_{SQ}$.
Equivalently, the bound can be written as
\begin{align}\label{degeneratefaist}
	c(\widehat{\map M},\widehat{\Pi}_{SQ}) & \ge  \left\|  \widehat{ \map M}  ( \widehat\Pi_{SQ}  ) \right\|  \, .        
\end{align}
Comparing the r.h.s. with Eq. (\ref{easyfaist}) one can see that the bound holds with the equality sign.

The rest of our analysis  follows an argument of Ref. \cite{faist2015minimal}, which is reproduced here for completeness  in the scenario where two approximation parameters $\epsilon$ and $\delta$ are used.      Let $\widehat  V:  \spc H_S \otimes \spc H_Q \to \spc H_S\otimes \spc H_Q  \otimes \spc H_E$ be a Stinespring isometry for the channel $\widehat  {\map M}$, so that
$\widehat  {\map M}  ( \widehat\Pi_{SQ} ) =  \Tr_E   [  \widehat V \widehat\Pi_{SQ}   \widehat  V^\dag]$.   Here we choose the environment to have dimension larger than the dimension of system $S$, so that the environment $E$ can also be used to implement the ideal erasure operation by  embedding  the state of system $S$ into $E$. 

In terms of the Stinespring isometry, the bound (\ref{degeneratefaist}) becomes 
\begin{align}
	\nonumber c(\widehat{\map M},  \widehat \Pi_{SQ}) & \ge  \left\|  \Tr_E  [   \widehat V \widehat \Pi_{SQ}   \widehat  V^\dag] \right \|\\
	&  \equiv    H_0   (E|SQ)_{  \widehat V \widehat \rho_{SQ}   \widehat  V^\dag}  \, .        
\end{align}
Since the R\'enyi entropy of order 0 is at least as large as the R\'enyi entropy of order 1/2, one also has
\begin{align}\label{environment}
	c(\widehat{\map M}, \widehat \Pi_{SQ}) & \ge   H_{1/2}   (E|SQ)_{  \widehat V \widehat \rho_{SQ}   \widehat  V^\dag}  \, .        
\end{align}

Now, let $ \widehat \Psi_{EQR} $   be the state obtained from  $ \widehat \Psi_{SQR}  $ by embedding system $S$ into the environment $E$.  Uhlmann's theorem implies that the isometry $V$ can be chosen in such a way that  the fidelity between the pure state $( \widehat V \otimes I_R)    \widehat \Psi_{SQR}     (\widehat  V^\dag \otimes I_R)$ and the pure state $\eta_S\otimes \widehat \Psi_{EQR}   $ is equal to the fidelity between their marginal states  $(\widehat {\map M}\otimes \map I_R)  (\widehat \rho_{SQR})$ and $\eta_S \otimes\widehat \rho_{QR}$, respectively.  Then, Eq.~(\ref{distance}) and  the Fuchs-van de Graaf inequality implies that the fidelity is    at least $(1-\delta)^2$.  Hence, the Fuchs-van de Graaf inequality implies that  the purified distance between the states  $( \widehat V \otimes I_R)    \widehat \Psi_{SQR}     (\widehat  V^\dag \otimes I_R)$ and $\eta_S\otimes \widehat \Psi_{EQR}   $  is at most $\sqrt {2\delta}$. Since the purified distance is nonincreasing under partial trace, one also has   $P  ( \widehat    V   \widehat \rho_{SQ}  \widehat V^\dag,   \eta_S  \otimes \widehat  \rho_{QE} ) \le \sqrt{2 \delta}$, where $\widehat  \rho_{QE}$ is the state obtained from $\widehat  \rho_{SQ} $ by embedding system $E$ into the environment.

The triangle inequality for the purified distance yields the bound  $  P  ( \widehat    V   \widehat \rho_{SQ}  \widehat V^\dag,   \eta_S  \otimes  \rho_{QE} )  \le P  ( \widehat    V   \widehat \rho_{SQ}  \widehat V^\dag,   \eta_S  \otimes \widehat  \rho_{QE} )   +   P  (  \eta_S  \otimes \widehat  \rho_{QE} \otimes   \eta_S  \otimes \rho_{QE} )  \le \sqrt{  2\delta}  +  \epsilon$.     Using this fact, the bound Eq.~(\ref{environment}) can be relaxed to  
\begin{align}
	c(\widehat{\map M},  \widehat \Pi_{SQ}) & \ge     \min_{  \rho_{SQE} : \,     P(  \rho_{SQE},  \eta_S \otimes \rho_{QE} ) \le   \sqrt{  2\delta}  +  \epsilon}      H_{1/2}   (E|SQ)_{  \rho_{SQE}}\\
	& \equiv   H_{1/2}^{  \sqrt{  2\delta}  +  \epsilon  }   (E|SQ)_{  \eta_S \otimes \rho_{QE}}  \, ,        
\end{align}
where $H_{1/2}^{  \sqrt{  2\delta}  +  \epsilon  }$ is the smooth conditional R\'enyi entropy of order $1/2$. Note that  the above expression can be further simplified, as one has the equality $H_{1/2}^{  \sqrt{  2\delta}  +  \epsilon  }   (E|SQ)_{  \eta_S \otimes \rho_{QE}}  =  H_{1/2}^{  \sqrt{  2\delta}  +  \epsilon  }   (E|Q)_{ \rho_{QE}}   =  H_{1/2}^{  \sqrt{  2\delta}  +  \epsilon  }   (S|Q)_{ \rho_{SQ}}$  \cite{faist2015minimal}, which follows from the invariance under isometries of $H_{1/2}$ (cf. Section 5.3 of \cite{tomamichel2015quantum}). 
Since the nonequilibrium cost is a lower bound to the work cost, we obtained the relation  $ \frac {W  (\widehat{\map M},  \widehat{\Pi}_{SQ})}{kT\ln 2}  \le   H_{1/2}^{\epsilon}  (S|Q)_{\rho_{SQ}}$, which coincides with the upper bound  from Ref.
\cite{del2011thermodynamic}, up to logarithmic terms  and to a slight redefinition  of the approximation parameters. 

{  \section*{Supplementary Note 10: state transformation tasks vs individual state transitions}

	Here we discuss the relation between the nonequilibrium cost of a state transformation task $\rho_x\mapsto  \rho_x'$, $\forall x\in\set X$, and the nonequilibrium cost of the individual state transitions  $\rho_x\mapsto \rho_x'$ corresponding to fixed values of $x$.  In particular, we show an example where the cost of each individual state transition is $\le 0$, while the cost of the overall state transformation task is strictly positive.  
	
	\subsection*{Relation between the reverse entropies}
	
	We start by proving the bound on the reverse entropy given in Eq. (19) of the Methods section: 
	
	\begin{prop}
		Let  $\map T$ be an arbitrary state transformation task  $\rho_x\mapsto  \rho_x'$, $\forall x\in\set X$. The reverse entropy  $\kappa_{\map T}$ satisfies the inequality
		\begin{align}\label{mnbv}
			\kappa_{\map T}  \ge   \max_{x\in \set X}   - \log \Tr [  O_x \Gamma_B]    -  D_{\max}  (   \rho_x\|  \Gamma_A ) \, ,   
		\end{align} 
		where $O_x$ is the observable used in the accuracy measure $\map F_{\map T}    (\map M)  =  \min_x  \,  \Tr[  O_x \map M  (\rho_x)]$.
	\end{prop}
	{\bf Proof.}  One has 
	\begin{align}
		\nonumber \kappa_{\map T}     &=  \max_{\st p }   \,  H(A|B)_{\omega_{\map T , \st p}}       \\
		\nonumber  &\ge   H(A|B)_{\omega_{\map T , \st p =  (\delta_{x,  x_0})}}       \\
		\nonumber  &=   H(A|B)_{     \Gamma_A^{-\frac 12} \rho_{x_0}^T  \Gamma_A^{-\frac 12}  \otimes  \Gamma_B^{\frac 12}  O_{x_0} \Gamma_B^{\frac 12} }         \\
		\nonumber &  = -  \log   \min\left \{    \Tr[\Lambda_B]   ~|~      (   I_A \otimes \Lambda_B)  \ge       \Gamma_A^{-\frac 12} \rho_{x_0}^T  \Gamma_A^{-\frac 12} \otimes \Gamma_B^{\frac 12}  O_{x_0} \Gamma_B^{\frac 12}   \right\} \, .  \label{ultimissima}
	\end{align}    
	To evaluate the minimum over $\Lambda_B$, one can take the trace on both sides of the constraint  $(   I_A \otimes \Lambda_B)  \ge       \Gamma_A^{-\frac 12} \rho_{x_0}^T  \Gamma_A^{-\frac 12} \otimes \Gamma_B^{\frac 12}  O_{x_0} \Gamma_B^{\frac 12} $, thus obtaining the operator inequality  
	\begin{align}
		\Tr  [ \Lambda_B]\,  I_A    \ge  \Tr [   O_{x_0} \Gamma_B]  \,      \Gamma_A^{-\frac 12} \rho_{x_0}^T  \Gamma_A^{-\frac 12} \, ,
	\end{align} 
	which in turn implies 
	\begin{align}
		\Tr  [ \Lambda_B]  \ge  \Tr [   O_{x_0} \Gamma_B]  \,      \left\|  \Gamma_A^{-\frac 12} \rho_{x_0}^T  \Gamma_A^{-\frac 12}\right\|      =   \Tr [   O_{x_0} \Gamma_B]  \,      \left\|  \Gamma_A^{-\frac 12} \rho_{x_0}  \Gamma_A^{-\frac 12}\right\|  =       \Tr [   O_{x_0} \Gamma_B]  \,    2^{D_{\max}  (  \rho_{x_0} \|  \Gamma_A)} \, .
	\end{align} 
	The bound is attained with the equality sign by setting $\Lambda_B    = 2^{D_{\max}  (  \rho_{x_0} \|  \Gamma_A)}   \,   \Gamma_B^{\frac 12}  O_{x_0} \Gamma_B^{\frac 12}$.  Hence,   Eq. (\ref{ultimissima})
	becomes 
	\begin{align}
		\nonumber \kappa_{\map T}   & \ge  -  \log   \Tr [   O_{x_0} \Gamma_B]  \,    2^{D_{\max}  (  \rho_{x_0} \|  \Gamma_A)} \\
		&  =  -\log  \Tr [   O_{x_0} \Gamma_B]    -  D_{\max}  (  \rho_{x_0} \|  \Gamma_A)  \qquad \forall x_0 \in \set X \, ,
	\end{align}
	which implies  Eq. (\ref{mnbv}).  \qed  
	
	\medskip 
	
	In the Methods section of the main text, we have seen  a  choice of observables $O_x$ for which Eq.   (\ref{mnbv})  reduces to the entropic  inequality    
	\begin{align}\label{dmaxbound}
		\kappa_{\map T}  \ge \max_{x\in\set X}   D_{\max}   (\rho_x' \|  \Gamma_B)  - D_{\max}   (\rho_x \|  \Gamma_A)    \, .
	\end{align}    
	This inequality has an intuitive physical interpretation, relating the nonequilibrium cost of the task $\map T$ to the deviation of the input and output states from the  equilibrium state.   However, it is important to stress that, in general,   the deviation of the input/output states from the equilibrium state is not sufficient to evaluate the  nonequilibrium cost of a given task. This fact is evident in the state transposition task $\rho_x\mapsto \rho_x^T$,  where $\rho_x$ is an arbitrary pure state:  since $ D_{\max}   (\rho^T \|  \Gamma)  =  D_{\max}   (\rho \|  \Gamma)$ for every quantum state $\rho$, Eq. (\ref{dmaxbound}) yields the trivial inequality $\kappa_{\rm transpose}  \ge 0$,  which is not tight, as we know that   $\kappa_{\rm transpose}  =  \log [(d+1)/2]$ in the fully degenerate case (cf. the Results part in the main text).    
	
	\subsection*{The example of ideal transposition}
	
	For transposition, the nonequilibrium cost is  nonnegative whenever the Hamiltonian is not fully degenerate (cf. Eq.  (\ref{cftransp})).  Hence, the quantity  $\max_x  \Big[D_{\max}  (  \rho_x'  \|  \Gamma)  -  D_{\max}  (  \rho_x  \|  \Gamma) \Big]$  fails to detect the presence of a strictly positive nonequilibrium cost.

	The same issue arises for other quantum versions of the relative entropy, including all the quantum R\'enyi relative entropies 
	\begin{align}
		D_\alpha  (  \rho  \|  \sigma)   =  \frac{ \log  \Tr [  \rho^{\alpha}  \sigma^{1-\alpha}]}{\alpha-1}   \,,  \qquad \alpha  \ge 0  ,  \alpha\not = 1 \, ,
	\end{align} 
	as well as the ``sandwiched" R\'enyi relative entropies  \cite{muller2013quantum,wilde2014strong}
	\begin{align}
		\widetilde D_\alpha (\rho\|\sigma)   =  \frac{\log \Tr  [   (  \sigma^{\frac {1-\alpha}{2\alpha}}  \rho \sigma^{\frac {1-\alpha}{2\alpha}}  )^{\alpha}   ]}{\alpha-1}  \, ,  \qquad \alpha  \ge 0  ,  \alpha\not = 1 \, ,
	\end{align}
	which feature in the quantum second laws of Ref. \cite{brandao2015second} for $\alpha \ge 1/2$.

	In  these cases, one still has the equalities 
	\begin{align}
		D_\alpha  (  \rho^T  \|  \Gamma)  =  D_\alpha  (  \rho  \|  \Gamma)\qquad {\rm and}\qquad \widetilde D_\alpha  (  \rho^T  \|  \Gamma)  =  \widetilde D_\alpha  (  \rho  \|  \Gamma)\quad \forall \alpha\, , \forall \rho \, ,
	\end{align} 
	valid whenever the transpose is defined in the eigenbasis of the Hamiltonian.   Even more generally, the relation  $\Delta (  \rho^T  \|  \Gamma)  =  \Delta (  \rho  \|  \Gamma) \, ,\forall \rho$ holds for every function $\Delta   (\rho\| \sigma)$ that is invariant under state space symmetries in the sense of Wigner's theorem (see e.g. \cite{chiribella2021symmetries}).  

	\subsection*{The example of approximate transposition} 
	
	A natural question is whether the failure of the relative entropy  to characterise the nonequilibrium cost of the transpose task is due to the fact that the ideal transposition  $\rho_x \mapsto\rho_x^T$ is forbidden by quantum mechanics \cite{bu2000universal,horodecki2003limits,buscemi2003optimal}.   Instead of ideal transposition, one could consider some physical approximation of this impossible transposition task.  The approximate transposition would then correspond to a valid quantum channel $\map M$.   The question is whether the change of relative entropy from the input to the output  provides the correct value of the nonequilibrium cost.   Here we show that, in general, the answer is negative.

	To facilitate the evaluation of the relative entropies, we consider a restricted transposition task, where the input states are pure qubit states in the set $\set S   = \{  |0\>\<0| \, , |1\>\<1| \}  \cup  \{   |e_\theta\>\<e_\theta|  \}_{\theta \in  [0,2\pi)}$, consisting of the computational basis states $|0\>$ and $|1\>$ and of the equatorial states $|e_\theta\>:  =   (|0\>  +  e^{i\theta})/\sqrt 2$.  In this case, the nonequilibrium cost is still positive for every nondegenerate Hamiltonian: for $d=2$,   Eq. (\ref{cftransp}) indicates a positive value of the nonequilibrium cost for every value of the fidelity above $F_{\min}$. In particular, we will focus on the maximum value $F_{\max}  = 2/3$, which is achieved by the channel   $\map M$ defined by \cite{bu2000universal,horodecki2003limits,buscemi2003optimal}
	\begin{align}\label{optimaltransposechannel}
		\map M  (\rho)   =  \frac { I+  \rho^T }3  \qquad \forall \rho \, .
	\end{align} 
	We will restrict our attention to the case where $  \Gamma  =  \frac 23  |0\>\<0|  + \frac 13  |1\>\<1|$, which considerably simplifies the calculations.   In this case, one has 
	\begin{align}
		D_\alpha  ( \map M( |0\>\<0| )  \,    \| \,  \Gamma)  -  D_\alpha  (  |0\>\<0|   \, \| \, \Gamma)    & =   -  D_\alpha  (  |0\>\<0|   \, \| \, \Gamma)   =  \log \frac 23     <0     \\
		D_\alpha  ( \map M( |1\>\<1| )  \,    \| \,  \Gamma)  -  D_\alpha  (  |1\>\<1|   \, \|  \,\Gamma)    &   = \frac{  \log  [ \left( \frac 23\right)^{\alpha}  + \left( \frac 13\right)^{\alpha} \,  2^{1-\alpha}   ]}{  \alpha-1}  \le   -  H_\alpha  \left(\frac 23 , \frac 13\right) < 0   \\
		D_\alpha  ( \map M( |e_\theta\>\<e_\theta| )  \,    \| \,  \Gamma)  -  D_\alpha  (  |e_\theta\>\<e_\theta|   \, \|\,  \Gamma)  &=  -  H_\alpha  \left(\frac 23 , \frac 13\right) < 0   \, ,  \qquad \forall \theta \in  [0,2\pi)\,,
	\end{align}
	where $H_\alpha  \left(\frac 23 , \frac 13\right) :=(1-\alpha)^{-1} \,  \log  \left[ \left( \frac 23\right)^{\alpha}  + \left( \frac 13\right)^{\alpha} \right] $ is the  R\'enyi entropy of the binary probability distribution $(2/3,1/3)$.  Hence, the quantum R\'enyi relative entropies  do not detect the positive work cost of the approximate transposition $\rho_x  \mapsto  \map M  (\rho_x)$ for all the states in   $\set S$. 
	
	Let us now consider the sandwiched R\'enyi relative entropy   $\widetilde D_\alpha$. For the states $|0\>$ and $|1\>$ the sandwiched R\'enyi relative entropies coincide with the quantum R\'enyi entropies. 
	Hence, we only need to consider the case of the equatorial states $|e_\theta\>$.   For $0\le \alpha <1$, we use the monotonicity of  $ \widetilde D_\alpha$ with respect to $\alpha$ \cite{muller2013quantum}  and the convergence to the von Neumann relative entropy  $\lim_{\alpha \to  1}  \widetilde D_\alpha   (\rho\|\sigma)  =  D(\rho\|\sigma)  :=  \Tr [\rho\log \rho] - \Tr[\rho \log \sigma] \, \forall \rho, \sigma$ \cite{muller2013quantum}, which yield the bound 
	\begin{align}
		\widetilde D_\alpha   (    \map M  (|e_\theta\>\<e_\theta|)   \,  \| \,  \Gamma)   &\le  D   (    \map M  (|e_\theta\>\<e_\theta|)   \,  \| \,  \Gamma)   = \frac 16 \, 
		\, .
	\end{align}
	On the other hand, one has 
	\begin{align}
		\nonumber  - \widetilde D_\alpha   (    |e_\theta\>\<e_\theta|   \,  \| \,  \Gamma)     &=      \log  \left[ \frac{\left( \frac  23\right)^{\beta }  +  \left( \frac  13\right)^{\beta}}{2} \right]^{\frac1{\beta}}  \qquad \beta  :  =  \frac{  1-\alpha}\alpha\\
		&  \le \log \frac 23  \, .
	\end{align}
	Hence, we obtained the bound   
	\begin{align}
		\widetilde D_\alpha   (    \map M  (|e_\theta\>\<e_\theta|)   \,  \| \,  \Gamma)    - \widetilde D_\alpha   (    |e_\theta\>\<e_\theta|   \,  \| \,  \Gamma)    \le \frac  16   +  \log \frac 23 \approx  -0.418  < 0   \, .
	\end{align}

	For $\alpha >  1$, we use the fact that the sandwiched divergence $\widetilde D_\alpha$ satisfies the data processing inequality  $\widetilde D_\alpha  ( \map C  (\rho) \, \|  \,  \map C  (\sigma) )\le \widetilde D_\alpha (\rho \|  \sigma)$ for every quantum channel $\map C$ and for every pair of states $\rho$ and $\sigma$ \cite{muller2013quantum,wilde2014strong,beigi2013sandwiched,frank2013monotonicity}.    
	In particular, choosing $\map C $ to be the  partial dephasing channel   $\map C (\rho)  :  =  \frac 23  \,\rho +   \frac 13  \sigma_z  \rho  \sigma_z$  (with $\sigma_z :  =  |0\>\<0|  -  |1\>\<1|$),   we obtain  the bound
	\begin{align}
		\nonumber \widetilde D_\alpha   (    \map M  (|e_\theta\>\<e_\theta|)  \,  \| \,  \Gamma)    & =   \widetilde D_\alpha   (    \map C  (|e_\theta\>\<e_\theta| ) \,  \| \,  \map C(\Gamma)  )  \le  \widetilde D_\alpha   (    |e_\theta\>\<e_\theta|  \,  \| \, \Gamma )  \,,
	\end{align}
	and therefore  $\widetilde D_\alpha   (    \map M  (|e_\theta\>\<e_\theta|)   \,  \| \,  \Gamma)    -    \widetilde D_\alpha   (    |e_\theta\>\<e_\theta|   \,  \| \,  \Gamma)   \le  0$. 
	
	Summarising, the change of sandwiched R\'enyi relative entropy from the input to the output  is negative for every possible pair of states in $\set S$.  Hence, the sandwiched R\'enyi relative entropies cannot  detect the positive nonequilibrium cost of the approximate transposition $\rho_x  \mapsto  \map M  (\rho_x)$.  
	
	The above conclusions hold also if one considers the entropy of the Gibbs state relative to the input/output states, instead of the entropy  of the input/output states relative to the Gibbs state.  In this case, some care is required when dealing with the relative entropy $D_\alpha   (  \rho \|  |\psi\>\<\psi|)$ for a pure state $|\psi\>$.    Here we adopt the definition
	\begin{align}
		D_\alpha   (  \rho \|  |\psi\>\<\psi|)     :=   \lim_{\epsilon  \to 0 }   \,   D_\alpha   \left(  \rho \left\|   (1-\epsilon)\,   |\psi\>\<\psi|+  \epsilon  \, \frac{I-  |\psi\>\<\psi|}{d-1} \right)   \right.   &  =  
		\left\{
		\begin{array}{ll}  
			\frac{ \log \<\psi|  \rho^\alpha |\psi\> }{\alpha-1}  \qquad \qquad  &  0\le\alpha  <  1  \\
			\infty  \qquad  & \alpha  >  1  \, . 
		\end{array}
		\right.
	\end{align}
	
	For the quantum R\'enyi relative entropies, one has the relations 
	\begin{align}
		D_\alpha  ( \Gamma \,  \|\,   \map M( |0\>\<0| ) )  -  D_\alpha  ( \Gamma\,  \| |0\>\<0|  )    & =   -  D_\alpha  (  \Gamma \, \| \,  |0\>\<0|   )    = 
		\left\{ 
		\begin{array}{ll}   -   \frac{ \alpha }{1-\alpha}  \log \frac 32     \qquad  \qquad & 0\le  \alpha  <1\\
			-\infty   & \alpha  > 1  
		\end{array}  \right.   \\
		D_\alpha  ( \Gamma \,  \|\,   \map M( |1\>\<1| ) )  -  D_\alpha  ( \Gamma\,  \| |1\>\<1|  )    & =   
		\left\{ 
		\begin{array}{ll}    -   \log \frac32   -  \frac{  1}{1-\alpha}  \log  (1+ 2^{2\alpha})   \qquad   \qquad & 0\le  \alpha  <1\\
			-\infty   & \alpha  > 1  
		\end{array}  \right.   \\
		D_\alpha   (  \Gamma \|   \map M  (|e_\theta\>\<e_\theta|)  ) -  D_\alpha   (  \Gamma \|   |e_\theta\>\<e_\theta|  )  &=  
		\left\{ 
		\begin{array}{ll}   -  H_{1-\alpha}   \left(\frac 32, \frac 12\right) <  0  \qquad \qquad & 0\le \alpha  <  1  \\
			-\infty    & \alpha > 1  
		\end{array}
		\right.
	\end{align} All these values are strictly negative. 
	
	The same conclusion applies to the sandwiched R\'enyi relative entropies,  adopting the definition  
	\begin{align}
		\widetilde D_\alpha   (  \rho \|  |\psi\>\<\psi|)     :=   \lim_{\epsilon  \to 0 }   \,  \widetilde D_\alpha   \left(  \rho \left\|   (1-\epsilon)\,   |\psi\>\<\psi|+  \epsilon  \, \frac{I-  |\psi\>\<\psi|}{d-1} \right)   \right.   &  =  
		\left\{
		\begin{array}{ll}  
			\frac{ \alpha}{\alpha-1}\, \log \<\psi|  \rho |\psi\>   \qquad \qquad  &  0\le\alpha  <  1  \\
			\infty  \qquad  & \alpha  >  1    \, .
		\end{array}
		\right.
	\end{align}
	(A more common definition is $\widetilde D_\alpha   (  \rho \|  |\psi\>\<\psi|)  =  \infty$ whenever $\Supp (\rho) \not \subseteq \Supp (|\psi\>\<\psi|)$. In this case, the  difference  $\widetilde D_\alpha   (  \Gamma \|    \map M  (  |e_\theta\>\<e_\theta|)   \, )  - \widetilde D_\alpha   (  \Gamma \|    |e_\theta\>\<e_\theta|     \, ) $  is trivially $-\infty$, and therefore cannot detect the positive nonequilibrium cost of the transposition task.)  
	
	We focus on the case of the equatorial states $|e_\theta\>$, because for the states $|0\>$ and $|1\>$ the sandwiched entropies coincide with the quantum R\'enyi relative entropies.   For  $0\le \alpha <  1$, we use the equality 
	\begin{align}
		\widetilde D_\alpha   ( \Gamma \|    |e_\theta\>\<e_\theta|   )     =  \frac{ - \alpha}{\alpha-1} \, ,
	\end{align}
	and the inequality  
	\begin{align}
		\widetilde D_\alpha   ( \Gamma \|    \Sigma \, )     =    \frac{1}{\alpha-1}   \log \Tr [   \left(  \Sigma^{\frac{1-\alpha}{2\alpha}}  \,  \Gamma  \,   \Sigma^{\frac{1-\alpha}{2\alpha}}\right)^{\alpha}   ]   \le   \frac{1}{\alpha-1}   \log \Tr [   \left(\frac 13  \,    \Sigma^{\frac{1-\alpha}{\alpha}} \right)^{\alpha} ]  \qquad \forall \Sigma\, ,  
	\end{align}
	following from the operator monotonicity of the function $f(X) =  X^\alpha$ and from the operator inequality $ \Sigma^{\frac{1-\alpha}{2\alpha}}  \,  \Gamma  \,   \Sigma^{\frac{1-\alpha}{2\alpha}}  \ge \frac 13  \,  \Sigma^{\frac{1-\alpha}{\alpha}}$, and from the fact that $\alpha$ is smaller than 1.     Setting $\Sigma  =  \map M  (  |e_\theta\>\<e_\theta|)$ we obtain  
	\begin{align}
		\nonumber \widetilde D_\alpha   (  \Gamma \|    \map M  (  |e_\theta\>\<e_\theta|)   \, )  - \widetilde D_\alpha   (  \Gamma \|    |e_\theta\>\<e_\theta|     \, )   &\le \frac{1}{\alpha-1}  \log \left[  \frac 1{3^{\alpha}}  \,  \left(  \left(  \frac 23  \right)^{1-\alpha}    +  \left(  \frac 13  \right)^{1-\alpha}   \right)  \right]  +  \frac{  \alpha}{\alpha-1}  \\
		\nonumber &   =   \frac{1}{\alpha-1}  \log \left[  \left(\frac 23\right)^{\alpha}  \,  \left(  \left(  \frac 23  \right)^{1-\alpha}    +  \left(  \frac 13  \right)^{1-\alpha}   \right)  \right] \\
		\nonumber &   =   \frac{1}{\alpha-1}  \log \left[    \frac 23    +  2^{\alpha}   \,  \frac 13   \right] \\
		&  <0 \, .
	\end{align}
	For $\alpha  >1$,  one has        
	$\widetilde D_\alpha   (  \Gamma \|    \map M  (  |e_\theta\>\<e_\theta|)   \, )   < \infty$ and  $\widetilde D_\alpha   ( \Gamma \|    |e_\theta\>\<e_\theta|   ) =  \infty$, and and therefore the difference is $-\infty$. }

\end{document}